\documentclass[twocolumn]{aastex631}

\usepackage{amsmath}

\usepackage{multirow}
\usepackage{makecell}
\usepackage{array}

\usepackage{lmodern}
\usepackage{booktabs}
\usepackage{rotating}
\begin{document}
\defcitealias{Johnstone2021}{J21}
\defcitealias{KingWheatley2021}{KW21}

\title{Separating Super-Puffs vs.~Hot Jupiters Among Young Puffy Planets}

\author[0000-0003-0525-1805]{Amalia Karalis}
\affiliation{Department of Physics and Trottier Space Institute, McGill University, 3600 rue University, H3A 2T8 Montreal QC, Canada}
\affiliation{Trottier Institute for Research on Exoplanets (iREx), Universit\'e de Montr\'eal, Canada}

\author[0000-0002-1228-9820]{Eve J. Lee}
\affiliation{Department of Physics and Trottier Space Institute, McGill University, 3600 rue University, H3A 2T8 Montreal QC, Canada}
\affiliation{Trottier Institute for Research on Exoplanets (iREx), Universit\'e de Montr\'eal, Canada}
\affiliation{Department of Astronomy \& Astrophysics, University of California, San Diego, La Jolla, CA 92093-0424, USA}

\author[0000-0002-5113-8558]{Daniel P. Thorngren}
\affiliation{Department of Physics \& Astronomy, Johns Hopkins University, Baltimore, MD, USA}

\begin{abstract}

Discoveries of close-in young puffy (R$_{\rm p} \gtrsim$ 6 R$_\oplus$) planets raise the question of whether they are bona fide hot Jupiters or puffed-up Neptunes, potentially placing constraints on the formation location and timescale of hot Jupiters. Obtaining mass measurements for these planets is challenging due to stellar activity and noisy spectra. Therefore, we aim to provide independent theoretical constraints on the masses of these young planets based on their radii, incident fluxes, and ages, benchmarking to the planets of age $<$1 Gyr detected by \textit{Kepler}, \textit{K2} and \textit{TESS}. Through a combination of interior structure models, considerations of photoevaporative mass loss, and empirical mass-metallicity trends, we present the range of possible masses for 22 planets of age $\sim$10-900 Myr and radii $\sim$6-16 R$_\oplus$.
We generally find that our mass estimates are in agreement with the measured masses and upper limits where applicable. There exist some outliers including super-puffs Kepler-51 b, c and V1298 Tau d, b, e, for which we outline their likely formation conditions.
Our analyses demonstrate that most of the youngest planets ($\lesssim$ 100 Myr) tend to be puffed-up, Neptune-mass planets, while the true hot Jupiters are typically found around stars aged at least a few hundred Myr, suggesting the dominant origin of hot Jupiters to be late-stage high eccentricity migration.

\end{abstract}

\section{Introduction} \label{sec:intro}
While hot Jupiters are one of the first detected exoplanets \citep{Mayor1995}, their origin remains elusive. Three hypotheses have been proposed---in situ formation, disk-induced migration, and high-eccentricity migration---of varying degree of success \citep[see, e.g.,][for review]{Dawson2018}.
In situ formation postulates that the entire formation process of the Jupiter-sized planet would occur close to where it is observed \citep[e.g.,][]{Batygin2015}. 
Forming such a large planet requires solid coagulation into a core of at least $\sim$10 M$_\oplus$ \citep[see, e.g.,][]{Lee2014, Piso2015} to trigger runaway gas accretion well before the disk gas dissipates. While core growth timescales are very short for planets within $\sim$ 0.1 AU, small feeding zones and low surface densities limit the maximum core size. \citet{Dawson2018} estimate that for a typical feeding zone width of 7 Hill radii and disk surface density of $10^3$ g/cm$^2$ \citep[e.g.,][]{Greenzweig1990} around a solar-mass star, the maximum core mass within 0.1 au is $\sim$0.05 M$_\oplus$. Even under pebble accretion, pebble isolation mass at such close-in orbits is expected to be $\lesssim$2$M_\oplus$ \citep{Bitsch2018, Fung2018}. It is therefore unlikely that hot Jupiters form close-in.

The other two theories propose that the planet forms at much larger distances and then migrates close to its host star.
Under the theory of disk-induced migration, planets are expected to excite waves as they tidally interact with the surrounding disk gas and the perturbed gas torques back on the planet \citep{Goldreich1980, Lin1986}. While the direction of the net torque depends sensitively on the thermal state of the disk, in typical disks, the torque is such that it drives the net inward motion of the planet \citep{Kley2012}. More massive planets drive greater perturbation so much so that for massive planets like gas giants, they are expected to carve out a deep gap in their vicinity \citep[e.g., ][]{Fung2014}, muting the gas feedback torque on the planet and slowing down migration \citep{Duffell2014, Kanagawa2018}. By contrast, high-eccentricity migration proposes that a post-disk era dynamical perturbation by a companion excites the gas giant into a highly elliptical orbit followed by tidal circularization and orbital decay \citep[e.g.,][]{Wu2003}.

Using orbital properties of hot and warm Jupiters to rule out one theory from another has been challenging. For example, non-zero obliquities of hot Jupiters may be a signature of Kozai-Lidov oscillation in driving high-eccentricity migration but migration in a warped disk may also produce the same non-zero obliquities (\citealt{Batygin2012}; but see \citealt{Zanazzi2018}). If warm Jupiters are en route to becoming hot Jupiters, we may expect many of them to be on an eccentric orbit under the high-eccentricity migration scenario but the observed eccentricity distribution of warm Jupiters is closer to being uniform. While the warm Jupiters on circular orbits can naturally be explained by those that underwent disk-induced migration (via gas dynamical friction), it has also been shown that a broad eccentricity distribution is realized in warm Jupiters undergoing secular eccentricity oscillations excited by neighboring planets \citep{Petrovich2016}. Another puzzling aspect of warm Jupiters is that approximately half of them are found with companion sub-Neptunes \citep{Huang2016}, a configuration that is difficult to explain by high-eccentricity migration without triggering orbital instabilities \citep{Antonini2016}.

Looking at more general observed properties such as demographic trends, the upper edge of the sub-Jovian desert has been shown to be well-described by the expectations of high-eccentricity migration whereby planets park at twice their periapse and therefore twice the Roche-lobe radius to ensure the survival of planets \citep{Matsako2016}. \citet{Owen2018} further showed that the subsequent tidal decay can explain a subset of hot Jupiters that lie between 1--2 of their Roche lobe radius. Focusing on the more general orbital period distribution, \citet{Hallatt2020} found that disk-induced migration can explain the occurrence rate of hot and warm Jupiters as a function of their orbital periods, including the observed peak of hot Jupiters at $\sim$3 days. They note however that the corresponding mass function of hot Jupiters is too bottom-heavy compared to what is observed.

In this paper, we instead turn our attention to the ages of hot Jupiters to test whether most hot Jupiters have underwent disk-induced migration or high-eccentricity migration. Disk-induced migration would halt when the disk gas dissipates, over timescales of a few Myr \citep[see, e.g.,][]{Mamajek2009, Michel2021}. By contrast, high eccentricity tidal migration occurs over longer dynamical timescales of $\gtrsim$100 Myr \citep[see, e.g., ][]{Naoz2011, Petrovich2015}. Given the clear divide in formation timescales, measuring the relative population of hot Jupiters across the system age of $\sim$100 Myr would provide a strong test of whether they arrived at their present location by disk migration or high-eccentricity migration.
The detection of young ($<$ 1 Gyr), puffy (R $\gtrsim$ 6 R$_\oplus$) planets from the \textit{Transiting Exoplanet Survey Satellite (TESS)} and \textit{Kepler/K2} indicates the existence of a candidate population of Jupiter-sized planets.
At these young ages, however, care must be taken in inferring the mass of the planet from their size as the planetary envelopes are still undergoing contraction by cooling \citep[e.g.,][]{Owen2020}.

This paper is organized as follows. Section \ref{sec:methods} describes our target selection and methods used to obtain theoretical mass estimates for the planets in our sample using a combination of interior structure models, considerations of photoevaporative mass loss, and empirical mass-metallicity trends. In Section \ref{sec:results}, we present the expected range of masses for our target planets and how they compare with the measured masses for 20 for which such measurements exist. In Section \ref{sec:discussion}, we discuss the outliers, including Kepler-51 and V1298 Tau planets, identifying their potential formation conditions. A summary and conclusions are presented in Section \ref{sec:conclusion}. 

\section{Methods} \label{sec:methods}
Our goal is to constrain the masses of the observed young Jupiter-sized planets to determine whether they are bonafide Jupiter-mass objects or puffy low-mass objects. In Section \ref{sec:sample}, we describe our target list of young planets and the selection criteria we use to build our list. The interior structure model we adopt to identify a plausible mass range for each planet in our sample is detailed in Section \ref{sec:int_struc_model}.
Often, our interior models converge onto two {\it families} of solutions, one at the lower mass end and the other at the higher mass end.
In Section \ref{sec:BD_sol}, we describe how we rule out the high mass family of solutions by comparing to the brown dwarf mass-radius-age curves. 
We further narrow the range of plausible mass in our low mass family of solution by constraining its lower limit against photoevaporative mass loss, as described in Section \ref{sec:ML}, and its upper limit against the known planetary and stellar metallicity trends, as described in Section \ref{sec:met_constraint}.

\begin{figure}
    \includegraphics[width=.473\textwidth]{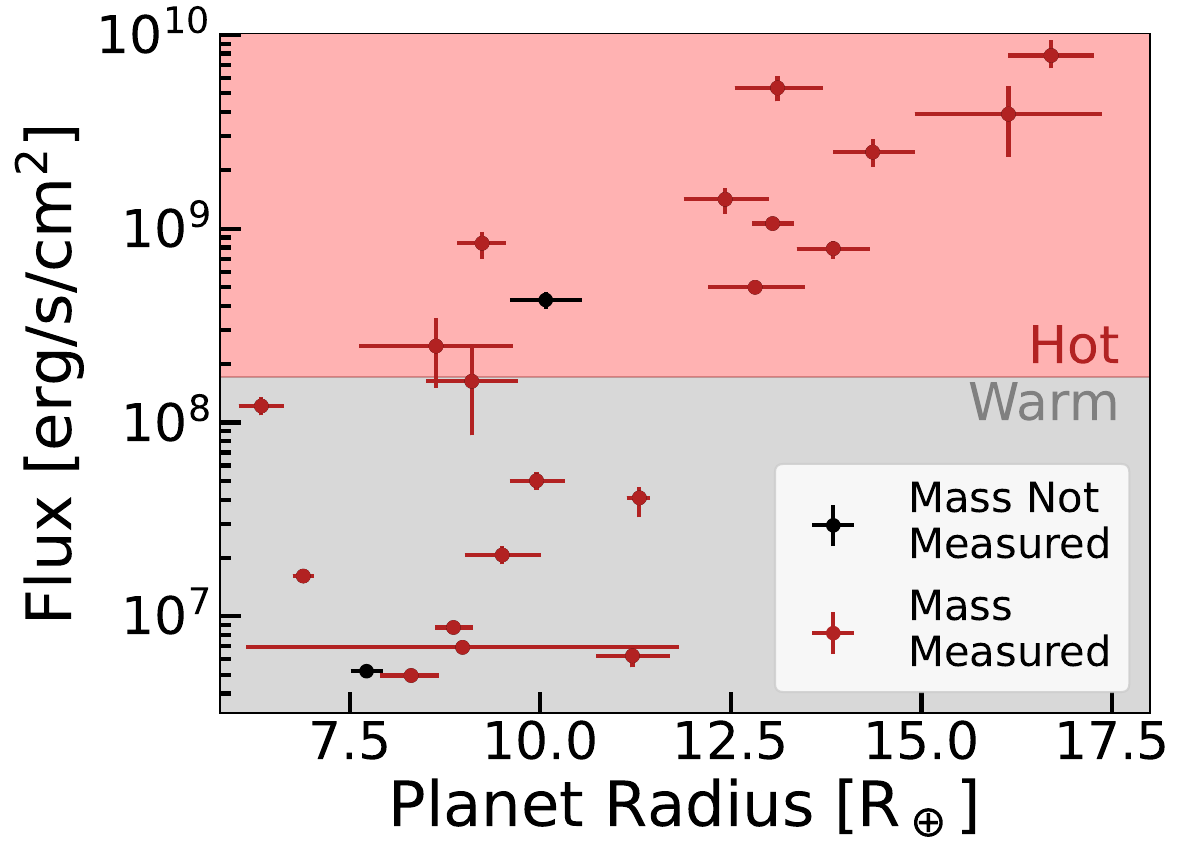}
        \centering
    \caption{The incident bolometric fluxes and radii of the sample of planets studied in this work. The planets shown in red have measured masses, while the ones shown in black do not. The grayed-out region indicates the planets that could potentially be warm Jupiters, while the red region indicates those that could be hot Jupiters. We consider hot and warm Jupiters to be planets 
    with radii $\geq$6$R_\oplus$ and
    with an incident flux above that from a Sun-like star at a period of 10 days ($f_\star \approx 1.72\times10^8 \mathrm{erg/s/cm}^2$) and between 10 and 200 days ($f_\star \approx 3.17\times10^6 \mathrm{erg/s/cm}^2$), respectively.}
    \label{fig:flux_rpl}
\end{figure}

\subsection{Target Selection} \label{sec:sample}
 
We select confirmed planets younger than 1 Gyr with radii larger than 6$R_\oplus$ and incident bolometric flux above 3.17$\times10^6 \mathrm{erg/s/cm}^2$ ($\sim$2.3$\times$ Earth insolation flux), observed by {\it TESS} \citep{Ricker2015_TESS}, {\it Kepler} \citep{Borucki2016}, and {\it K2} \citep{Howell2014} from \citet{TESS_archive, K2_archive, Kepler_archive}, hereafter `Archive'. We further remove Kepler-1976 and Kepler-302 from our sample. Their default ages are drawn from the nominal value in the Q1--Q8 KOI table \citep{Burke14}. The age of Kepler-302 has since been updated to 6.61 Gyr \citep{Morton16}. Recently, \citet{Bouma24} revisited the ages of planetary systems in the {\it Kepler} field younger than 4 Gyr and found no detection of stellar rotation for Kepler-1976 and Kepler-302, suggesting that their ages are likely beyond a few Gyr, too old for our selection criteria.

Of the 19 planet-hosting stars in our sample, 3 have age estimates below 100 Myr: V1298 Tau, HIP 67522, and TOI-837. These youngest stars are the ones for which it is most important to have a reliable age estimate. 
These young stars all belong to young clusters with known ages.
V1298 Tau, part of the stellar association Group 29, also has lithium content and rotational period measurements that confirm and further constrain the stellar age \citep{V1298Tau_met}.
For HIP 67522, part of the Scorpius-Centaurus OB association, the effective temperature and luminosity of the star are fit to stellar isochrones PARSEC 1.2 s \citep{Bressan2012} and BHAC15 \citep{Baraffe2015} to determine the stellar age and mass \citep{HIP67522_pl_st}. 
The age estimate for TOI-837, part of the open cluster IC 2602, is taken to be an absolute range encompassing the results from age estimates obtained through different stellar isochrone fittings and lithium aging techniques \citep[see Table 3 of][]{TOI-837_pl_st}. 
Given that the quoted age estimate for each of the above systems is verified against multiple independent techniques,
we consider the ages of these stars to be reliable and continue with our analysis.

Some planets have multiple parameter sets available, so we prioritize data sets with planetary mass measurements and otherwise choose the default parameter from the Archive.
We further require that all planets studied have the necessary information available to obtain incident bolometric flux values such as the bolometric luminosity of the host star and the planet's semi-major axis. 
If these two parameters are not available, then we estimate the stellar luminosity from the stellar effective temperature and radius using the Stephan-Boltzmann law. Likewise, we calculate the planet's semi-major axis from the orbital period and stellar mass using Kepler's third law.
We obtain an uncertainty on all flux values by propagating the errors on the measured values.

Based on our selection criteria, we retrieve 20 planets with known mass measurements and 2 planets either without a mass measurement or only with upper limits on their mass,
as illustrated in Figure \ref{fig:flux_rpl}.
We label hot Jupiters as those with incident bolometric flux $\gtrsim$1.72$\times 10^8\,{\rm erg\,s^{-1}\,cm^{-2}}$, equivalent to solar insolation flux at 10 days. We additionally label warm Jupiters as those with incident bolometric flux that range 1.72$\times 10^8$ and 3.17$\times 10^6 \,{\rm erg\,s^{-1}\,cm^{-2}}$, equivalent to solar insolation flux at 200 days.

\begin{table*}[ht]
    \centering
    \renewcommand{\arraystretch}{1.15}
    \begin{tabular}{l c c c c c c c}
        \toprule
        \textbf{Planet} & \textbf{M$_p$ [M$_\oplus$]} & \textbf{R$_p$ [$R_\oplus$]} & \textbf{f$_\star$ [erg/s/cm$^2$]} & \textbf{a [au]} & \textbf{P [days]} & \textbf{Extra Reference} \\
        \midrule
        V1298 Tau b & $13.10^{+5.30}_{-5.30}$ & $9.95^{+0.37}_{-0.35}$ & ${5.01^{+0.53}_{-0.52}\times 10^7}$ & ${0.172}$ & $24.14$ & \multirow{3}{4cm}{\centering Livingston, et al.,\\in preparation} \\
        V1298 Tau d & $6.00^{+0.70}_{-0.70}$ & $6.34^{+0.30}_{-0.30}$ & ${1.22^{+0.13}_{-0.13}\times 10^8}$ & ${0.110}$ & $12.40$ & \\
        V1298 Tau e & $15.30^{+4.20}_{-4.20}$ & $9.50^{+0.51}_{-0.49}$ & ${2.07^{+0.22}_{-0.22}\times 10^7}$ & ${0.267}$ & $46.77$ & \\
        HATS-36 b & $1022.14^{+19.71}_{-19.71}$ & $13.84^{+0.48}_{-0.48}$ & ${7.89^{+0.79}_{-0.88}\times 10^8}$ & ${0.054}$ & $4.18$ &  --\\
        HIP 67522 b & $<$20 & $10.07^{+0.47}_{-0.47}$ & ${4.28^{+0.41}_{-0.42}\times 10^8}$ & ${0.076}$ & $6.96$ &  \citet{Thao2024}\\
        TOI-837 b & $120.62^{+18.46}_{-19.41}$ & $8.63^{+1.01}_{-1.01}$ & ${2.48^{+0.97}_{-0.97}\times 10^8}$ & ${0.083}$ & $8.32$ & \citet{Barragan2024}\\
        TOI-1268 b & $96.40^{+8.20}_{-8.30}$ & $9.10^{+0.60}_{-0.60}$ & ${1.63^{+0.78}_{-0.77}\times 10^8}$ & ${0.071}$ & $8.16$ &  --\\
        TOI-1431 b & $991.62^{+57.21}_{-57.21}$ & $16.70^{+0.56}_{-0.56}$ & ${7.82^{+1.56}_{-1.09}\times 10^9}$ & ${0.046}$ & $2.65$ &  --\\
        TOI-2046 b & $731.01^{+88.99}_{-88.99}$ & $16.14^{+1.23}_{-1.23}$ & ${3.90^{+1.55}_{-1.55}\times 10^9}$ & ${0.027}$ & $1.50$ &  --\\
        TOI-4087 b & $232.01^{+44.50}_{-44.50}$ & $13.05^{+0.28}_{-0.27}$ & ${1.06^{+0.07}_{-0.06}\times 10^9}$ & ${0.045}$ & $3.18$ &  --\\
        TOI-2152 A b & $899.45^{+120.77}_{-117.60}$ & $14.36^{+0.56}_{-0.52}$ & ${2.48^{+0.40}_{-0.40}\times 10^9}$ & ${0.051}$ & $3.38$ &  --\\
        TOI-201 b & $133.49^{+15.89}_{-9.53}$ & $11.30^{+0.14}_{-0.17}$ & ${4.09^{+0.57}_{-0.83}\times 10^7}$ & ${0.300}$ & $52.98$ &  --\\
        TOI-622 b & $96.30^{+21.93}_{-22.88}$ & $9.24^{+0.31}_{-0.33}$ & ${8.41^{+1.25}_{-1.41}\times 10^8}$ & ${0.071}$ & $6.40$ &  --\\
        KOI-1783.01 & $71.00^{+11.20}_{-9.20}$ & $8.86^{+0.25}_{-0.24}$ & ${8.76^{+0.38}_{-0.38}\times 10^6}$ & ${0.513}$ & $134.46$ &  -- \\
        Kepler-51 b & $6.60^{+2.70}_{-2.70}$ & $6.89^{+0.14}_{-0.14}$ & ${1.61^{+0.07}_{-0.07}\times 10^7}$ & ${0.251}$ & $45.15$ &  \multirow{2}{4cm}{\centering \citet{Masuda24}$^\star$}\\
        Kepler-51 c & $6.07^{+0.42}_{-0.42}$ & $8.98^{+2.84}_{-2.84}$ & ${6.92^{+0.30}_{-0.28}\times 10^6}$ & ${0.384}$ & $85.31$ &    \\
        KOI-351 g & -- & $7.72^{+0.21}_{-0.21}$ & ${5.21^{+0.16}_{-0.16}\times 10^6}$ & ${0.710}$ & $210.61$ &  \citet{Kepler-351_plR}$^{\dagger \dagger}$\\
        Kepler-76 b & $638.84^{+117.60}_{-111.24}$ & $13.11^{+0.59}_{-0.56}$ & ${5.32^{+0.80}_{-0.79}\times 10^9}$ & ${0.027}$ & $1.54$ & \citet{Kepler-302+76_plR}$^{\dagger \dagger}$\\
        Kepler-289 c & $^{\dagger \dagger \dagger}$ $157.00^{+7.00}_{-7.00}$& $11.21^{+0.50}_{-0.47}$ & ${6.26^{+0.80}_{-0.81}\times 10^6}$ & ${0.510}$ & $125.85$ & \citet{Kepler-302+76_plR}$^{\dagger \dagger}$\\
        Kepler-43 b & $998.00^{+27.70}_{-28.00}$ & $12.43^{+0.57}_{-0.54}$ & ${1.42^{+0.21}_{-0.22}\times 10^9}$ & ${0.046}$ & $3.02$ &  \citet{Kepler-302+76_plR}$^{\dagger \dagger}$\\
        Kepler-74 b & $192.00^{+28.00}_{-30.00}$ & $12.82^{+0.65}_{-0.61}$ & ${4.98^{+0.28}_{-0.26}\times 10^8}$ & ${0.078}$ & $7.34$ &  \citet{Kepler-302+76_plR}$^{\dagger \dagger}$\\
        Kepler-539 b & $308.30^{+92.17}_{-92.17}$ & $8.30^{+0.36}_{-0.41}$ & ${4.95^{+0.28}_{-0.29}\times 10^6}$ & ${0.499}$ & $125.63$ &  \citet{Kepler-302+76_plR}$^{\dagger \dagger}$\\

        \bottomrule
    \end{tabular}
    \caption{The mass ($M_p$), radius ($R_p$), incident flux ($f_\star$), semi-major axis ($a$), and period ($P$) for the target planets in our study. All planetary parameters are obtained from the \citet{TESS_archive, K2_archive, Kepler_archive}, with individual references specified in Table \ref{tab:stellar_params}, with the exception of $f_\star$ which is calculated. Period values are rounded to 3 decimal places. The extra reference is for the planet mass, except for $^{\dagger \dagger}$, where it is instead for planet radius. $^{\dagger \dagger \dagger}$Updated mass from \citet{Kepler-289c_plM}. $^\star$The values tabulated here are slightly different but within the 1-$\sigma$ error of what is reported in \citet{Masuda24} because our analyses use the values that were communicated to us by Masuda, Libby-Roberts in private prior to their publication. Since the numbers agree well within 1-$\sigma$ we keep our original values of Kepler-51 planets.}
    \label{tab:planet_params}
\end{table*}

\begin{table*}[ht]
    \centering
    \renewcommand{\arraystretch}{1.15}
    \begin{tabular}{l c c c c c c c c}
        \toprule
        \textbf{Star} & \textbf{[M/H] [dex]} & \textbf{t$_*$ [Myr]} & \textbf{L$_*$ [$\log_{10}(L_\odot)$]} & \textbf{T$_*$ [K]} & \textbf{R$_*$ [$R_\odot$]} & \textbf{M$_*$ [$M_\odot$]} & \textbf{Reference} \\
        \midrule
        V1298 Tau & ${0.10^{+0.15}_{-0.15}}$ & $25^{+5}_{-5}$ & ${0.018^{+0.044}_{-0.043}}$ & ${5050^{+100}_{-100}}$ & $1.35^{+0.03}_{-0.03}$ & $1.16^{+0.06}_{-0.06}$ & \citet{V1298Tau_pl} \\
        HATS-36 & ${0.28^{+0.04}_{-0.04}}$ & $620^{+550}_{-550}$ & ${0.215^{+0.043}_{-0.048}}$ & ${6149^{+76}_{-76}}$ & $1.16^{+0.04}_{-0.04}$ & $1.22^{+0.03}_{-0.03}$ & \citet{HATS-36_pl_st} \\
        HIP 67522 & ${0.00}$ & $17^{+2}_{-2}$ & ${0.243^{+0.022}_{-0.023}}$ & ${5675^{+75}_{-75}}$ & $1.38^{+0.06}_{-0.06}$ & $1.22^{+0.05}_{-0.05}$ & \citet{HIP67522_pl_st} \\
        TOI-837 & ${-0.07^{+0.04}_{-0.04}}$ & $35^{+11}_{-5}$ & ${0.084^{+0.164}_{-0.164}}$ & ${6047^{+162}_{-162}}$ & $1.02^{+0.08}_{-0.08}$ & $1.12^{+0.06}_{-0.06}$ & \citet{TOI-837_pl_st} \\
        TOI-1268 & ${0.36^{+0.06}_{-0.06}}$ & $245^{+135}_{-135}$ & ${-0.234^{+0.193}_{-0.193}}$ & ${5300^{+100}_{-100}}$ & $0.92^{+0.06}_{-0.06}$ & $0.96^{+0.04}_{-0.04}$ & \citet{TOI-1268_pl_st} \\
        TOI-1431 & ${0.09^{+0.03}_{-0.03}}$ & $290^{+320}_{-190}$ & ${1.068^{+0.078}_{-0.047}}$ & ${7690^{+400}_{-250}}$ & $1.92^{+0.07}_{-0.07}$ & $1.90^{+0.10}_{-0.08}$ & \citet{TOI-1431_pl_st} \\
        TOI-2046 & ${-0.06^{+0.15}_{-0.15}}$ & $450^{+430}_{-21}$ & ${0.290^{+0.092}_{-0.092}}$ & ${6250^{+140}_{-140}}$ & $1.21^{+0.07}_{-0.07}$ & $1.13^{+0.19}_{-0.19}$ & \citet{TOI-2046_pl_st} \\
        TOI-4087 & ${0.24^{+0.08}_{-0.08}}$ & $800^{+1200}_{-600}$ & ${0.176^{+0.025}_{-0.020}}$ & ${6060^{+74}_{-67}}$ & $1.11^{+0.02}_{-0.02}$ & $1.18^{+0.04}_{-0.04}$ & \citet{TOI-4087_pl_st} \\
        TOI-2152 A & ${0.28^{+0.07}_{-0.08}}$ & $830^{+1100}_{-580}$ & ${0.653^{+0.069}_{-0.067}}$ & ${6630^{+300}_{-290}}$ & $1.61^{+0.06}_{-0.05}$ & $1.52^{+0.09}_{-0.10}$ & \citet{TOI-2152A_pl_st} \\
        TOI-201 & ${0.24^{+0.04}_{-0.04}}$ & $870^{+460}_{-490}$ & ${0.415^{+0.016}_{-0.017}}$ & ${6394^{+75}_{-75}}$ & $1.32^{+0.01}_{-0.01}$ & $1.32^{+0.03}_{-0.03}$ & \citet{TOI-201_pl_st} \\
        TOI-622 & ${0.09^{+0.07}_{-0.07}}$ & $900^{+200}_{-200}$ & ${0.474^{+0.010}_{-0.010}}$ & ${6400^{+100}_{-100}}$ & $1.42^{+0.05}_{-0.05}$ & $1.31^{+0.08}_{-0.08}$ & \citet{TOI-622_pl_st} \\
        KOI-1783& ${0.11^{+0.04}_{-0.04}}$ & $200^{+1300}_{-1300}$ & ${0.213}$ & ${5922^{+60}_{-60}}$ & $1.14^{+0.03}_{-0.03}$ & $1.08^{+0.04}_{-0.03}$ & \citet{KOI-1783_pl_st} \\
        Kepler-51 & ${0.05^{+0.04}_{-0.04}}$ & $500^{+250}_{-250}$ & ${-0.142}$ & ${6018^{+107}_{-107}}$ & $0.94^{+0.50}_{-0.50}$ & $1.04^{+0.12}_{-0.12}$ & \citet{Libby-Roberts2020} \\
        KOI-351 & ${-0.12^{+0.18}_{-0.18}}$ & $530^{+880}_{-880}$ & ${0.269}$ & ${6080^{+260}_{-170}}$ & $1.20^{+0.10}_{-0.10}$ & $1.20^{+0.10}_{-0.10}$ & \citet{Kepler-351_st} \\
        Kepler-76 & ${-0.10^{+0.20}_{-0.20}}$ & $620^{+480}_{-480}$ & ${0.451^{+0.015}_{-0.022}}$ & ${6409^{+95}_{-95}}$ & $1.32^{+0.08}_{-0.08}$ & $1.20^{+0.20}_{-0.20}$ & \citet{Kepler-76_pl_st} \\
        Kepler-289 & ${0.05^{+0.04}_{-0.04}}$ & $650^{+440}_{-440}$ & ${0.061^{+0.022}_{-0.023}}$ & ${5990^{+38}_{-38}}$ & $1.00^{+0.02}_{-0.02}$ & $1.08^{+0.02}_{-0.02}$ & \citet{Kepler-289_st_pl} \\
        Kepler-43 & ${0.33^{+0.11}_{-0.11}}$ & $690^{+540}_{-540}$ & ${0.326^{+0.043}_{-0.048}}$ & ${6041^{+143}_{-143}}$ & $1.42^{+0.07}_{-0.07}$ & $1.32^{+0.09}_{-0.09}$ & \citet{Kepler-43_pl_st} \\
        Kepler-74 & ${0.42^{+0.11}_{-0.11}}$ & $800^{+900}_{-500}$ & ${0.332}$ & ${6000^{+100}_{-100}}$ & $1.12^{+0.04}_{-0.04}$ & $1.18^{+0.04}_{-0.04}$ & \citet{Kepler-43_pl_st} \\
        Kepler-539& ${-0.01^{+0.07}_{-0.07}}$ & $910^{+840}_{-840}$ & ${-0.060^{+0.022}_{-0.023}}$ & ${5820^{+80}_{-80}}$ & $0.95^{+0.02}_{-0.02}$ & $1.05^{+0.04}_{-0.04}$ & \citet{Kepler-539_pl_st} \\
        \bottomrule
    \end{tabular}
    \caption{The stellar properties, including metallicity [M/H], age (t$_*$), luminosity (L$_*$), effective temperature (T$_*$), radius (R$_*$), and mass (M$_*$), for all the planet-hosting stars in our study. All stellar parameters are obtained from the \citet{TESS_archive, K2_archive, Kepler_archive}. 
    The metallicity for V1298 Tau 
    was obtained from \citep{V1298Tau_met}, since it was not available from the data source given in the table. The quoted metallicity for HIP 67522 is an assumed solar value and not a measurement.}
    \label{tab:stellar_params}
\end{table*}

\begin{figure*}
    \begin{minipage}{0.48\linewidth}
        \centering
        \includegraphics[width=\textwidth]{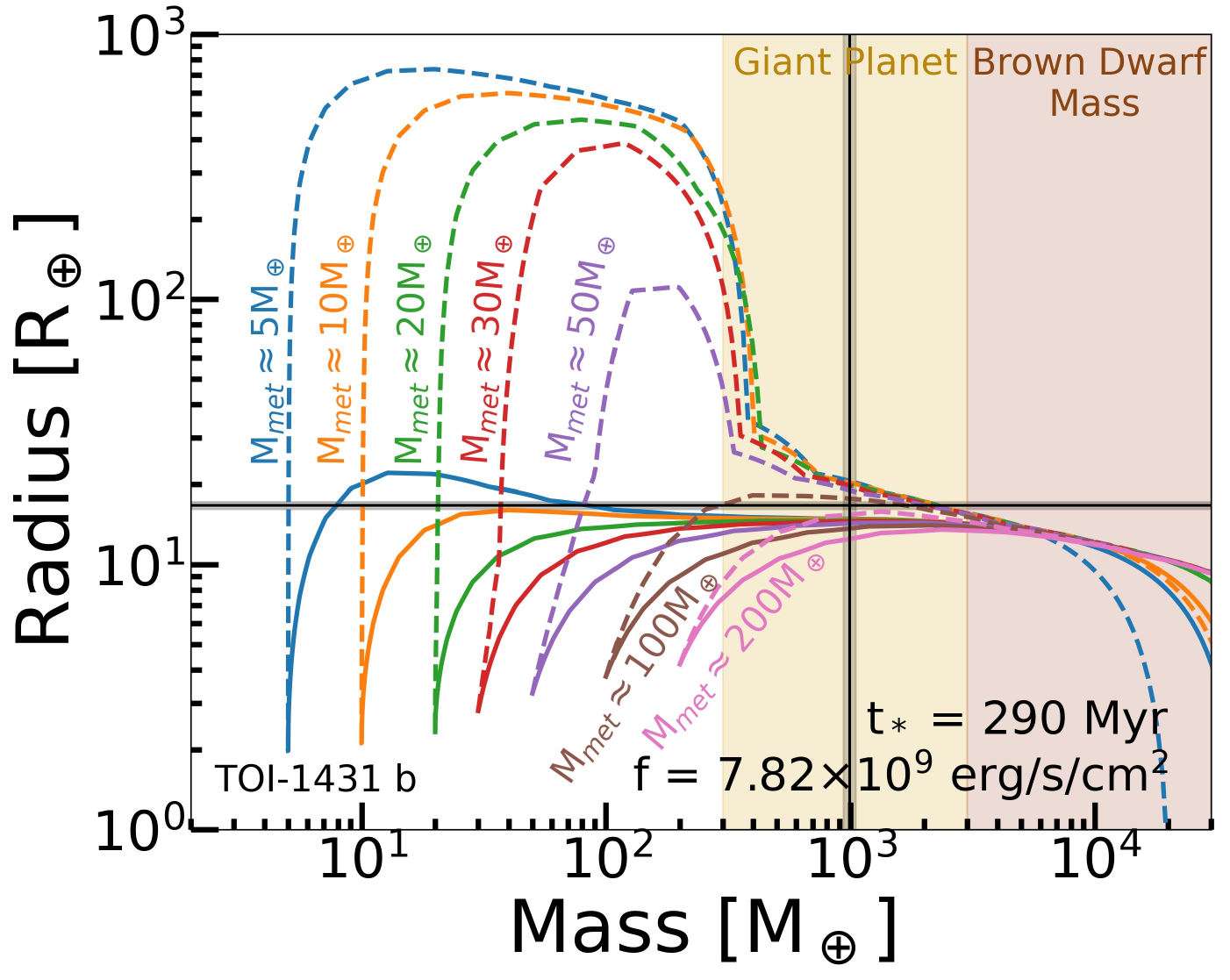}
    \end{minipage}
    \begin{minipage}{0.48\linewidth}
        \centering
        \includegraphics[width=\textwidth]{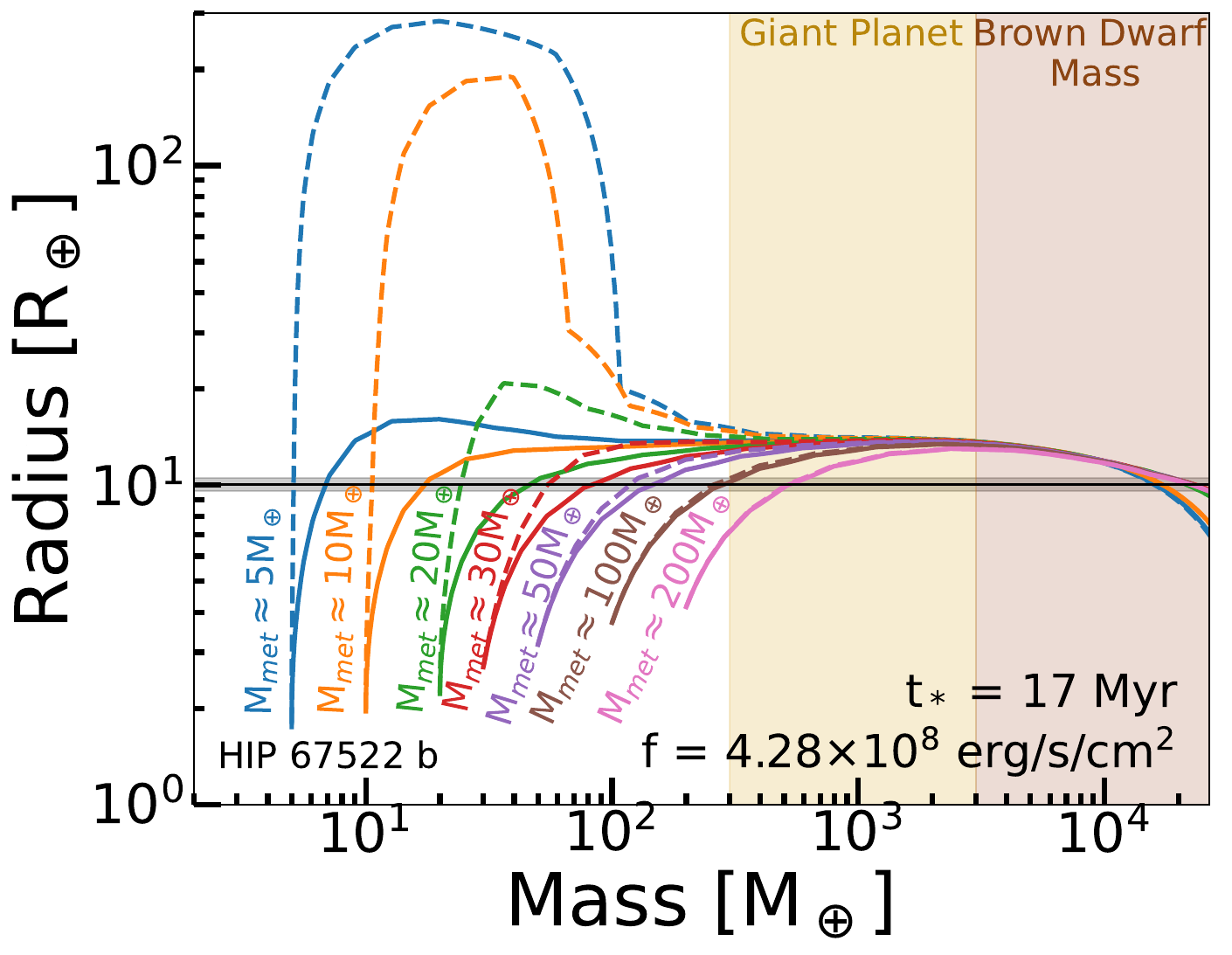}
    \end{minipage}
    
    \caption{Left: 
    radius vs.~mass for the age ($t_\star$) and incident flux ($f$) of TOI-1431 b.
    Each color corresponds to varying metal masses with dashed and solid lines illustrating radius with and without extra heating 
    from stellar irradiation (both at the same flux),
    respectively.
    The black lines and the surrounding grey zone show the measured radius and mass and their 1-$\sigma$ error for this planet.
    The yellow-shaded region indicates the range of masses for a Jupiter 
    class planet, while the brown region indicates the range of masses for a brown dwarf. Right: same as the left panel, but for HIP 67522 b, which does not have a measured mass (only an upper limit exists). 
    We see two families of solution (planetary mass vs.~brown dwarf/stellar mass) that explain the observed radius of the planet (see text for how we rule out the brown dwarf/stellar mass solution and further constrain the physically plausible range of planet mass).
    }
    \label{fig:grid_withMass}
\end{figure*}

\subsection{Interior Structure Model} \label{sec:int_struc_model}

With the target list ready, we now use thermal evolution models from \citet{Thorngren2016} to constrain the masses of the planets given their radius, incident bolometric flux, and age.
The thermal evolution of a planetary interior is constructed by solving the structure equations (hydrostatic equilibrium, mass conservation, and energy conservation, in that order):
\begin{equation}
    \frac{\partial P}{\partial m} = - \frac{Gm}{4 \pi r^4}
\end{equation}
\begin{equation}
    \frac{\partial r}{\partial m} = \frac{1}{4 \pi r^2 \rho}
\end{equation}
\begin{equation}
    \frac{\partial L}{\partial m} = - T \frac{\partial S}{\partial T}
\end{equation}
where $P$ is pressure, $m$ is mass, $r$ is radius, $\rho$ is density, $L$ is luminosity, $T$ is temperature, $S$ is entropy, and $G$ is the gravitational constant. 
We therefore solve for the radius-mass relationship for a given metal mass, gas-to-metal mass ratio, incident flux, and age.
We make the distinction between core mass and metal mass since not all heavy elements go into the planet's core. 
While giant planets require a $\sim$10 M$_\oplus$ core to 
nucleate by runaway gas accretion \citep[e.g.,][]{Pollack1996,Rafikov2006,Lee2014},
concomitant solid accretion and/or post-formation pollution by solid infall can increase the total metal mass content.
Furthermore, observational data indicate that sub-Neptunes have rocky cores with masses that can go up to $\sim$ 20 M$_\oplus$ \citep[e.g.,][]{Otegi2020}.
Therefore, if the planet is composed of less than or equal to 20 M$_\oplus$ of heavy elements, all metals go into the core and we consider a H/He envelope with solar metallicity. 
For planets with over 20 M$_\oplus$ of heavy elements, we put 20 M$_\oplus$ into the core and uniformly mix the remaining metals into the planet's H/He atmosphere using additive volumes.\footnote{At least one low mass sub-Neptune is reported to have highly metal-enriched atmosphere \citep{Benneke24}, which could be from core-envelope mixing \citep[e.g.,][]{Misener23} or post-formation pollution \citep[e.g.,][]{Vlahos24}. Among young puffy planets, both sub-stellar (V1298 Tau b; \citealt{Barat2023}) and superstellar atmospheric metallicity (HIP 67522 b; \citealt{Thao2024}) are reported although in the latter case, HIP 67522 was assumed to have solar metallicity. We keep our simple partitioning of metallic content to ensure tractability of our models and consider further structural complications in a follow-up study.} 
The final grid covers metal masses between 0.3 M$_\oplus$ and 200 M$_\oplus$ in 60 logarithmic bins, gas-to-metal mass ratio
between $10^{-6}$ and $10^3$ in 33 logarithmic bins, incident fluxes between $10^4$ and $10^{10}$ erg/s/cm$^2$ in 19 logarithmic bins, and ages from 10 Myr to 10 Gyr in 100 logarithmic bins.

Since many of our selected planets are under intense incident flux 
(see Table \ref{tab:planet_params}), 
we account for extra heating as described in \citet{Thorngren2018}, where some of the irradiation from the star is being converted into heating the planet. The effects of this extra heating are illustrated in Figure \ref{fig:grid_withMass}. When the incident flux is high, the extra heating effectively puffs up the planet, resulting in a larger radius for the same mass. This effect is more pronounced for smaller metal masses, since these planets have less heavy elements, the H/He envelope is more susceptible to expansion due to weaker gravitational potential.

Using the \texttt{RegularGridInterpolator} function from \texttt{scipy.interpolate} to interpolate between 
the grid points, we obtain the range of plausible total mass of a planet given their measured radius, incident flux, and age.\footnote{We consider the incident flux to be time-invariant for all stars in our sample. 
Such an assumption becomes less trustworthy for pre-main sequence stars such as V1298 Tau or K2-33. However, the radius dependence on incident flux is weak \citep{Lopez2014}, so for the purpose of deriving planet radius, we expect our result to be negligibly dependent on the time evolution of stellar irradiation. This weak radius dependence on incident flux is separate from the extra heating.} We lift this assumption when we calculate mass loss which is more strongly dependent on stellar flux.
For each of the 22 planets in our sample, we step through each of the 60 log-spaced metal mass grid points and use the \texttt{COBYLA} method of \texttt{scipy.optimize.minimize} to solve for the gas-to-metal mass ratio that produces a radius that matches the observed value within the tolerance of $10^{-5} R_\oplus$.
Our models find two different solutions for each metal mass, one at low total mass (usually planetary) and another at high total mass (usually brown dwarf or stellar; see the right panel of Figure \ref{fig:grid_withMass}). 
This degeneracy breaks and always favors planetary mass solution for planets with mass measurements (see the left panel of Figure \ref{fig:grid_withMass}). Kepler-74 b is a special case where both families of solutions are within the planetary mass regime, so we consider both in our analysis.

\subsection{Ruling out the Brown Dwarf Solution} \label{sec:BD_sol}

\begin{figure}[ht!]
    \centering
    \includegraphics[width=0.5\textwidth]{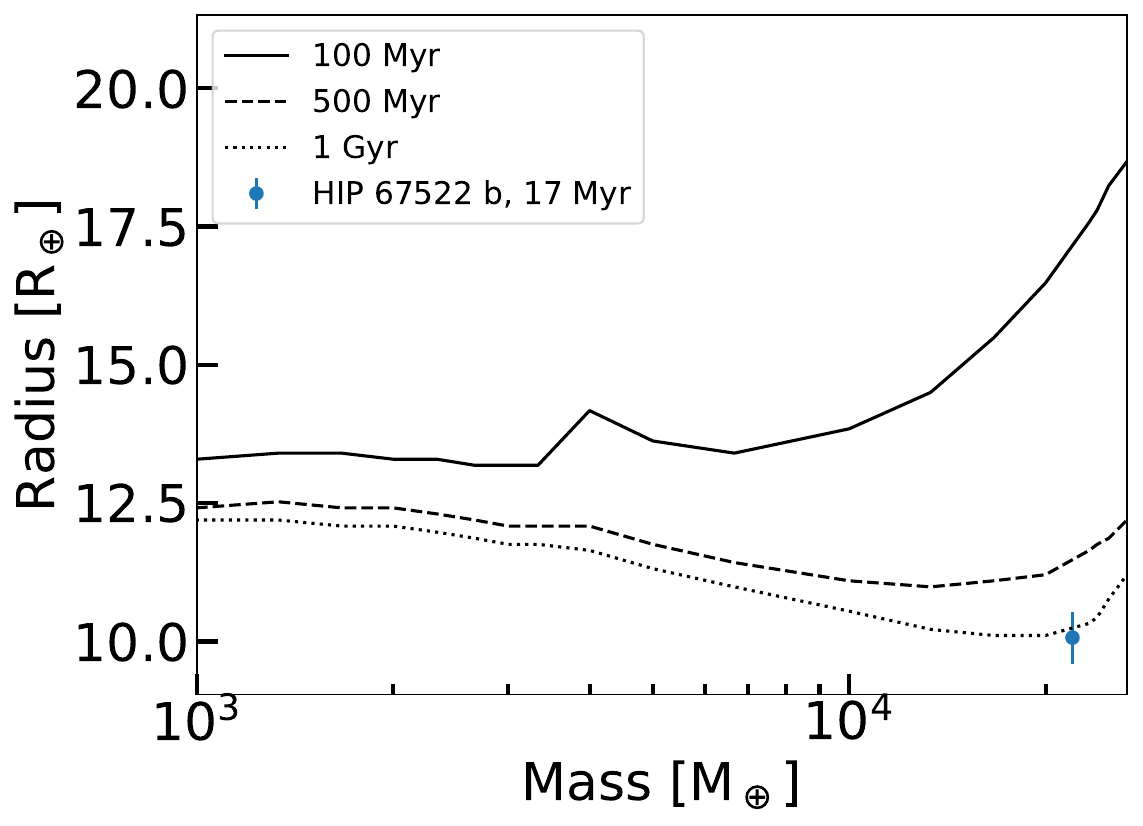}
    \caption{ 
    Verifying the validity of the high total mass solution.
    The black lines show the theoretically expected mass-radius relation at 100 Myr (solid), 500 Myr (dashed), and 1 Gyr (dotted) reported by \citet{Baraffe2003}. 
    For its youth, HIP 67522 b is way too dense to be considered a brown dwarf.}
    \label{fig:BD_Test}
\end{figure}

We now describe how we rule out the high-mass family of solutions for planets with no mass measurements.
For KOI-351 g we can rule out the high mass solution since it lies beyond 100 M$_J$, well within the stellar mass regime, implying a density that is unphysically large. 
For HIP 67522 b, we check the high-mass solution against predicted mass-radius curves at different ages for brown dwarfs from \citet{Baraffe2003}, as shown in Figure \ref{fig:BD_Test}. 
Compared to the expected brown dwarf cooling curve, the high-mass solution for HIP 67522 b is way too dense for its estimated age, so we can safely rule such solutions out. 
This lack of high mass solutions is consistent with the empirical paucity of short-period ($\lesssim$200 days) brown dwarf companions to main sequence stars \citep[e.g.,][]{Grether06,Duchene13,Kiefer21}.

\subsection{Mass Loss} \label{sec:ML}

We now place further mass constraints on the remaining low-mass family of solutions. We first consider photoevaporative mass loss to place a lower limit on the mass.
We integrate the following equation over the system age to determine the amount of mass lost:

\begin{equation}
    \dot{M} = -\eta\frac{ \pi R_p^3 f_{XUV}}{G K M_p}
    \label{eq:mdot}
\end{equation} 
where $\eta$ is the efficiency parameter for which we adopt the empirical fit to numerical simulations computed in Appendix A of \citet{Caldiroli2022}, $f_{XUV}$ is the XUV flux on the planet from the host star, $R_p$ is the planet radius, $M_p$ is the planet mass, and $K$ is the reduction factor which accounts for stellar tidal forces \citep{Erkaev2007}, given by

\begin{equation}
    K = 1 - \frac{3}{2} \left(\frac{R_{Hill}}{R_p} \right)^{-1} + \frac{1}{2} \left(\frac{R_{Hill}}{R_p} \right)^{-3}
\end{equation}
where $R_{\rm Hill} = a \left(\frac{M_p}{3M_\star}\right)^{1/3}$ is the Hill radius, $a$ is the orbital distance of the planet, and $M_\star$ is the mass of the host star. 
While the form of equation \ref{eq:mdot} resembles that of energy-limited approximation \citep[see, e.g.,][]{Erkaev2007, Sanz-Forcada2011}, $\eta$ is a non-trivially varying parameter with respect to $f_{XUV}$ and the gravitational potential of the planet that fully captures the non-energy limited regime \citep[see][for detail]{Caldiroli2022}.

\subsubsection{Comparing and Choosing a Source for the XUV Flux Evolution} \label{sec:XUV_check}

To calculate mass loss, we need the time evolution of the XUV luminosity of the planet's host star over the lifetime of the system. We consider two stellar evolution grids which provide the bolometric luminosities throughout the star's lifetime, \citet{Johnstone2021} (hereafter \citetalias{Johnstone2021}) which provides grids for stars ranging from 0.1 to 1.2 $M_\odot$, and the MESA Isochrones \& Stellar Tracks (MIST) \citep{MIST0, MIST1}, which provides grids for stars with masses between 0.1 and 300 $M_\odot$ and metallicities from -4 to 0.5 dex, where zero corresponds to solar metallicity. 
We ultimately choose MIST and describe why below.

First, we take the bolometric luminosity for a given stellar mass from the corresponding stellar evolution track.
We note a slight discrepancy between the present-day measured values for the luminosity and the values from the stellar evolution grid at the age of the star. To account for this, we scale the grid luminosity to the measured value at system age.
We then use the empirical relation to obtain the time-dependent X-ray luminosity: 

\begin{equation} \label{eq:Fx_F*}
    L_{X} =
    \begin{cases}
      L_{sat} & t < 100\, {\rm Myr}\\
      L_{sat} \left( \frac{t}{100 Myr} \right)^{-1.42} & t \geq 100\, {\rm Myr}\\
    \end{cases} 
\end{equation}
where the value of the luminosity during the saturation interval is $L_{sat} = 10^{-3.6} L_*(t)$ \citep[see, e.g.,][]{Vilhu1987, Wright2011, KingWheatley2021}, $L_*(t)$ is the time-dependent bolometric luminosity over the lifetime of the star, and 100 Myr corresponds to a typical saturation timescale. The power-law index describing the time dependence of X-ray luminosity is taken to be -1.42, which corresponds to the median value obtained by \citet{Tu2015} who use rotational evolution models to predict how stellar X-ray luminosity decays with time.
Typically, this value is found to be between -1.5 and -1.2 \citep[see, e.g.,][]{Jackson2012, Claire2012, Ribas2005}.
The EUV flux evaluated at the stellar surface, $F_{EUV}$, is then obtained from the following ratio: 

\begin{equation} \label{eq:Feuv_Fx}
    \frac{F_{EUV}}{F_X} = \beta {F_X}^{\gamma}
\end{equation}
where $F_{X}$ is the X-ray flux evaluated at the surface of the star, $\gamma$ is the power-law index and $\beta$ is the scaling factor.

\begin{figure*}[ht!]
    \begin{minipage}{0.49\linewidth}
        \centering
        \includegraphics[width=\textwidth]{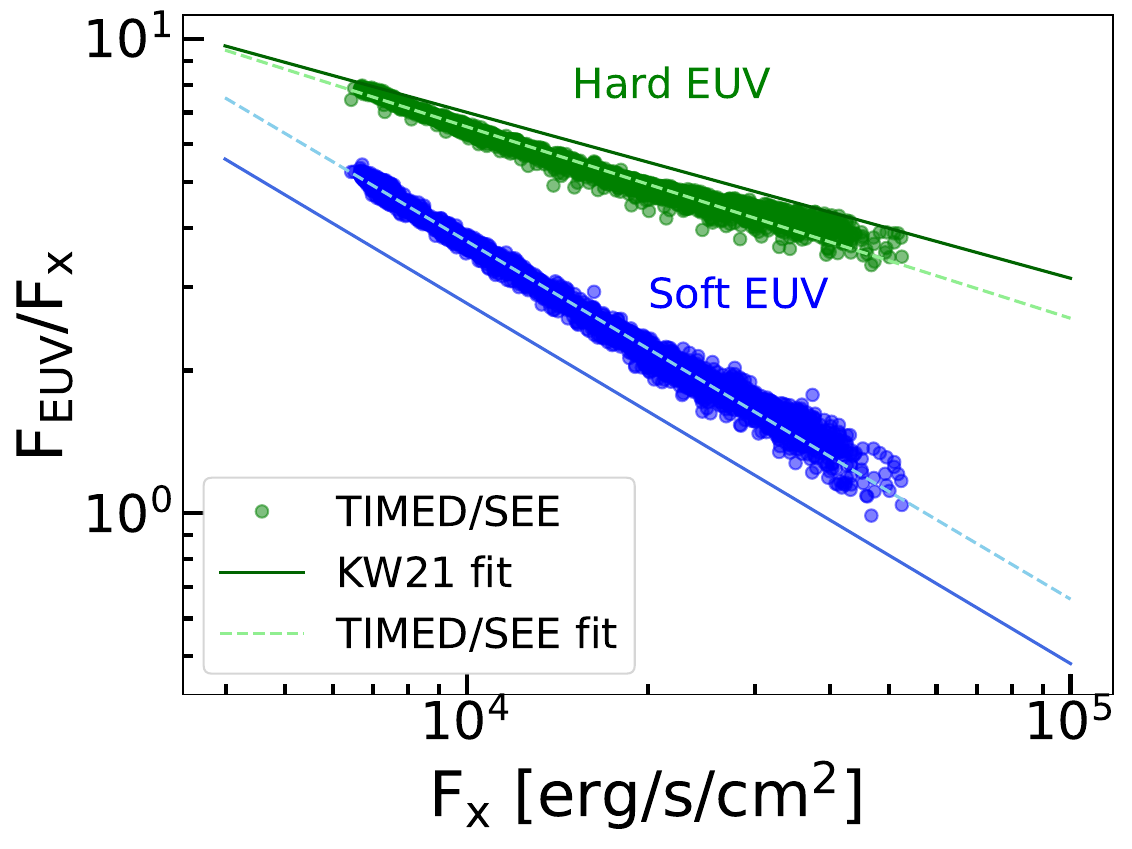}
    \end{minipage}
    \begin{minipage}{0.49\linewidth}
        \centering
        \includegraphics[width=\textwidth]{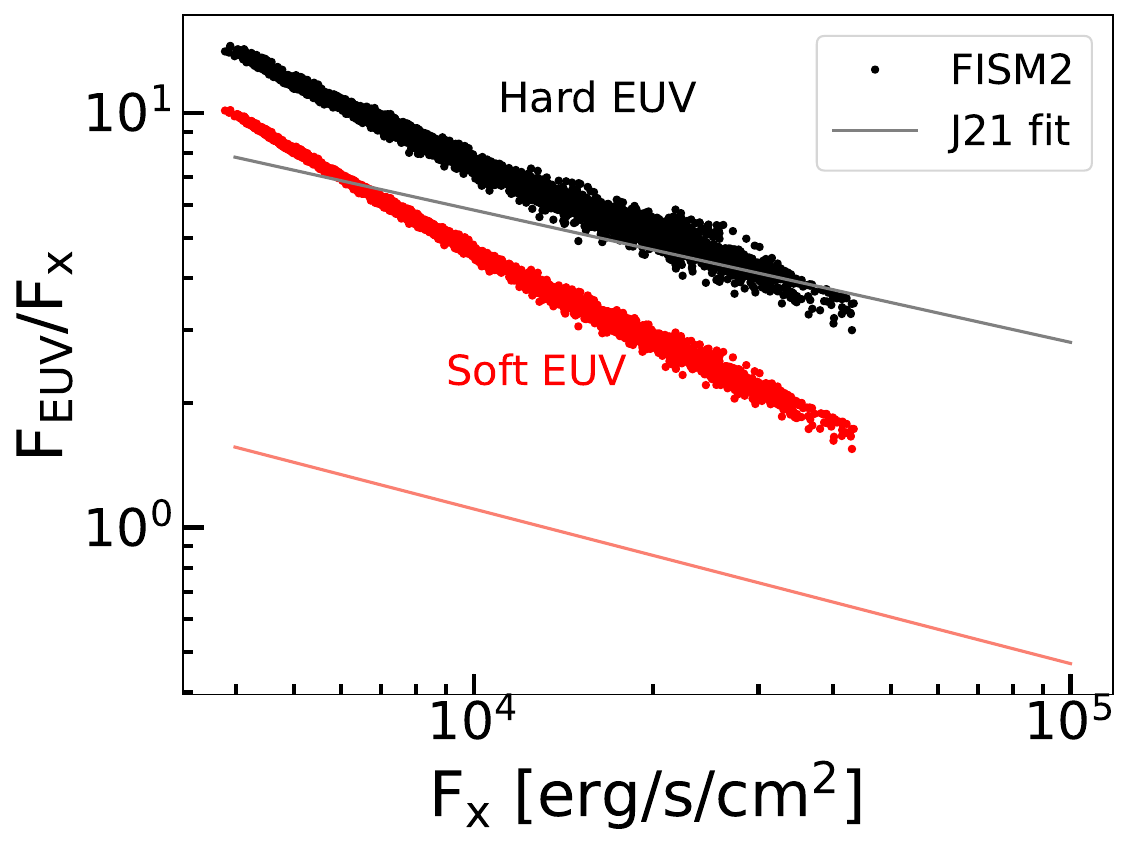}
    \end{minipage}
    \caption{Left: the X-ray to EUV surface flux ratio for the solar data and scaling equations presented in \citet{KingWheatley2021}. The TIMED/SEE solar data \citep{Woods2005} are shown as green and blue points, for hard and soft EUV, respectively. The dashed lines show the result of our own power law fit, given by Equation \ref{eq:Feuv_Fx}, to the TIMED/SEE data.
    The solid lines illustrate the best-fit parameters from \citet{KingWheatley2021}, corrected for the typo in the text ($\beta_{\rm hard}=176$, not 116).
    Right: same as the left panel, using the equations and data presented in \citet{Johnstone2021}. The black and red dots show the solar data from FISM2 \citep{Chamberlin2020}, for hard and soft EUV, respectively. 
    The solid lines show the scaling relations presented in Equations 20 and 22 of \citet{Johnstone2021}. While we can closely reproduce the best-fit results reported by \citet{KingWheatley2021}, we are unable to reproduce that by \citet{Johnstone2021}.}
    \label{fig:EUV_data_check}
\end{figure*}

We compare the X-ray to EUV surface flux scaling equations from \citetalias{Johnstone2021}  to the equations presented in \citet{KingWheatley2021} (hereafter \citetalias{KingWheatley2021}).
Comparing the best-fit parameters from \citetalias{Johnstone2021} and \citetalias{KingWheatley2021}, we find that the slopes and the general trends of the EUV flux as a function of X-ray flux are very different. 
The best-fit parameters for the ratio of hard (100-360 $\AA$) and soft (360-920 $\AA$) EUV to X-ray surface flux obtained by \citetalias{KingWheatley2021} are the power-law indices $\gamma_{\rm hard} = -0.35^{+0.07}_{-0.15}$ and $\gamma_{\rm soft} = -0.76^{+0.16}_{-0.04}$, with corresponding $\beta_{\rm hard} = 176$ and $\beta_{\rm soft}=3040$.\footnote{Note that in the text of \citetalias{KingWheatley2021}, we believe a typo is made where $\beta_{\rm hard}$ is given there as 116, rather than the correct value of 176 which agrees with the best-fit line shown in Figure 2 of \citetalias{KingWheatley2021}.} 
The best-fit parameters presented by \citetalias{Johnstone2021} are $\gamma_{\rm hard} = -0.319$ and $\gamma_{\rm soft} = -0.373$ with $\beta_{\rm hard} = 110$ and $\beta_{\rm soft}=34.4$.\footnote{Once again, we believe that a typo was made in the \citetalias{Johnstone2021} paper where a calculation error was made going from their Equation 21 to Equation 22.}
\citetalias{KingWheatley2021} generally find a steeper slope than \citetalias{Johnstone2021}, particularly in the soft EUV. \citetalias{KingWheatley2021} also find that the hard EUV and soft EUV have a different slope (factor of $\sim$2 difference), while \citetalias{Johnstone2021} find that the two have similar slopes. In an attempt to better understand the equations presented in each paper and to get a better understanding of which of the two is more reliable, we attempt to reproduce the results of \citetalias{Johnstone2021} and \citetalias{KingWheatley2021} using the source data.

For \citetalias{KingWheatley2021}, we use TIMED/SEE data \citep{Woods2005} as described in \citetalias{KingWheatley2021} to obtain the results shown in their Figure 2.\footnote{The data used in \citetalias{KingWheatley2021} is publicly available at \url{https://lasp.colorado.edu/see/data/daily-averages/level-3/}.} 
Using \texttt{scipy.optimize.curvefit}, we fit Equation \ref{eq:Feuv_Fx} to the TIMED/SEE data and obtain  $\gamma_{\rm hard} = -0.4039157\pm0.0000006$ and $\gamma_{\rm soft} = -0.7551290\pm0.0000009$, with corresponding $\beta_{\rm hard} = 270\pm4$ and $\beta_{\rm soft}=4000\pm1000$. The left panel of Figure \ref{fig:EUV_data_check} shows the resulting TIMED/SEE data and fit we obtain compared to the results from \citetalias{KingWheatley2021}. We find that 
our $\gamma_{\rm hard}$ and $\gamma_{\rm soft}$ agree with theirs within their quoted error while our $\beta$'s differ by factors of order unity.
We perform two independent mass loss calculations using each set (ours vs.~\citetalias{KingWheatley2021}) of parameters and find that they produce almost the same final result.
Therefore, given that our final results are not significantly affected by the small difference in the parameters, we perform all calculations using the corrected best-fit parameters from \citetalias{KingWheatley2021}: the power-law indices $\gamma_{\rm hard} = -0.35^{+0.07}_{-0.15}$ and $\gamma_{\rm soft} = -0.76^{+0.16}_{-0.04}$, and corresponding $\beta_{\rm hard} = 176$ and $\beta_{\rm soft}=3040$.

We perform a similar exercise on the \citetalias{Johnstone2021} data. Their EUV fluxes are also obtained from solar data, using the Flare Irradiance Spectrum Model (FISM) \citep{Chamberlin2007}. Since FISM is not publicly available, we use the updated version of this model, FISM2 \citep{Chamberlin2020}. FISM2 is an improved, more reliable version of the original FISM, so we consider this a valid comparison to verify whether the parameters obtained by \citetalias{Johnstone2021} match up with solar data. In their paper, \citetalias{Johnstone2021} find that their models agree with the solar data from FISM.
However, we find that FISM2 does not agree with the scaling relations presented in \citetalias{Johnstone2021}, particularly in the soft EUVs, where the EUV surface flux is almost an order of magnitude larger at low X-ray surface fluxes. The slopes of the FISM2 data are steeper (factor of two larger) than what is given by \citetalias{Johnstone2021}. The results of this exercise are shown in the right panel of Figure \ref{fig:EUV_data_check}. We see that for both hard and soft EUV flux, the equations presented in \citetalias{Johnstone2021} do not agree with the FISM2 data. The discrepancy between the data and the fit is significantly more severe for \citetalias{Johnstone2021} than for \citetalias{KingWheatley2021}.

We conclude, given the discrepancy between the \citetalias{Johnstone2021} equations and the FISM2 data that we cannot fully explain, that the \citetalias{Johnstone2021} grid and scaling equations are not reliable enough for our purposes. 
Since our final mass ranges are robust to the difference in the fit parameters between our analysis and the results from \citetalias{KingWheatley2021}, we use the scaling equations from \citetalias{KingWheatley2021} for all mass loss calculations.
We also find that we are limited by the range of stellar masses for which stellar evolution tracks are available from \citetalias{Johnstone2021} since some of our stars have masses larger than 1.2 $M_\odot$. Therefore, we use MIST instead of the \citetalias{Johnstone2021} stellar evolution tracks. We perform all calculations using the time-varying bolometric luminosities from MIST (equivalent evolutionary points tracks for 0.4 $\times$ the critical spin, rounded to the closest stellar mass and metallicity values for which a grid is available---where the stellar metallicity measurement is not available, we use solar metallicity), scaled to match the observed luminosities where necessary, Equation \ref{eq:Fx_F*} to obtain the X-ray luminosity, and Equation \ref{eq:Feuv_Fx} with the best-fit parameters from \citetalias{KingWheatley2021} to obtain the EUV surface fluxes. We then calculate the incident X-ray and EUV fluxes on the planet from the star and add them to obtain the incident high-energy flux, $f_{XUV}$.

\begin{figure}
    \gridline{\fig{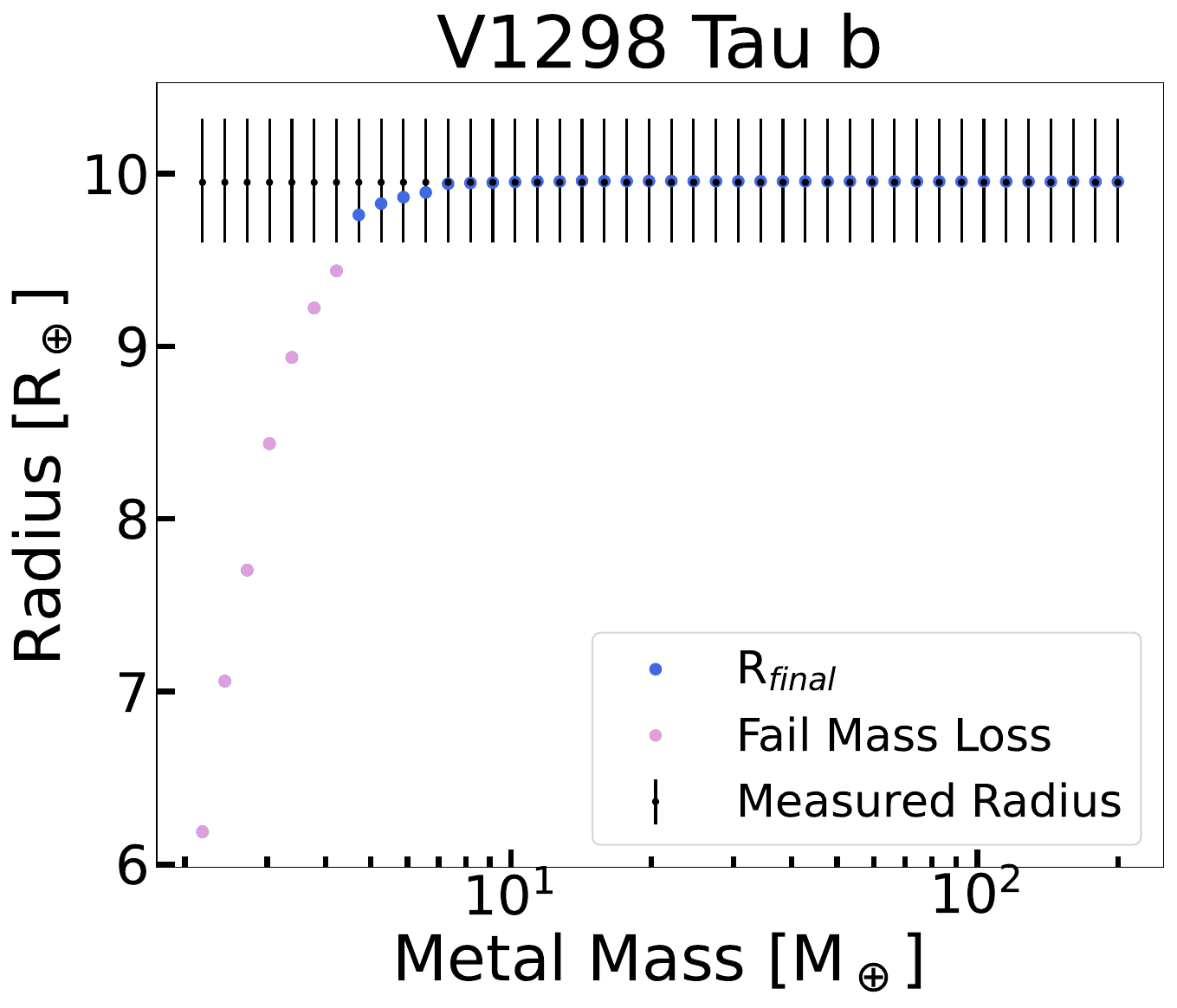}{0.45\textwidth}{}}
    \vspace{-0.7cm}
    \gridline{\fig{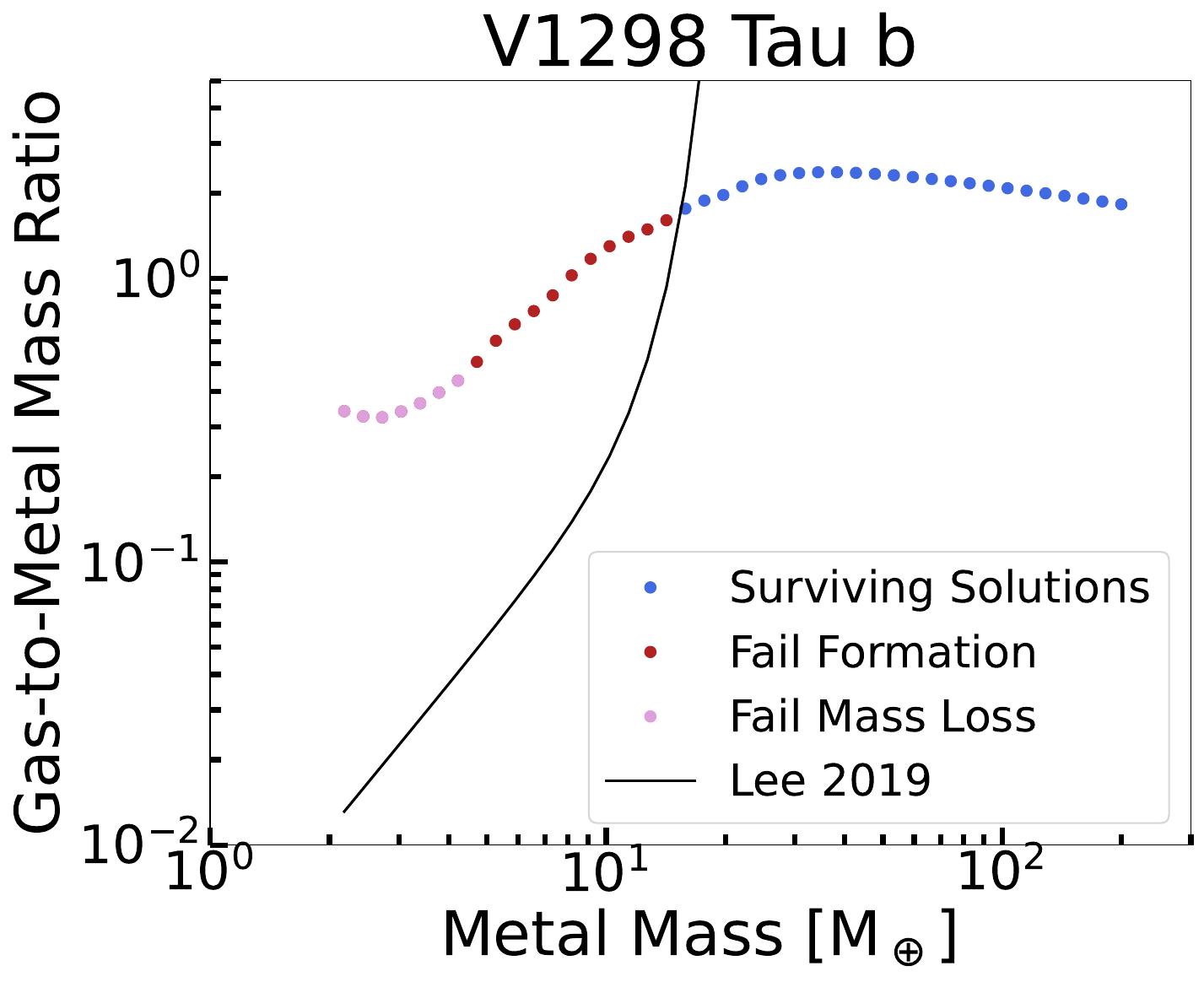}{0.45\textwidth}{}}
    \vspace{-0.7cm}
    \caption{Ruling out solutions based on mass loss and fiducial formation constraints for V1298 Tau b. Top: the result of the mass loss constraint, explained in Section \ref{sec:calc_ML}. The black triangles show the measured radius, $R_p=9.95R_\oplus$, with 1-$\sigma$ error, while the circles show the final, post-mass loss radius. The solutions that fail this mass loss constraint are shown as purple circles, while the ones that survive are shown as blue circles. Bottom: the result of the backward mass loss and formation constraint, described in Section \ref{sec:back_ML}. The black line shows the maximum (i.e., over the full disk lifetime $\sim$10 Myr) gas mass that can be accreted for a given metal mass under dusty accretion in situ, from \citet{Lee2019}. Solutions above this line are ruled out by formation consideration and are shown in red. Solutions that fail both the mass loss test (Section \ref{sec:calc_ML}) and the formation test are shown in purple. The solutions that survive both tests are shown in blue.}
    \label{fig:ML_nomass_check}
\end{figure}

\subsubsection{Calculating Mass Loss} \label{sec:calc_ML}

For each metal mass, we solve for the gas-to-metal mass ratio given the radius, incident flux, and age of the planet, as described in Section \ref{sec:int_struc_model}. We consider this gas-to-metal mass ratio as the initial ratio and numerically integrate Equation \ref{eq:mdot} using the trapezoid method, going forward in time starting at 5 Myr up to the age of the system. We take the timesteps from the MIST grid, which gives 454 log-spaced times from 0.032 yr to 1.806 Gyr, and slice the array to correspond to times $5 \;\mathrm{Myr} < t < t_*$. We have experimented with twice as coarse and twice as fine time resolution and found our final result to converge at the adopted MIST grid timestep. For the initial condition, we recalculate the radius of the planet given this gas-to-metal mass ratio at 5 Myr. At each iteration, we update the radius, mass, and incident flux, then the efficiency $\eta$ and the reduction factor $K$. The radius is updated using the radius grid 
as described in Section \ref{sec:int_struc_model}. By doing so, we implicitly neglect the adiabatic expansion of the envelope which can underestimate the amount of mass loss \citep[e.g.,][]{Thorngren2023}.

If we find that the entire envelope is lost, we end the calculation and consider the solution unstable to mass loss. Otherwise, we iterate until $t=t_*$, the stellar age, and check if the final radius after mass loss is within the error bounds of the observed radius. If it is, we consider the solution stable against mass loss, and otherwise, we rule it out. The top panel of Figure \ref{fig:ML_nomass_check} shows the result of this test for V1298 Tau b, where all solutions with a final radius below the lower limit set by the 1-$\sigma$ uncertainty on the measured radius are ruled out.  We perform this calculation for each of the 60 metal mass and gas-to-metal mass ratio solutions obtained. This allows us to narrow down the mass range, constraining the lower limit, since solutions with small metal masses are particularly susceptible to atmospheric escape due to weak surface gravity.

\subsubsection{`Backward' Mass Loss and Formation Constraint} \label{sec:back_ML}
Planets may very well have undergone significant mass loss in the early times so that their initial envelope mass is larger than that inferred from their present-day measurements. We therefore examine whether such an initial envelope is consistent with our fiducial formation model given by dusty accretion at close-in orbital periods presented in \citet{Lee2019}.
We use a similar process as described in Section \ref{sec:calc_ML}, except we begin at the system age, solve for the gas-to-metal mass ratio given the present-day parameters, and iterate backward in time up to 5 Myr, cumulatively adding the amount of mass lost at each backward timestep, where we consider the same timesteps as in Section \ref{sec:calc_ML}, to the current mass obtained from our radius grid. We repeat this calculation for each of the 60 metal masses in the radius grid, where we calculate the initial mass and radius of the planet for every solution obtained. 
Once the initial envelope mass is computed, we check the corresponding initial gas-to-metal mass and the metal mass against Figure 6 of \citet{Lee2019}. 
If our result lies above the maximum gas mass (i.e., accretion for the full disk lifetime $\sim$10 Myr) for the given metal mass, we rule out the solution. 
The result of this test for V1298 Tau b is shown in the bottom panel of Figure \ref{fig:ML_nomass_check}, where all solutions with gas mass above the limit set by \citet{Lee2019} for a given metal mass are ruled out.
A caveat to this test is that the maximum gas mass can be larger than what is presented in Figure 6 of \citet{Lee2019} for subsolar metallicity gas and/or dust-free accretion \citep{Lee2015}, which will be discussed in a case-by-case in Section \ref{sec:discussion}.

It is possible for a solution to survive the mass loss constraint in Section \ref{sec:calc_ML} but be ruled out by the formation constraint from \citet{Lee2019}.  For example, in the case of V1298 Tau b (R$_p$=9.95R$_\oplus$, $f_\star$=5.01$\times 10^7$ erg/s/cm$^2$), the 7 solutions with the smallest metal mass are ruled out by mass loss in Section \ref{sec:calc_ML} but the next 12 lowest metal mass solutions are still too puffy to be consistent with our fiducial formation constraint.

\begin{figure*}
    \begin{minipage}{0.48\linewidth}
        \centering
        \includegraphics[width=\textwidth]{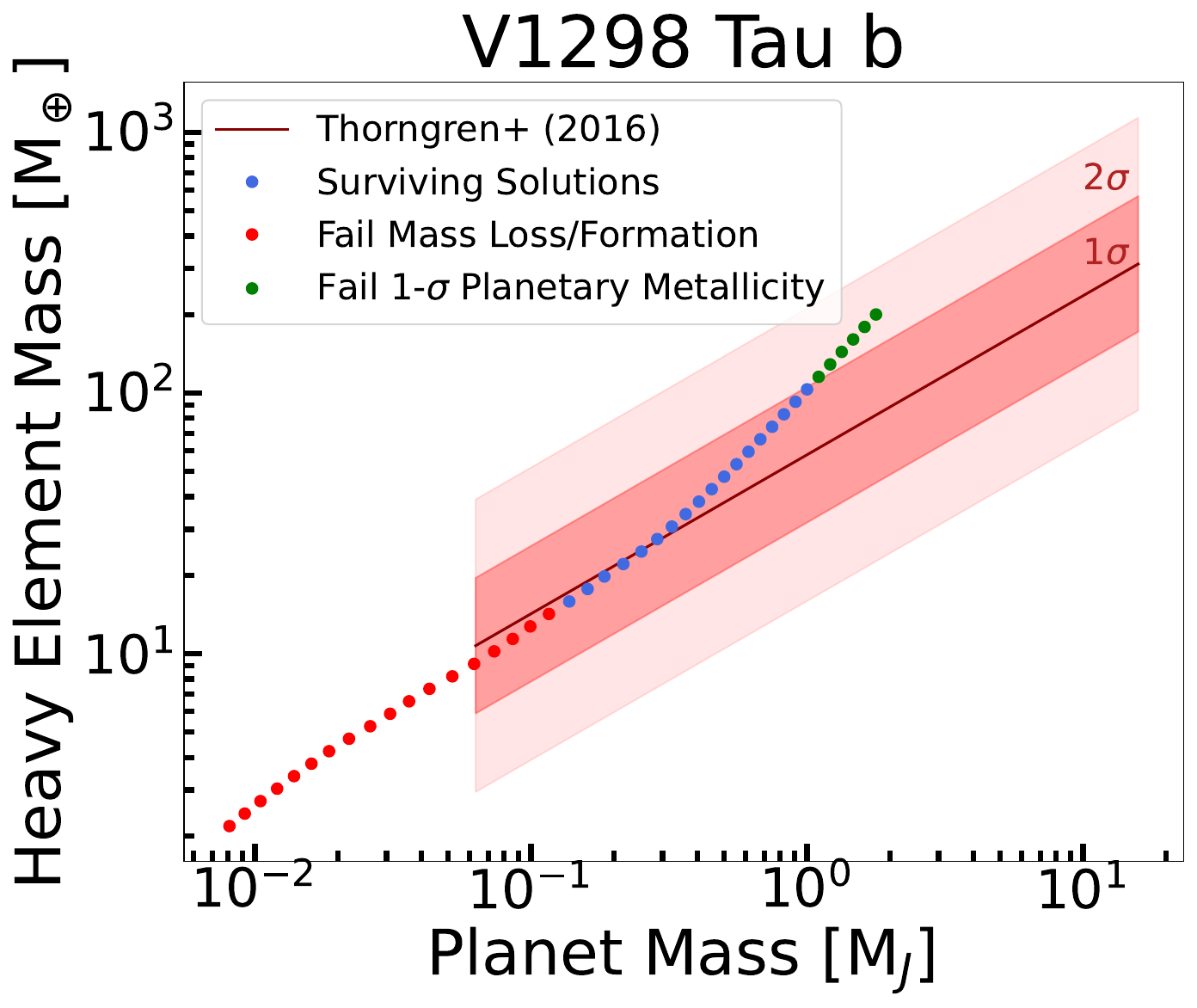}
    \end{minipage}
    \begin{minipage}{0.48\linewidth}
        \centering
        \includegraphics[width=\textwidth]{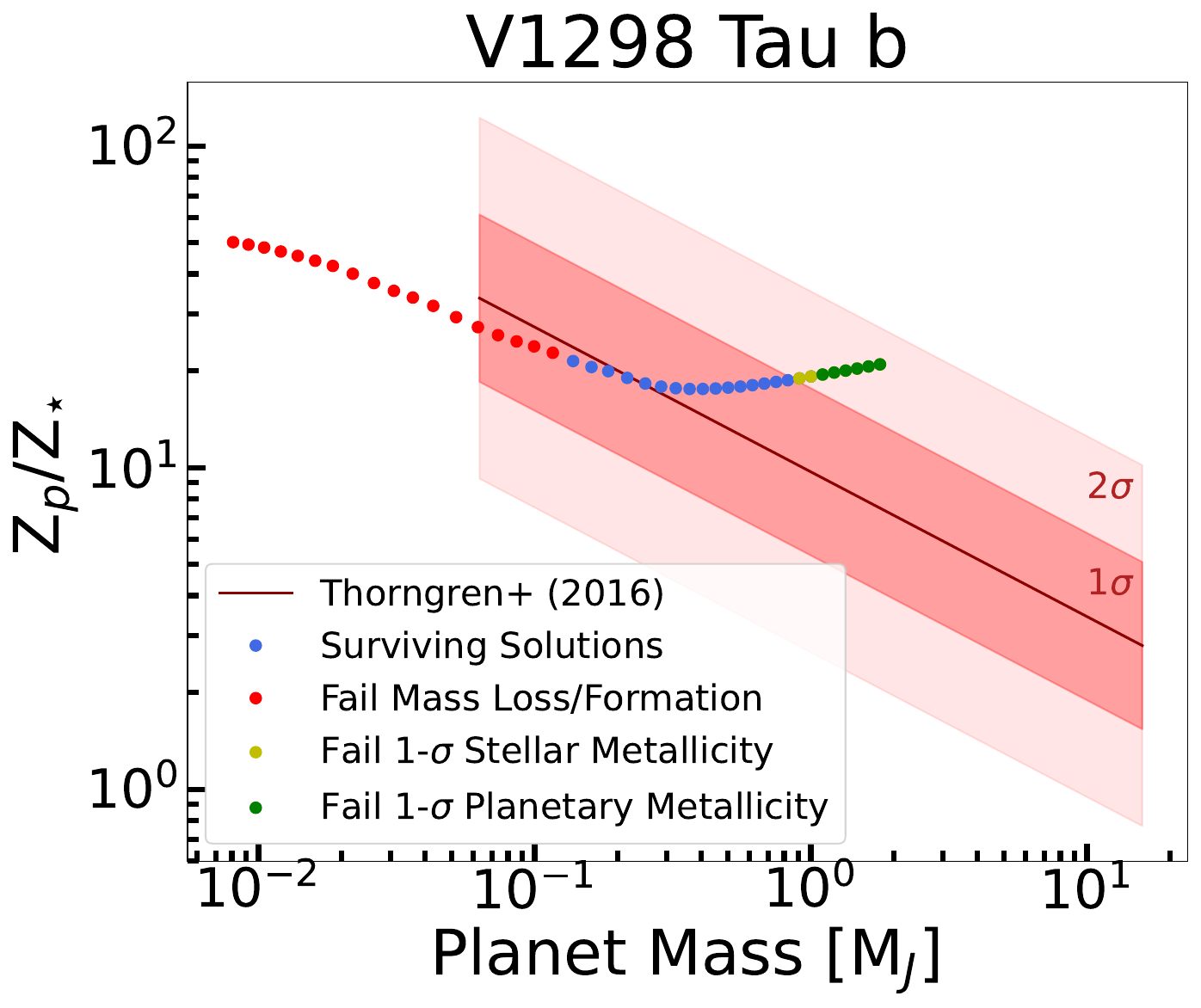}
    \end{minipage}
    \caption{Ruling out solutions based on planetary and stellar metallicity constraints described in Section \ref{sec:met_constraint} for V1298 Tau b. Left: planetary metallicity trends from \citet{Thorngren2016}. The best fit is shown as a red line, and the 1-$\sigma$ and 2-$\sigma$ regions are shaded in red. We mark the solutions that fall outside the 1-$\sigma$ region in green. Solutions that fail the mass loss and/or formation constraints in Section \ref{sec:ML} are shown in red. The solutions in blue survive all constraints and make up the final favored mass range for the planet. Right: same as the left, for the stellar metallicity constraint, based on the correlation between planetary and stellar metallicity and total planet mass from \citet{Thorngren2016}. The solutions lie beyond 1-$\sigma$ are marked in yellow.}
    \label{fig:Met_nomass_check}
\end{figure*}

\subsection{Planetary and Stellar Metallicity Constraints} \label{sec:met_constraint}

Finally, we check our model-inferred metals-to-gas ratio against empirical mass-metallicity trends from \citet{Thorngren2016}, which presents a study of the bulk compositions of transiting giant planets with masses $20 M_\oplus < M_p < 20 M_J$. We compare our results to the relationship they obtain between the planet's heavy element mass $M_{met}$ and total mass $M_p$, given by
\begin{equation}
    M_{met} = (57.9 \pm 7.03) M_p^{(0.61 \pm 0.08)}.
\end{equation}
They further analyzed the correlation between the planet's bulk heavy-element enrichment and the host star's metallicity, finding
\begin{equation}
    \frac{Z_{p}}{Z_{\star}} = (9.7 \pm 1.28) M_p^{(-0.45 \pm 0.09)}
\end{equation}
where $Z_p$=$M_{met}/M_{p}$ is the planet metallicity and $Z_{\star} = 0.0134 \times 10^{\rm [M/H]}$ is the stellar metallicity. For each metal mass and gas-to-metal mass ratio solution, we verify whether the planet lies within the 1- and 2-$\sigma$ region from the best fit. The intrinsic spread is given by $10^\sigma = 1.82 \pm 0.09$, so our 1-$\sigma$ region corresponds to the area within a factor of 1.82 from the best-fit line.

We consider these constraints only for solutions with total mass $M_p > 20 M_\oplus$ since \citet{Thorngren2016} studied giant planets and their results do not extend to lower mass planets. An example is shown in Figure \ref{fig:Met_nomass_check}
which illustrates how the check on planet and stellar metallicity places an upper limit on the plausible planet and heavy element mass.

\begin{figure*}[ht!]
    \centering
    \includegraphics[width=\textwidth]{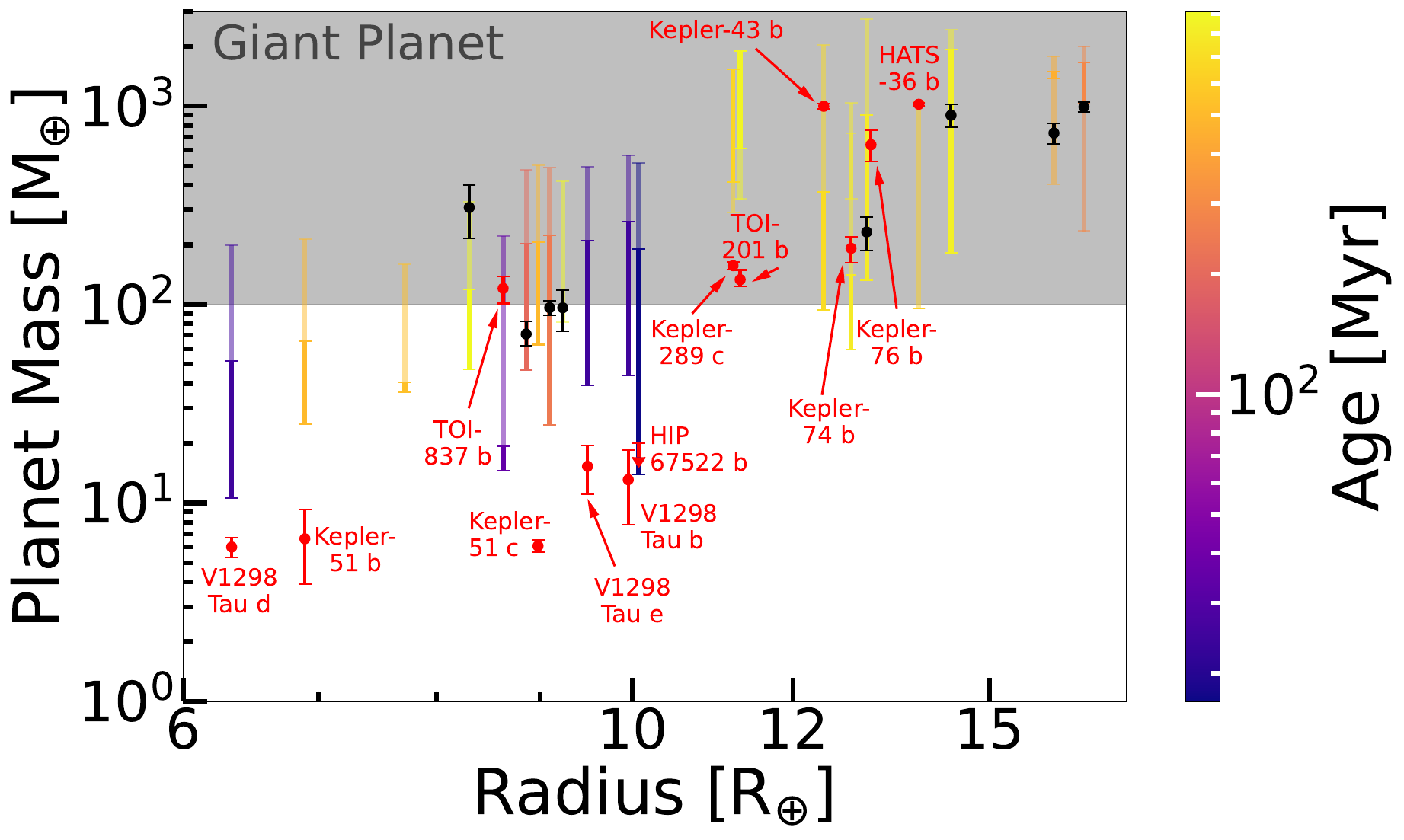}
    \caption{The total mass and radius of all 22 planets studied. Our mass estimates are shown with error bars color-mapped to the planet's age. The opaque and transparent error bars correspond to the mass range obtained from the 1-$\sigma$ and 2-$\sigma$ metallicity constraints, respectively. The masses for the 20 planets for which we have mass measurements are shown in black and red, with their 1-$\sigma$ uncertainties. The planets shown in red are labelled and are discussed in more detail in a case-by-case. HIP 67522 b, also shown in red, only has an upper limit, denoted by an arrow. The youngest planets ($\lesssim$ 100 Myr) tend to be lighter, Neptune mass planets, while the bonafide hot Jupiters are mostly found around older stars.}
    \label{fig:Mtot_Rpl_age}
\end{figure*}

\section{Results} \label{sec:results}

\subsection{Planets With Mass Measurements} \label{sec:result_mass}
Our default analysis produces final mass estimates for 19 of the 20 planets with observed masses. Figure \ref{fig:Mtot_Rpl_age} shows our final mass ranges given all the constraints described in Section \ref{sec:methods}, where we also show the results of extending the metallicity constraint out to 2-$\sigma$. When comparing the planetary and stellar metallicities to trends from \citet{Thorngren2016}, as described in Section \ref{sec:met_constraint}, we find that for only 5 planets the measured mass agrees with our metallicity constraint within the 1-$\sigma$ range. While the 1-$\sigma$ limit is a good reference, it is not unreasonable to find planets beyond that. Figures 7 and 11 of \citet{Thorngren2016} show the best-fit lines for the planetary and stellar metallicity trends, respectively, compared with the planets used to obtain the trends. Although most of the planets used to obtain the trends lie within the 1-$\sigma$ region, a significant number have metallicities outside the 1-$\sigma$ region. In each case, $\sim$40-50\% of the planets studied by \citet{Thorngren2016} lie outside the 1-$\sigma$ region. Therefore, we can safely extend our metallicity constraint out to 2-$\sigma$ without invalidating our results within which we find 6 planets with solutions consistent with their measured masses.

We now discuss planets that fail even the 2-$\sigma$ metallicity constraint.
Kepler-76 b does not have any masses that survive our constraints, either by mass-loss or by metallicity constraints; for example, the mass solutions within 1-$\sigma$ agreement with the measured mass lie beyond the 2-$\sigma$ metallicity constraints. Similarly, for Kepler-289 c and TOI-201 b, all the solutions which agree with the measured mass fall outside the 2-$\sigma$ range of the metallicity trends and are therefore ruled out (unlike Kepler-76 b, these planets have solutions that survive our mass loss constraints). In case of HATS-36 b, a subset of the solutions that are consistent with its measured mass lie just at the edge of 2-$\sigma$ metallicity constraint. We note that while the results from \citet{Thorngren2016} provide a good foundation for general metallicity trends obeyed by short-period gas giants, we cannot entirely disregard the possibility of outliers. 

The planets in the Kepler-51 and V1298 Tau systems are too puffy and require special formation constraints, where the planets likely accreted dust-free envelope farther out and migrated in to their present-day orbits (in contrast to our fiducial in situ dusty accretion). The Kepler-51 and V1298 Tau planets are discussed in more detail in Sections \ref{sec:Kepler51} and \ref{sec:V1298Tau}, respectively. 

For Kepler-43 b, the bounds of our interior structure grid (Section \ref{sec:int_struc_model}) do not allow for any mass solutions in agreement with the measured mass. Therefore, for this planet, we extend our analysis to include solutions which fall within the 1-$\sigma$ bounds of the measured radius to obtain a final mass range which encompasses the measured mass. Kepler-74 b has two sets of solutions since both the high-mass and low-mass family of solutions fall below the brown dwarf regime.

\subsection{Planets Without Mass Measurements} \label{sec:result_nomass}
Having shown that our methods can recover a mass estimate for 19 of the 20 planets with mass measurements, given their measured radius, age, and incident flux, we now move on to estimating the plausible mass range of the 2 planets without mass measurements: KOI-351 g and HIP 67522 b. 
The final mass range obtained from our methods is shown in Figure \ref{fig:Mtot_Rpl_age}, where the error bars without markers show the planets without mass measurements. We find that our method produces a solution which survives all constraints for both planets, straddling the boundary between low-mass and giant planets. However, HIP 67522 b has an upper limit of 20 M$_\oplus$ \citep{Thao2024}, which agrees with the lower end of our mass range, making it a puffed-up Neptune mass planet. We discuss this planet in more detail in Section \ref{sec:discussion}. 

\section{Discussion} \label{sec:discussion}

In general, we find planets larger than $\sim$10$R_\oplus$ tend to be consistent with the mass of a giant planet. For planets smaller in size, sub-Jovian mass becomes a greater possibility, particularly for very young ($<$100 Myr) planets. In this section, we discuss specific cases where the measured masses do not align with our fiducial analysis and require a special formation channel (Kepler-51, V1298 Tau planets). We additionally discuss HIP 67522 b, a planet likely undergoing rapid mass loss, and TOI-837 b, the only planet aged $<100$ Myr with a measured mass that is consistent with our definition of giant planet.

\subsection{Kepler-51 planets} \label{sec:Kepler51}

\begin{figure*}
    \gridline{\fig{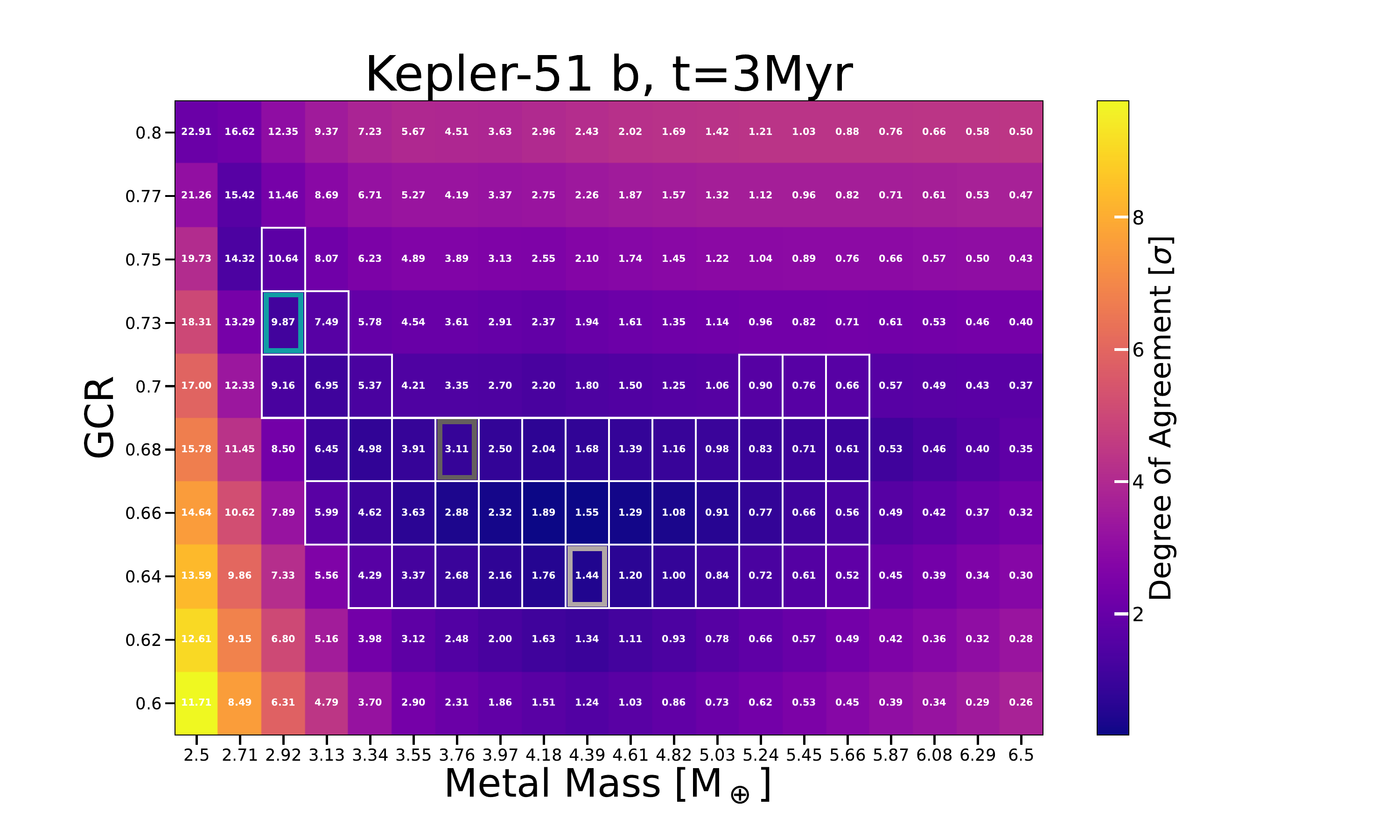}{0.5\textwidth}{} \fig{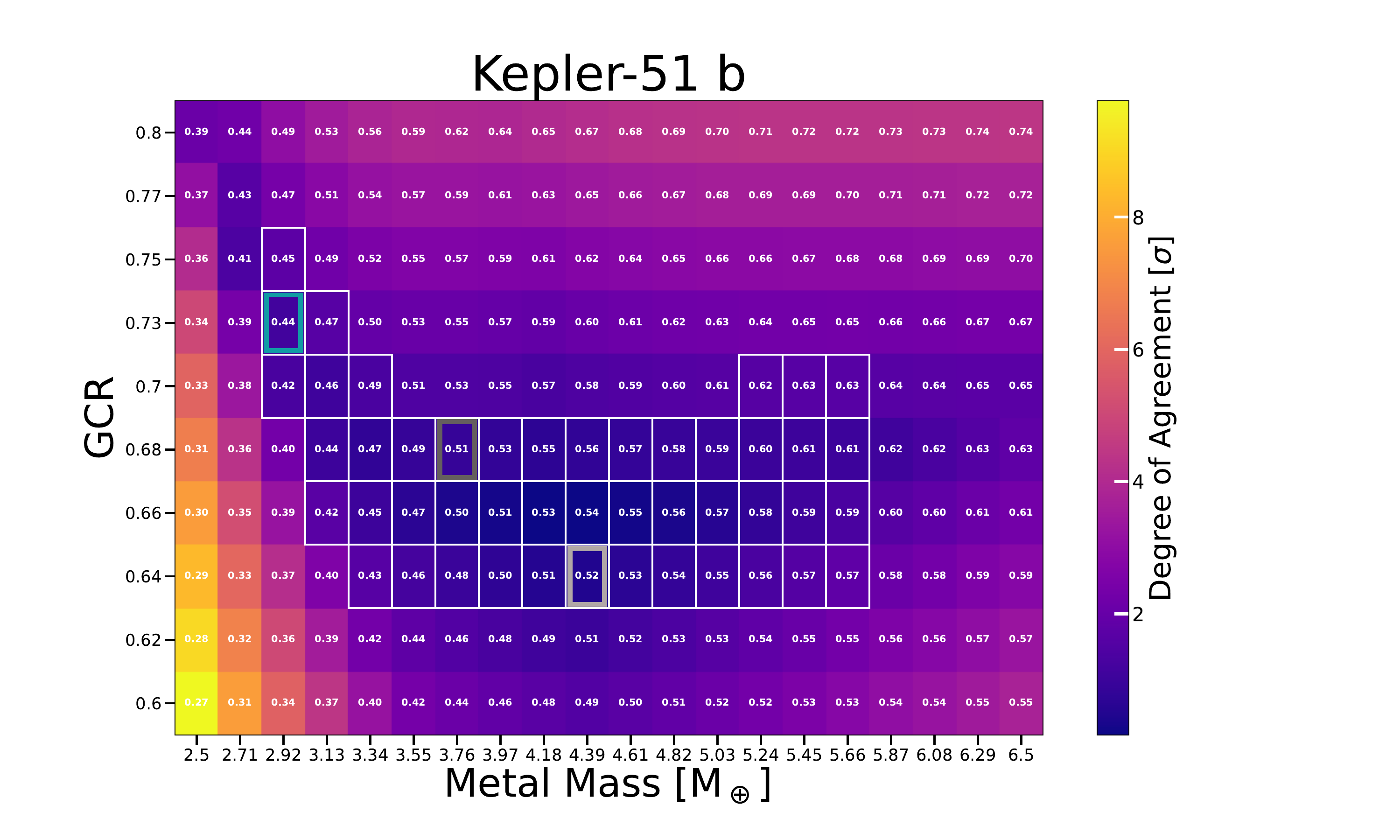}{0.5\textwidth}{}}
    \vspace{-1cm}    \gridline{\fig{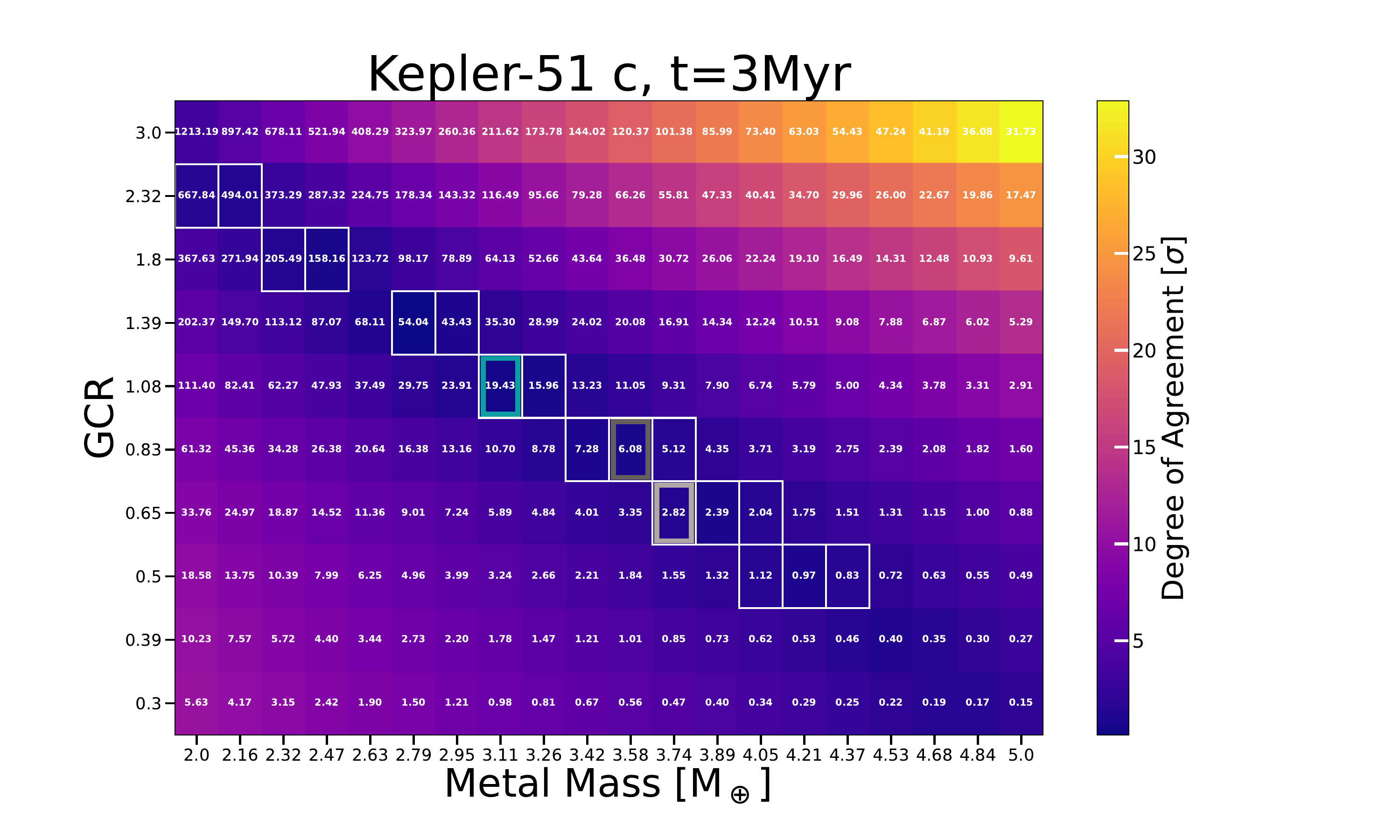}{0.5\textwidth}{}\fig{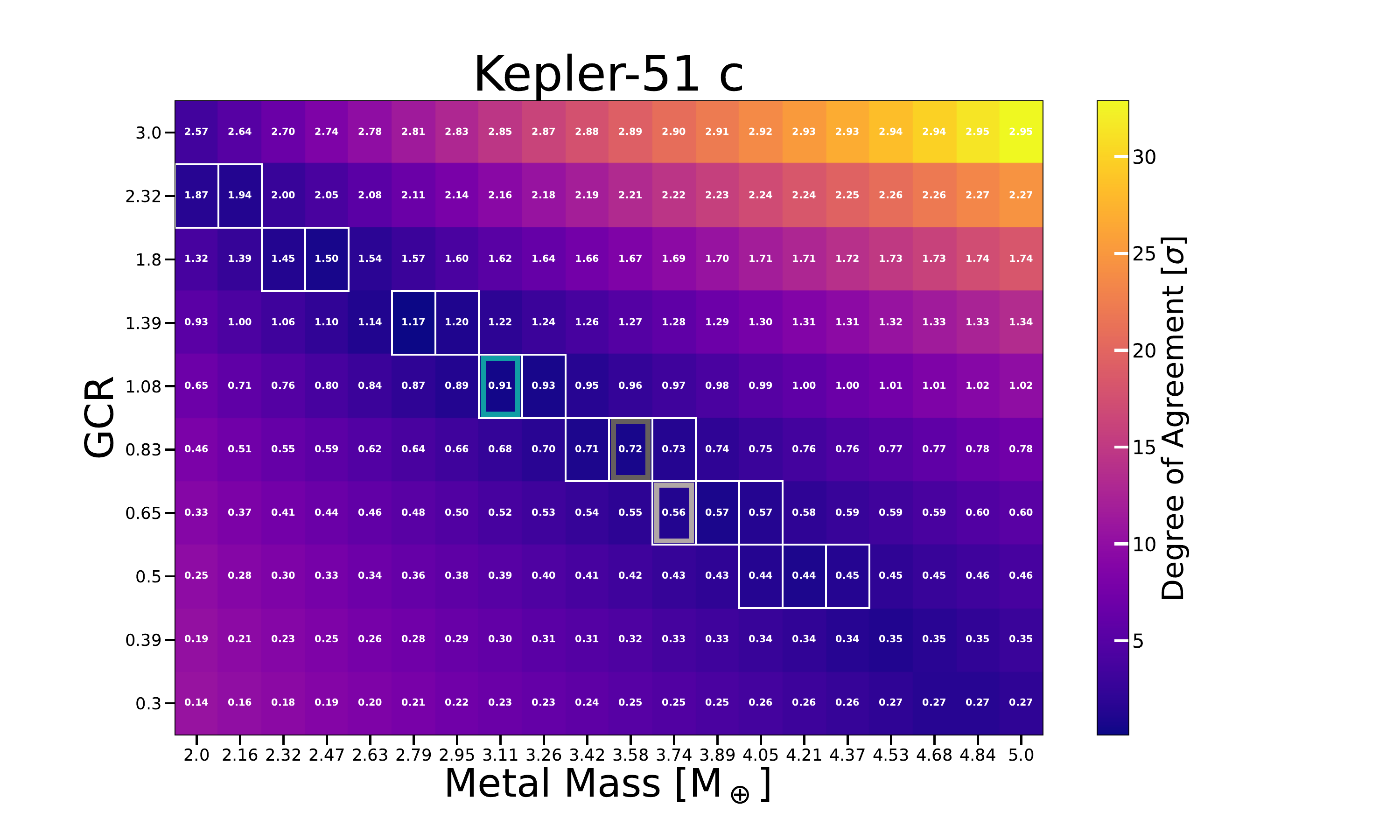}{0.5\textwidth}{}}
    \vspace{-1cm}
    \caption{Left: the difference between the measured and final radius and mass for Kepler-51 planets in terms of measurement error $\sigma$ (colormap) over the initial GCR and metal mass grid, assuming initial gas accretion for 3 Myr and mass loss at the planets' present locations over the age of the system. The corresponding initial formation locations are indicated in white text in units of 1000 days. The white contour line outlines the region of the parameter space where both the post-mass-loss radius and mass agree with the measured values within 1-$\sigma$. Grid points are boxed where the planets b and c are near 2:1 resonance within 5\% (corresponding pairs in same color box outline). Right: likewise as the left column but the white text denotes the final GCR.}
    \label{fig:Kepler51}
\end{figure*}

Kepler-51 is a well-known system harboring multiple super-puffs \citep{Masuda14,Libby-Roberts2020}. While the revised masses of member planets generally increase their bulk density \citep{Masuda24}, they remain puffier than expected for their size and age (see Figure \ref{fig:Mtot_Rpl_age}). \citet{Lee2016} hypothesized that the super-puffs likely formed beyond the water ice line where the rate of gas accretion under dust-free accretion greatly accelerates so that low mass cores can reach the gas-to-core mass ratio (GCR) well beyond 10\%. These planets can also appear spuriously large due to high-altitude hazes \citep{Wang19,Gao20}, planetary rings \citep{Piro20}, or tidal heating \citep{Millholland19}. Of these alternate scenarios, tidal heating is likely not the main cause of puffing up Kepler-51 planets as they are too far from the star. Neither high-altitude hazes nor planetary rings are mutually exclusive with initially high GCR (albeit lowering the initial GCR by some amount) so we proceed with the hypothesis of dust-free gas accretion followed by inward migration.

First, over a range of initial GCR and metal mass, we deduce the orbital period at which a planet of given metal mass would have been to reach a given GCR (since the metal mass relevant for Kepler-51 planets are always less than 20$M_\oplus$, their metal mass is equivalent to core mass and their gas-to-metal mass ratio is identical to GCR). We adopt the scaling relationship for dust-free accretion reported in \citet[][their equation 7]{Lee2022},

\begin{align} \label{eq:form_dust-free}
    {\rm GCR} &= 0.27 \left(\frac{t}{0.1 {\rm Myr}} \right)^{0.4} \left(\frac{M_{\rm metals}}{20 M_\oplus} \right)^{1.7} \nonumber \\
    &\times \left(\frac{T_{d}}{1000 K} \right)^{-1.5} {\rm Exp}\left(\frac{t}{2.2 {\rm Myr}} \left(\frac{M_{\rm metals}}{20 M_\oplus} \right)^{4.2} \right),
\end{align}
where the normalization is taken from visual inspection of Figure 3 of \citet{Lee2022}, $t$ is the accretion time varied over 1--10 Myr, the exponential term mimics runaway accretion (\citealt{Lee2019}, see also \citealt{Ginzburg19}), and $T_d = 1000\,{\rm K}\,(a/0.1\,{\rm au})^{-3/7}$ is the disk midplane temperature for an irradiated disk \citep{Chiang1997}. The dependence on disk gas surface density is insignificantly weak \citep[see also][]{Lee2016} so we ignore it. We note that the steep dependence of GCR on $T_d$ arises from the temperature-dependence of gaseous opacity which is benchmarked to \citet{Freedman14} and \citet{Lee18}, the latter of which is based on the calculations outlined in \citet{Ferguson05}. Our chosen opacities are apt to test the hypothesis of dust-free gas accretion; however, any opacity that drops at colder temperatures---generally expected for gaseous opacities from the freeze-out of internal degrees of freedom---would produce similar results.

Next, over the same grid of initial GCR and metal mass, we compute atmospheric mass loss as described in Section \ref{sec:calc_ML} at the present location of each planet over the system age. We evaluate the final mass and radius of the planet using our interior structure model (Section \ref{sec:int_struc_model}) and compare them to the measured mass and radius. Figure \ref{fig:Kepler51} summarizes our calculation where we illustrate in colors the degree of agreement between our model and observations in units of 1-$\sigma$ measurement error. For ease of visualization, we sum the mass and radius deviation; i.e., each color represents $|{\rm measured\,\,mass - model\,\,mass}|/\sigma_{\rm mass} + |{\rm measured\,\,radius - model\,\,radius}|/\sigma_{\rm radius}$. Highlighted in thin white lines are the GCR-metal mass pairs that are consistent with both the present-day mass and radius within 1-$\sigma$ after mass loss. We find a strictly negative correlation between the plausible initial GCR and metal mass for Kepler-51 c as expected for a planet that undergoes minimal level of mass loss (i.e., to explain the same total mass and radius, we need lower GCR for higher metal mass). Kepler-51 b is closer to the star so the higher irradiation flux contributes to extra heating, puffing up the planet especially for lower metal mass so we require slightly lower initial (and final; see right column of Figure \ref{fig:Kepler51}) GCR for lower metal mass down to $\sim$3.5$M_\oplus$.

Among the GCR-metal mass pairs that lead to final radii and masses that agree with the measured values within 1-$\sigma$ error, we locate the most likely formation location for each planet assuming that the two planets are within 5\% of 2:1 orbital period ratio, their present-day near-resonant configuration. We find four sets of solutions, highlighted in colored boxes in Figure \ref{fig:Kepler51}. Specifically, these sets of solutions are:
\begin{itemize}
    \item Kepler-51 b 
    \begin{itemize}
        \item GCR$_{\rm init}$ = [0.64, 0.68, 0.73]
        \item GCR$_{\rm fin}$ = [0.52, 0.51, 0.44]
        \item M$_{\rm metal}$ = [4.39, 3.76, 2.92] $M_\oplus$
        \item $a_{\rm init}$ = [2.53, 4.22, 9.12] au
    \end{itemize}
    \item Kepler-51 c
    \begin{itemize}
        \item GCR$_{\rm init}$ = [0.65, 0.83, 1.08]
        \item GCR$_{\rm fin}$ = [0.56, 0.72, 0.91]
        \item M$_{\rm metal}$ = [3.74, 3.58, 3.11] $M_\oplus$
        \item $a_{\rm init}$ = [3.96, 6.61, 14.3] au
    \end{itemize}
\end{itemize}
where GCR$_{\rm init}$ is the initial GCR, GCR$_{\rm fin}$ is the final GCR, and $a_{\rm init}$ is the formation location. While Kepler-51 d is not in our sample because it is too far from the host star, it is currently near 3:2 resonance with planet c so its corresponding $a_{\rm init} = [5.19, 8.66, 18.8]$ au.

All of our solutions indicate formation beyond the typical water ice line $\sim$1 au and therefore subsequent large-scale migration to bring both planets to their present locations $\sim$0.25 au (planet b) and $\sim$0.38 au (planet c), consistent with the hypothesis of \citet{Lee2016}. We note that our calculation is limited to two-layer planets with inner metallic core and H/He envelope on top with atmospheric metallicity that of solar. One consequence of formation beyond the water ice line is the deposition of water during and after formation. Our solutions for GCR$_{\rm init}$ and $a_{\rm init}$ will likely change if we allow for a significant amount of water deposition (\citealt{Vlahos24}; Gao et al.~in preparation).

The estimated age of Kepler-51 is 500$\pm$250 Myr. Over the full 1-$\sigma$ error in the age, we find that while the exact resonant pair solutions for Kepler-51 b \& c can vary, their $a_{\rm init}$ are always beyond 1 au with minimal change to the initial and final GCR (see Appendix \ref{sec:app-age}), demonstrating the robustness of our qualitative result.

\begin{figure*}
    \gridline{\fig{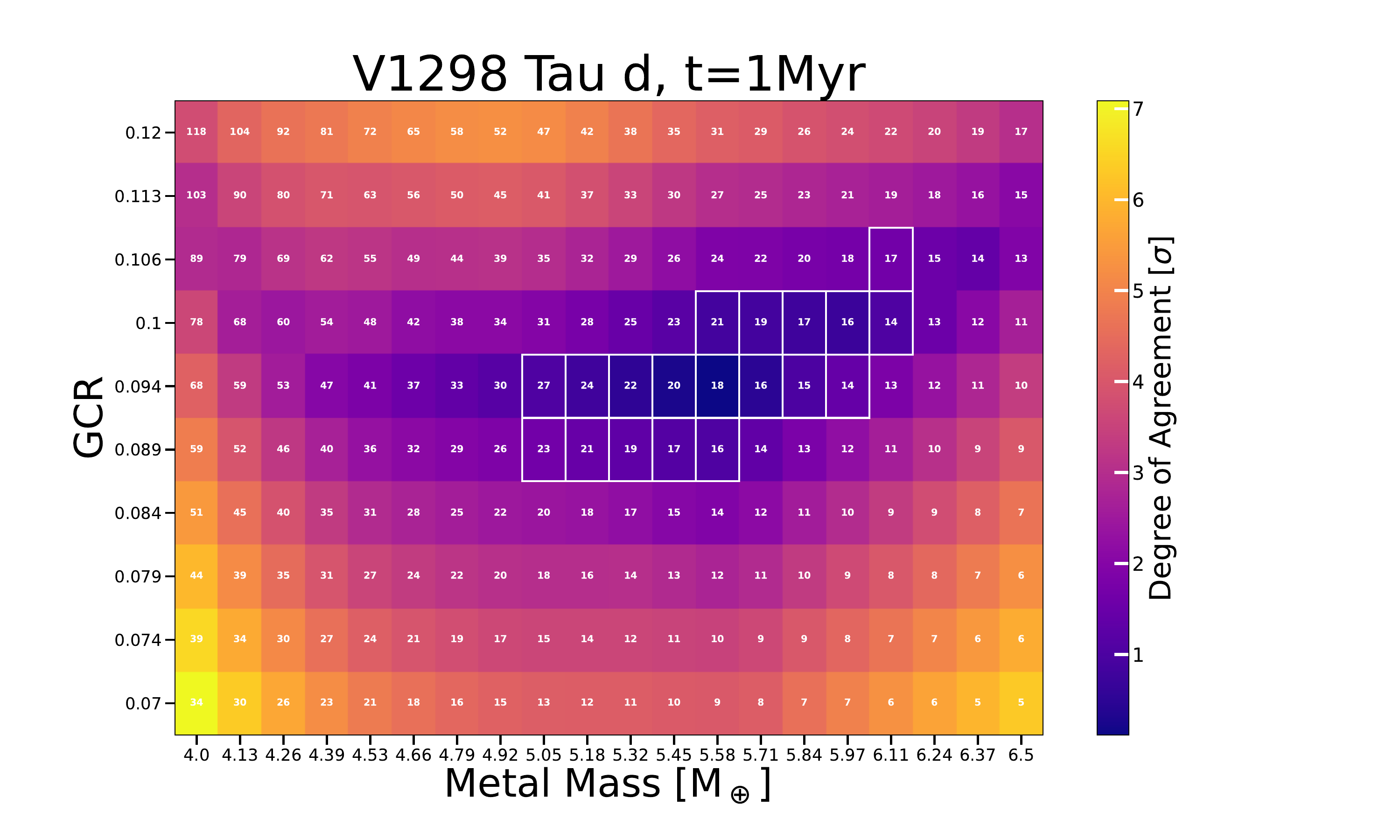}{0.5\textwidth}{}\fig{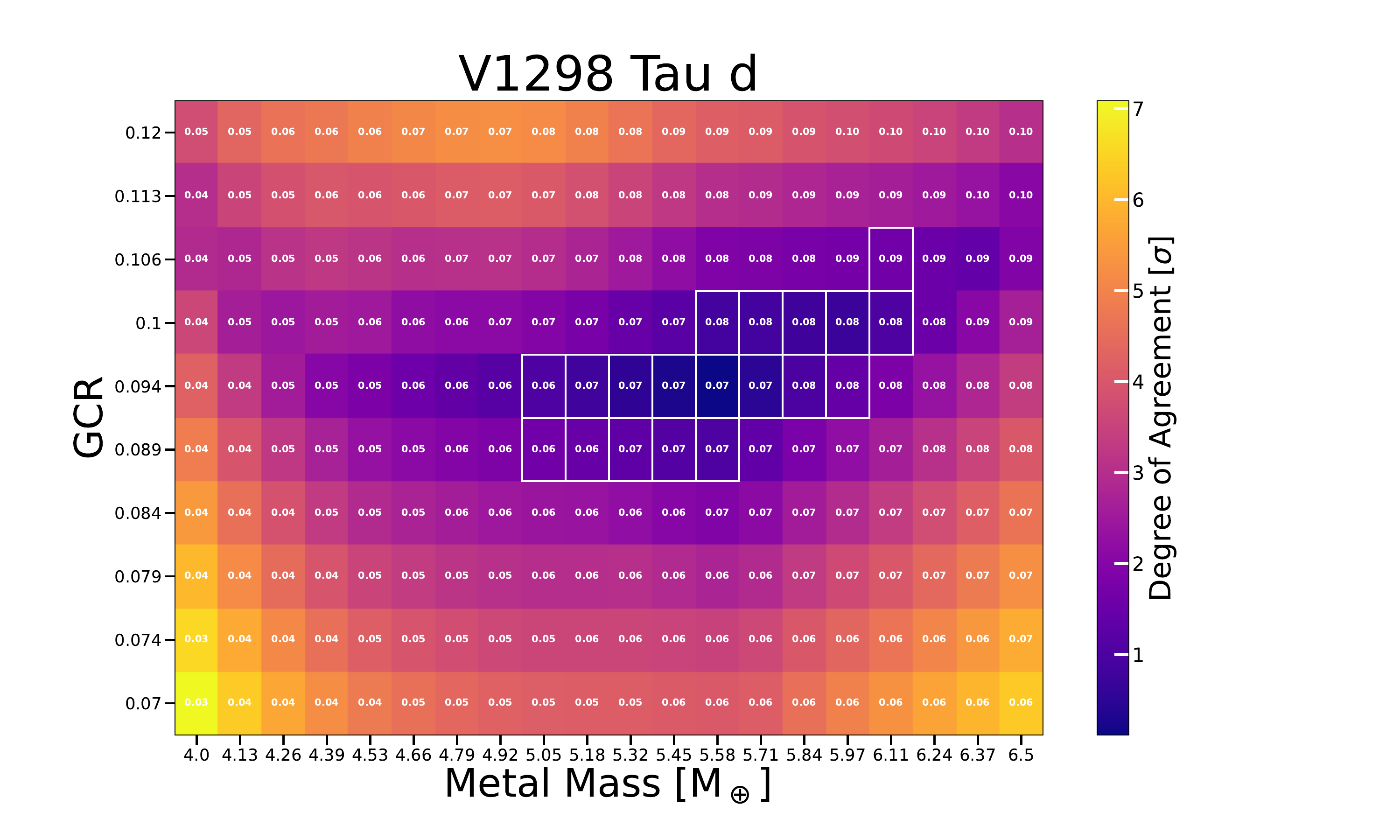}{0.5\textwidth}{}}
    \vspace{-0.9cm}
    \gridline{\fig{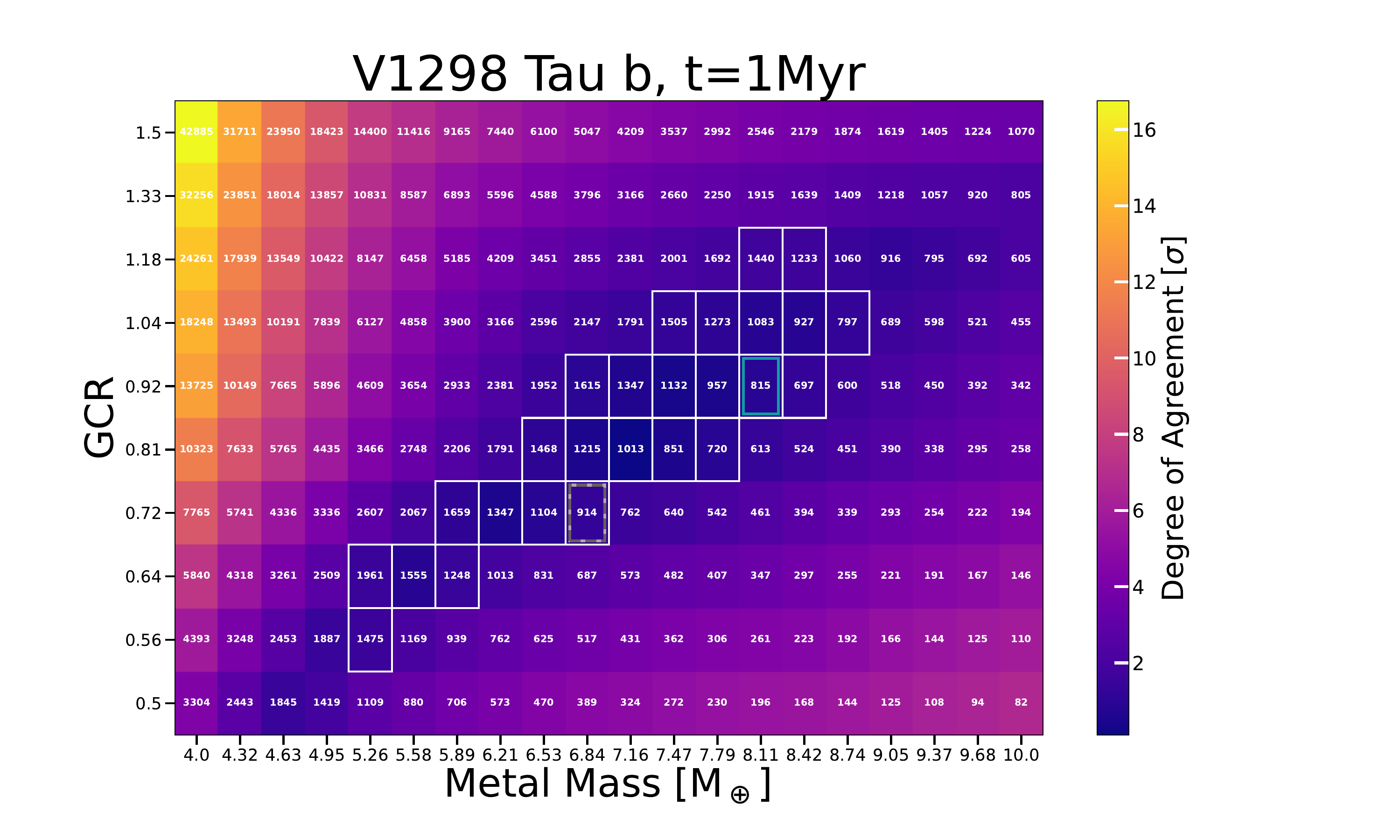}{0.5\textwidth}{}\fig{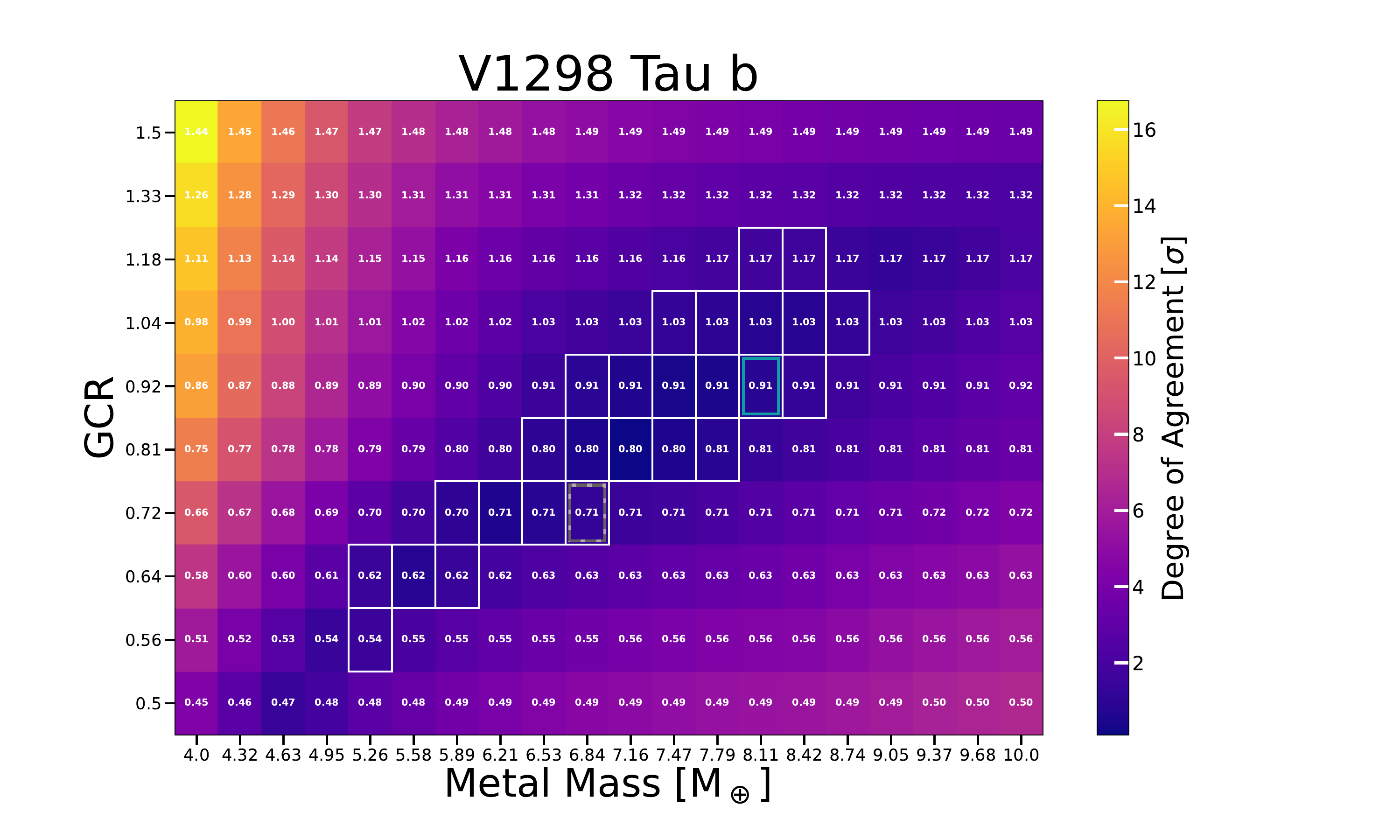}{0.5\textwidth}{}}
    \vspace{-0.9cm}
    \gridline{\fig{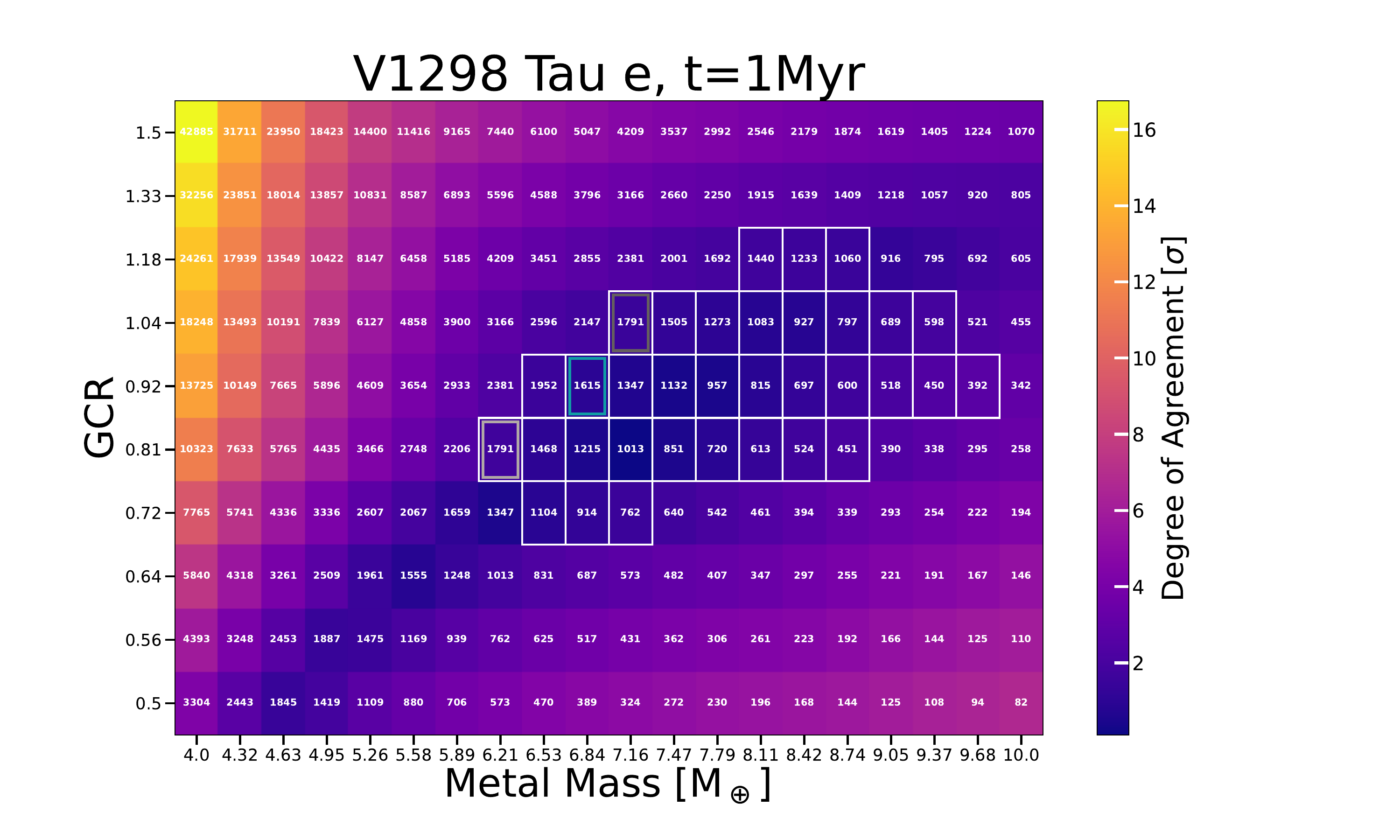}{0.5\textwidth}{}\fig{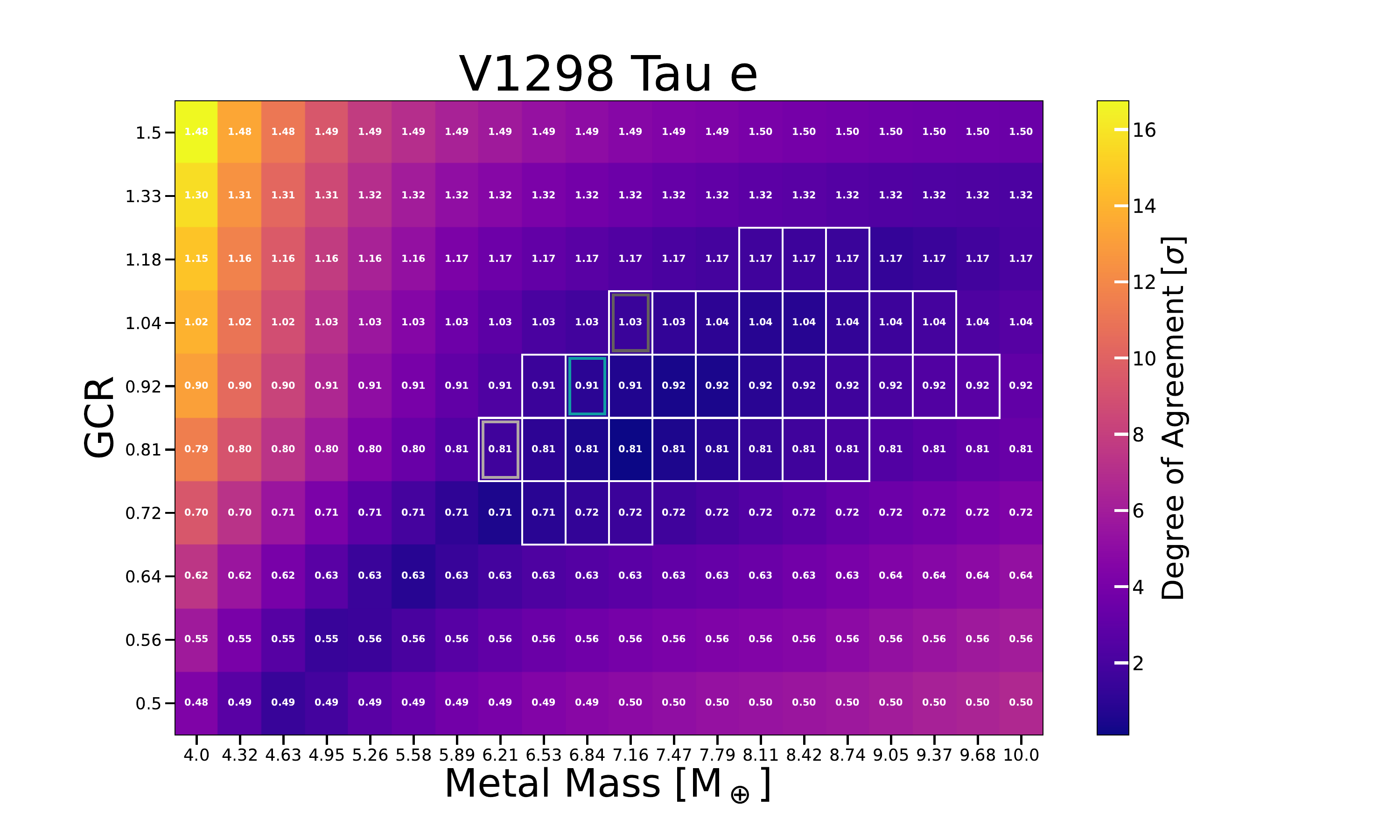}{0.5\textwidth}{}}
    \vspace{-0.9cm}
    \caption{Likewise as Figure \ref{fig:Kepler51} but for V1298 Tau planets and the formation locations are indicated in units of 1 day and gas accretion of 1 Myr. The initial GCR - metal mass pairing that provides radius and mass consistent with the measurement for planet d correspond to formation location that are already very close to its present location. For planets b and e, larger-scale migration is a possibility and we highlight the boxes that are near 2:1 period ratio between the two planets.}
    \label{fig:V1298Tau}
\end{figure*}

\subsection{V1298 Tau planets} \label{sec:V1298Tau}

We apply the same analysis as in Section \ref{sec:Kepler51} for V1298 Tau planets in our sample. We limit the gas accretion time to 1 Myr because any longer duration leads to the formation location of planet d to be inside its present location which we find to be unlikely. Physically, shorter gas accretion time implies later assembly of planetary cores.

Similar to Kepler-51 b (and unlike Kepler-51 c), a positive GCR-metal mass scaling that provides a set of solutions that agree with measured mass and radius within 1-$\sigma$ error is observed for V1298 Tau d. Owing to the extra puffing of planets by stellar irradiation (which drives up the rate of mass loss), the required initial GCR is low $\sim$0.09. Even for planets b and e which are slightly farther out from the host star, we see a generally {\it positive} correlation between the initial GCR and metal mass that matches the measured radius and mass. V1298 Tau is a much younger system compared to Kepler-51, and both planets b and e have shorter orbital periods as compared to Kepler-51 b. Consequently, all V1298 Tau planets are subject to higher irradiation flux, increasing the extra heating that enlarges the planets significantly towards lower metal mass, requiring lower GCR to explain the same radius (see dashed lines in Figure \ref{fig:grid_withMass}).

It has been argued that planets b and e are near 2:1 resonance \citep{Feinstein22}, although not necessarily formally in resonance \citep{Turrini23}. While the quoted orbital periods of planet d suggest it may be close to 2:1 resonance with planet b, whether this planetary system is in a resonant chain or not remains to be confirmed. Our inferred initial formation location of planet d is close to its present-day location, both of which are within $\sim$20 days and close to the typical spin periods of pre-main sequence stars \citep[e.g.,][]{Irwin08}. We surmise that planet d formed close to the inner edge of the disk set by the co-rotation radius with respect to the host stellar spin \citep{Ghosh79,Konigl91} whereas the outer planets formed much farther away and underwent a large-scale inward migration. Following the same analysis as Section \ref{sec:Kepler51}, the sets of solutions for planets b and e under the assumption that their initial locations were close to 2:1 resonance are:
\begin{itemize}
    \item V1298 Tau b
    \begin{itemize}
        \item GCR$_{\rm init}$ = [0.72, 0.72, 0.92]
        \item GCR$_{\rm fin}$ = [0.71, 0.71, 0.91]
        \item M$_{\rm metal}$ = [6.84, 6.84, 8.11] $M_\oplus$
        \item $a_{\rm init}$ = [1.90, 1.90, 1.76] au
    \end{itemize}
    \item V1298 Tau e
    \begin{itemize}
        \item GCR$_{\rm init}$ = [0.81, 1.04, 0.92]
        \item GCR$_{\rm fin}$ = [0.81, 1.03, 0.91]
        \item M$_{\rm metal}$ = [6.21, 7.16, 6.84] $M_\oplus$
        \item $a_{\rm init}$ = [2.97, 2.97, 2.78] au
    \end{itemize}
\end{itemize}

The level of mass loss is more muted for V1298 Tau planets compared to Kepler-51 planets (in spite of higher irradiation flux) because the former are more massive. Nevertheless, like Kepler-51, we find that a similar formation scenario---formation beyond $\sim$1 au followed by large-scale inward migration---can explain the current mass and radii of V1298 Tau b and e. At face value, such a scenario may be in tension with the reported substellar oxygen abundance \citep{Barat2023} for planet b. However, oxygen-rich solids can be locked into the central core and/or deep interior without being dredged up all the way to high atmospheric altitudes where we can probe the atmospheric composition. Furthermore, rapid inward migration would have prevented the planet from attaining an water-enriched envelope \citep{Vlahos24}.
Alternatively, in situ gas accretion for planet b is possible if the gas metallicity is as low as 0.01$\times$ solar value, which is more than an order of magnitude below the value reported by \citet{Barat2023} so we do not favor this scenario. By contrast, planet d is more robustly consistent with practically in situ gas accretion. Similar to Kepler-51 planets, these results---in situ gas accretion of planet d and large-scale migration of planets b and e---are robust to the uncertainties in the estimated age within 1-$\sigma$ error (see Appendix \ref{sec:app-age}).

\subsection{HIP 67522 b} \label{sec:HIP67522b}

At the measured radius and the upper limit on mass from atmospheric constraint $\lesssim$20$M_\oplus$ \citep{Thao2024}, HIP 67522 b has a bulk density that is below 0.1 g/cm$^3$, the lower limit below which very few planets are seen in nature \citep[see the bottom panel of Figure 1 from][]{Thorngren2023}. At its high irradiation flux (i.e., short orbital period), such low density is expected to spell runaway mass loss through an adiabatic expansion of the envelope during its photoevaporative loss.

Given the extremely young age of HIP 67522 b (t$_* \sim$ 17 Myr), we may be observing the planet in a semi-runaway state, having not yet lost its entire envelope. Figure 3 of \citet{Thorngren2023} illustrates planets at various ages undergoing mass loss and thermal evolution and shows that at 10 Myr, small planets ($\sim$20 M$_\oplus$, $\sim$11 R$_\oplus$) have begun to lose mass but still have most of their envelopes, consistent with the existence and the measured properties of HIP 67522 b. This planet may be an ideal target to study ongoing mass loss process.

\subsection{TOI-837 b}
Recent radial velocity measurements from the High Accuracy Radial
Velocity Planet Searcher \citep[HARPS;][]{Mayor2003} find that TOI-837 b has a mass of $\sim$ 120 M$_\oplus$ \citep{Barragan2024} which is in agreement with our proposed mass range with the 2-$\sigma$ metallicity constraint. This mass estimate would imply that TOI-837 b is a young ($\sim$ 35 Myr), hot Jupiter that migrated to its close-in orbit ($\sim$ 8.32 days) over a short timescale. This suggests that disk gas migration is a possible origin channel for hot Jupiters. However, it is likely not the dominant one given the rarity of gas giants around young stars ($\lesssim$ 100 Myr).

\subsection{Implications for Hot Jupiter Migration} \label{sec:disc_hotJup}

In general, we find that planets larger than 10$R_\oplus$ are much more likely to have masses $\gtrsim$100$M_\oplus$, consistent with being gas giants. In terms of system age, 
among the particularly young planets ($\lesssim$40 Myr), with the exception of TOI-837 b, all the planets in our sample (HIP 67522 and V1298 Tau) have either measured or upper limit on masses that place them well within the Neptune to sub-Neptune regime. 

We additionally analyze K2-33 b (5.04$^{+0.34}_{-0.37}R_\oplus$, 5.425 days around 1.02$M_\odot$ star aged 9 Myr; \citealt{K2-33_pl_st}), DS Tuc A b (5.7$\pm 0.17 R_\oplus$, 8.138 days around 1.01$M_\odot$ star aged 45 Myr; \citealt{DSTucA_pl_st}, metallicity from \citealt{DSTucA_met}), and TOI-1227 b (9.572$^{+0.751}_{-0.583}R_\oplus$, 27.364 days around 0.17$M_\odot$ star aged 11 Myr; \citealt{TOI-1227_pl_st}) that were excluded by our target selection (K2-33 b and DS Tuc A b are smaller than 6$R_\oplus$ while TOI-1227 b is around a very low mass star so its insolation flux was too low to be considered even a warm Jupiter). We find that the range of masses consistent with 1-$\sigma$ metallicity constraint are 12--30.60 $M_\oplus$, 14.38--18.40 $M_\oplus$, and 38.26--222.65$M_\oplus$ for K2-33 b, DS Tuc A b, and TOI-1227 b, respectively. Allowing for a 2-$\sigma$ metallicity constraint increases the upper limit to 148.77 $M_\oplus$, 73.29$M_\oplus$, and 486.37$M_\oplus$. These results are largely consistent with our main result showcased in Figure \ref{fig:Mtot_Rpl_age}. The smaller planets K2-33 b and DS Tuc A b are more consistent with being puffed-up Neptunes whereas the larger TOI-1227 b has a higher likelihood of being a gas giant. We note however that in terms of insolation flux, TOI-1227 b is considered a {\it cold} Jupiter and the disk temperature \citep[][their equation 18]{Chachan23} at its current location $\sim$200 K is cold enough to allow for rapid gas accretion in situ (GCR $>$1 over 0.1 Myr; see Equation \ref{eq:form_dust-free}). Given that TOI-1227 b likely does not require either disk or high eccentricity migration to explain its property, we do not consider it as part of our sample of young hot/warm Jupiter.

We further consider the possibility of young hot and warm Jupiters being inflated to a size large enough to be misclassified as an eclipsing stellar binary, and we consider this unlikely. Following the same procedure as \citet{Thorngren2023}, we find that young giants would cool down to $\lesssim$1.5 Jupiter radius within $\sim$20--40 Myr even with extra heating and so they would have been classified as a giant planet.

In the sample we study (plus K2-33 b and DS Tuc A b), we see no planet larger than 10$R_\oplus$ of age younger than $\sim$100 Myr.
This paucity of very young, large and massive planets provides tentative evidence that the dominant origin channel for hot Jupiters is likely high eccentricity migration.
If disk gas migration were dominant, we would find Jupiter mass planets around stars as young as $\sim$10 Myr, but we instead find that the youngest hot/warm Jupiter candidates are inflated lower mass planets. The one exception of TOI-837 b may be a rare case of a hot Jupiter that arrived at its present location by disk migration. Our result is consistent with obliquity measurements from \citet{Spalding2022}, who find evidence that most hot Jupiters arrive late ($\gtrsim$ 100 Myr) based on tidal theory and stellar evolution models. 

We argue that our finding is statistically significant against the general rarity of hot and warm Jupiters seen in the mature system, which reports a cumulative occurrence rate of giants at $\sim$4--5\% out to $\sim$200 days \citep[e.g.,][]{Fressin13,Petigura18}. We find $\sim$66 systems younger than 1 Gyr across K2, {\it Kepler}, and {\it TESS} detections at the time of writing this paper. If the hot/warm Jupiter occurrence rate is the same as that of main sequence stars, we expect to find $\sim$3 giants. We find more than 3 (albeit without any correction for detection bias/sensitivity, transit probability, etc.), with a clear trend towards lack of giant at ages $\lesssim$100 Myr.

\section{Conclusion} \label{sec:conclusion}

In this paper, we obtain a theoretical mass constraint on 22 young planets with ages $\sim$5-900 Myr and radii $\sim$6-16 R$_\oplus$ to determine whether they are massive hot Jupiters or merely puffy Neptunes. We use interior structure models to obtain an initial mass range which we narrow down using photoevaporative mass loss constraints and empirical mass-metallicity trends. 
Our predicted mass ranges are generally in agreement with the 21 planets for which there is a measured mass or upper limit. Exceptional cases such as Kepler-51 and V1298 Tau planets require special formation conditions consistent with dust-free accretion beyond $\sim$1 au followed by inward migration.
We find that the planets aged less than a few tens of Myr tend to be puffy, lower-mass planets, while the bona fide hot Jupiters are typically found around older stars, aged more than a few hundred Myr. Such a division in age suggests that hot Jupiters likely arrive at their short orbital periods predominantly through a process that operates over longer timescales, favoring high eccentricity migration.

\vspace{0.5cm}

The anonymous referee provided a positive and thoughtful report that helped improve the manuscript.
We thank Luke Bouma, Nicolas Cowan, Peter Gao, Heather Knutson, Jessica Libby-Roberts, John Livingston, Georgia Mraz, and Erik Petigura for useful discussions. This research has made use of the NASA Exoplanet Archive, which is operated by the California Institute of Technology, under contract with the National Aeronautics and Space Administration under the Exoplanet Exploration Program. A. K. acknowledges support from NSERC and FRQNT. E.J.L. gratefully acknowledges support by NSERC, by FRQNT, by the Trottier Space Institute, and by the William Dawson Scholarship from McGill University.  D.P.T. thanks the Allan C. and Dorothy H. Davis Postdoctoral Fellowship at Johns Hopkins University for support.

\appendix
\counterwithin{figure}{section}

\section{Effect of Age Uncertainty}
\label{sec:app-age}

\begin{figure*}
    \gridline{\fig{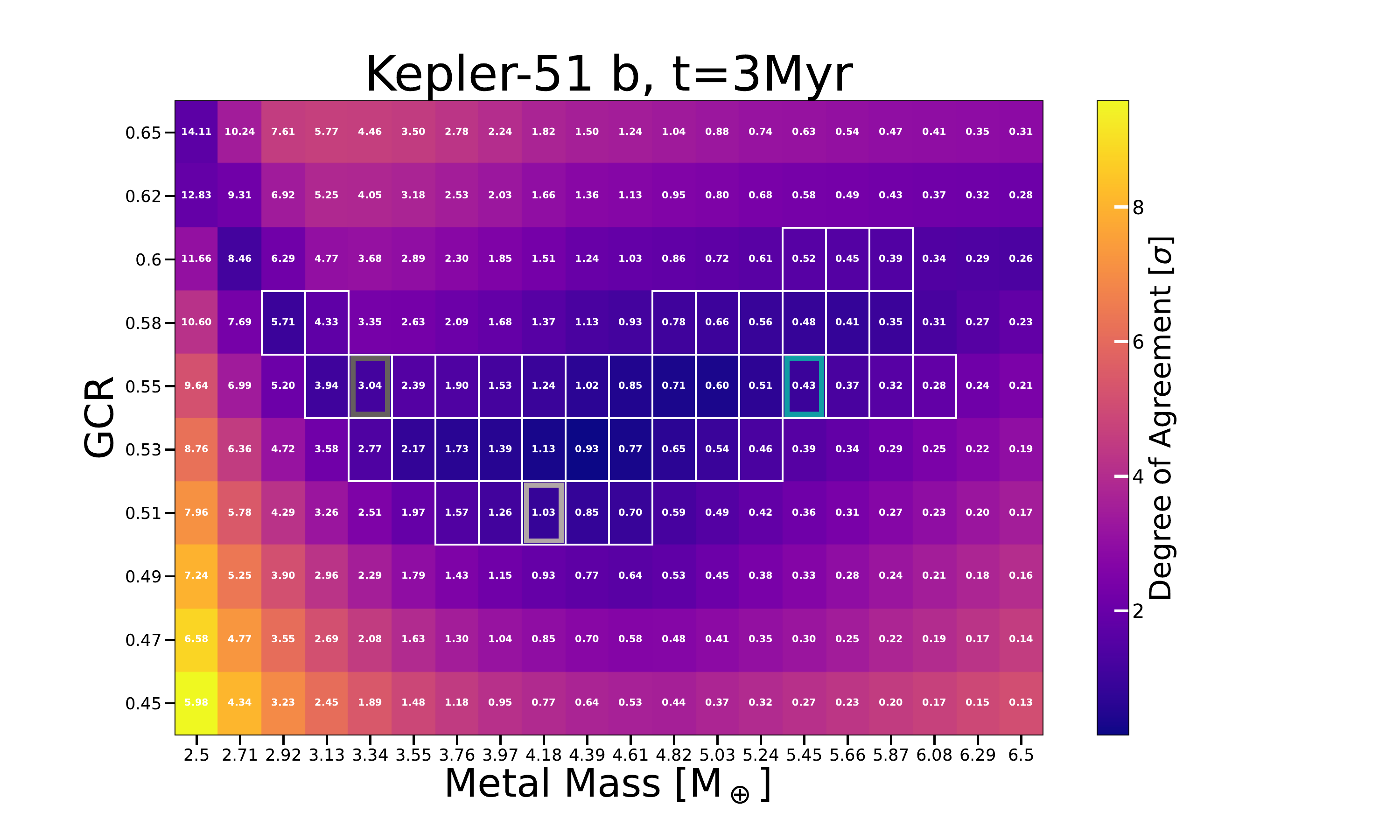}{0.5\textwidth}{} \fig{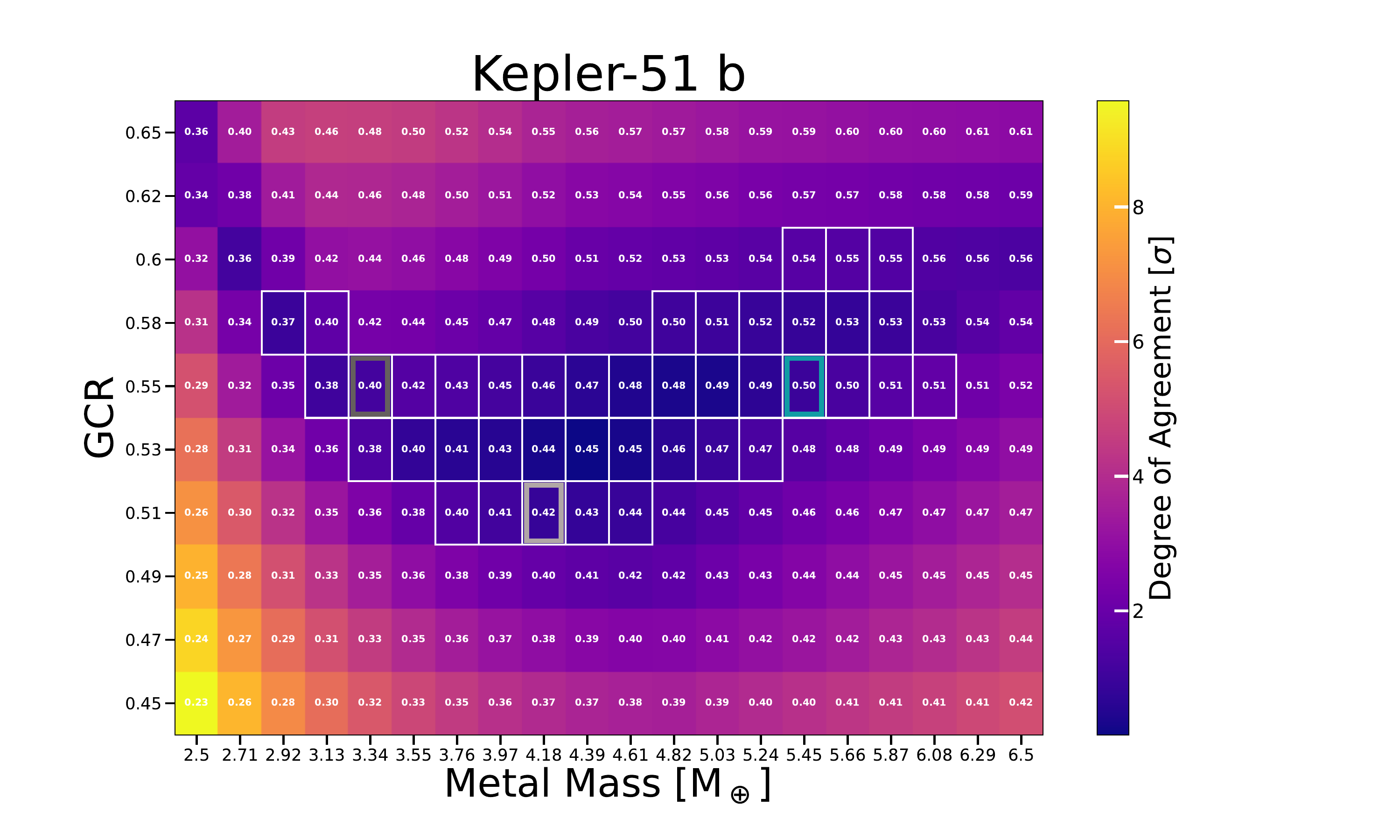}{0.5\textwidth}{}}
    \vspace{-1cm}    \gridline{\fig{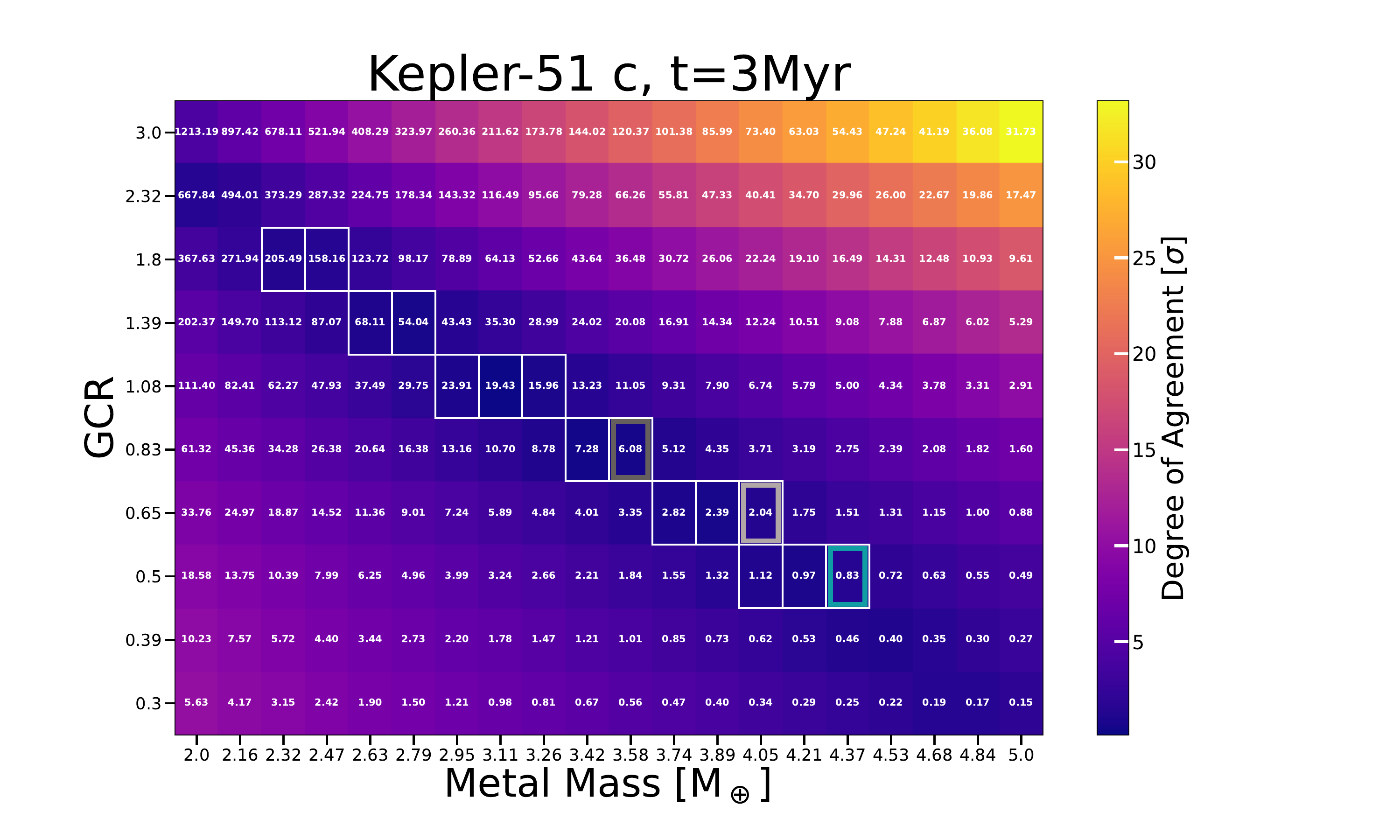}{0.5\textwidth}{}\fig{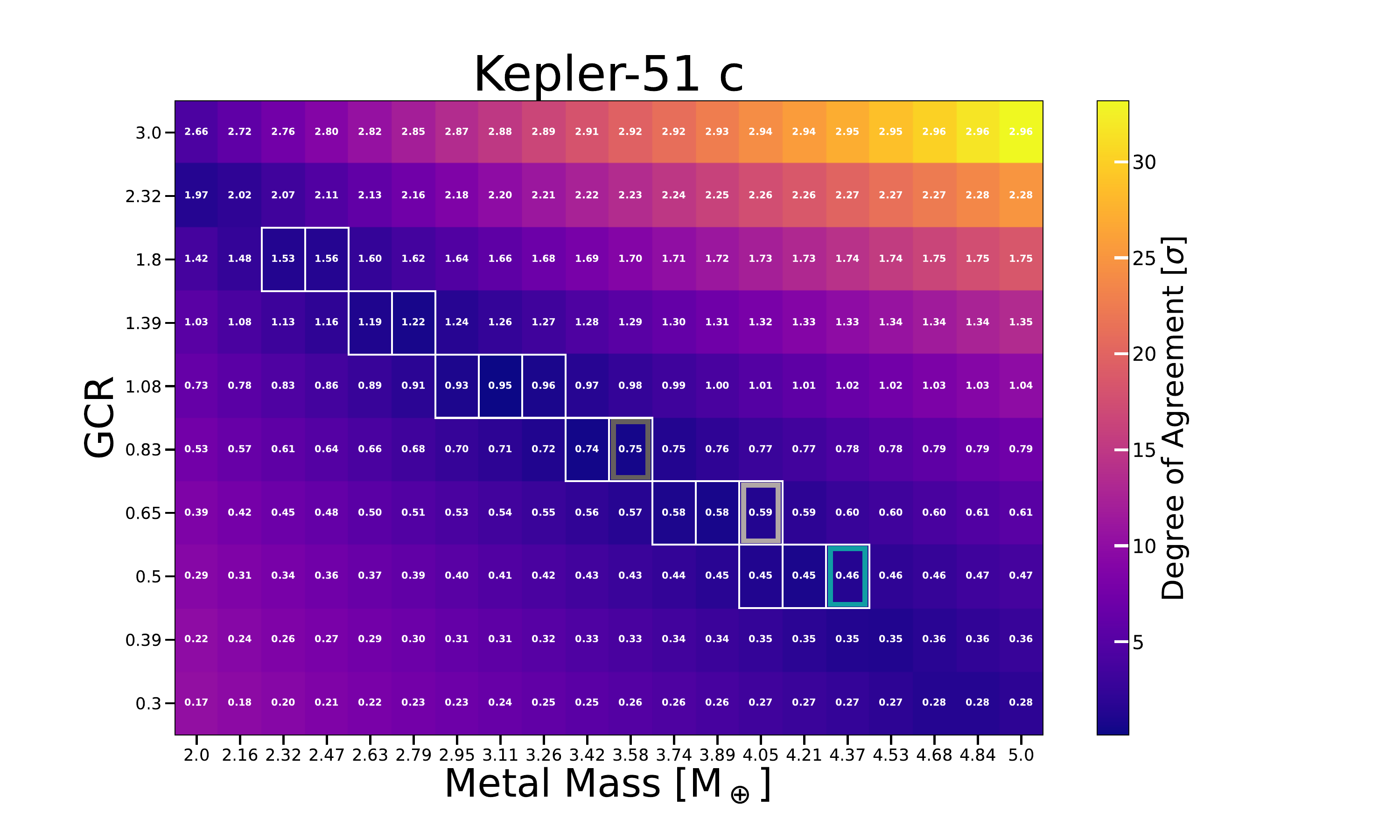}{0.5\textwidth}{}}
    \vspace{-1cm}
    \caption{Same as Figure \ref{fig:Kepler51} but at 250 Myr (lower 1-$\sigma$ limit of the estimated age).}
    \label{fig:Kepler51-agelerr}
\end{figure*}

\begin{figure*}
    \gridline{\fig{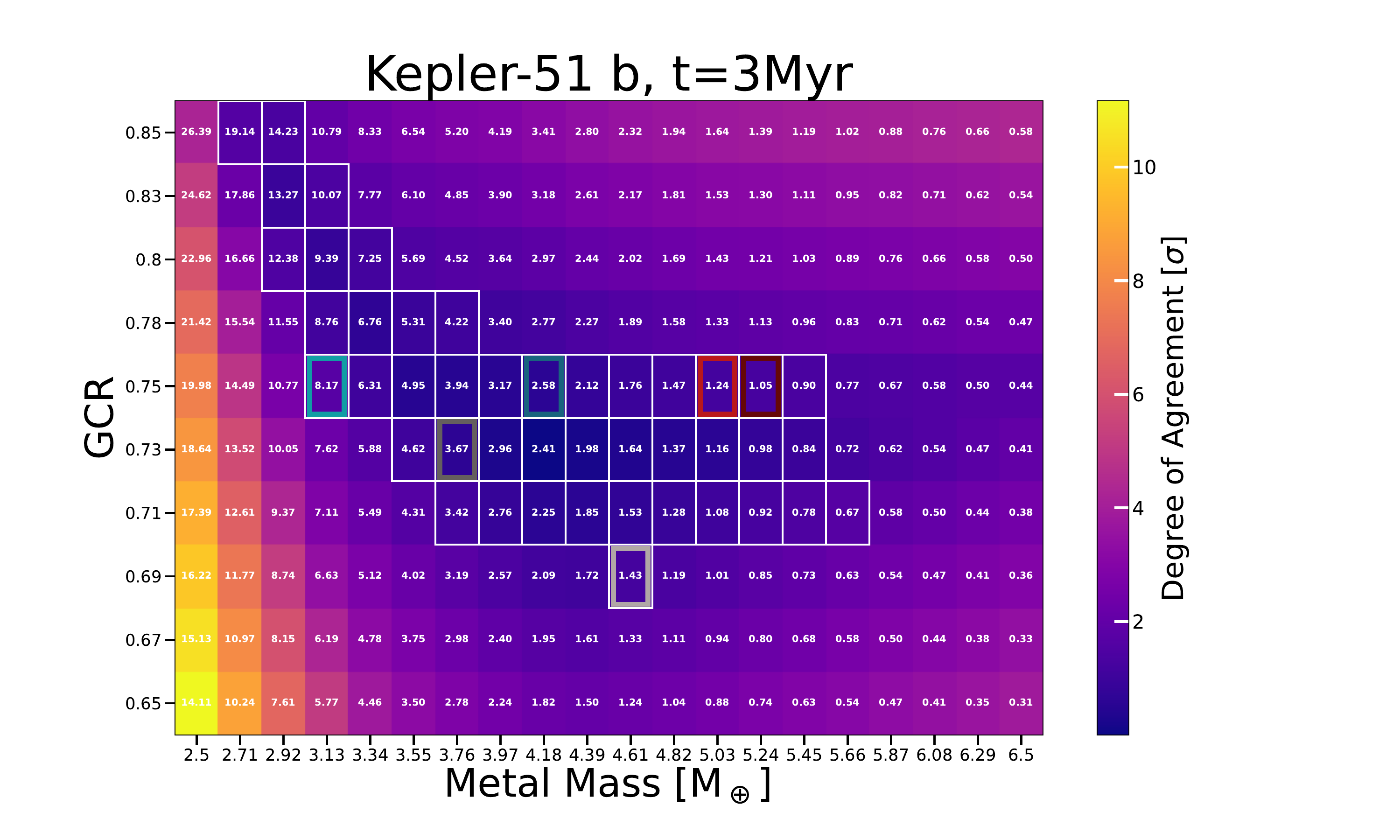}{0.5\textwidth}{} \fig{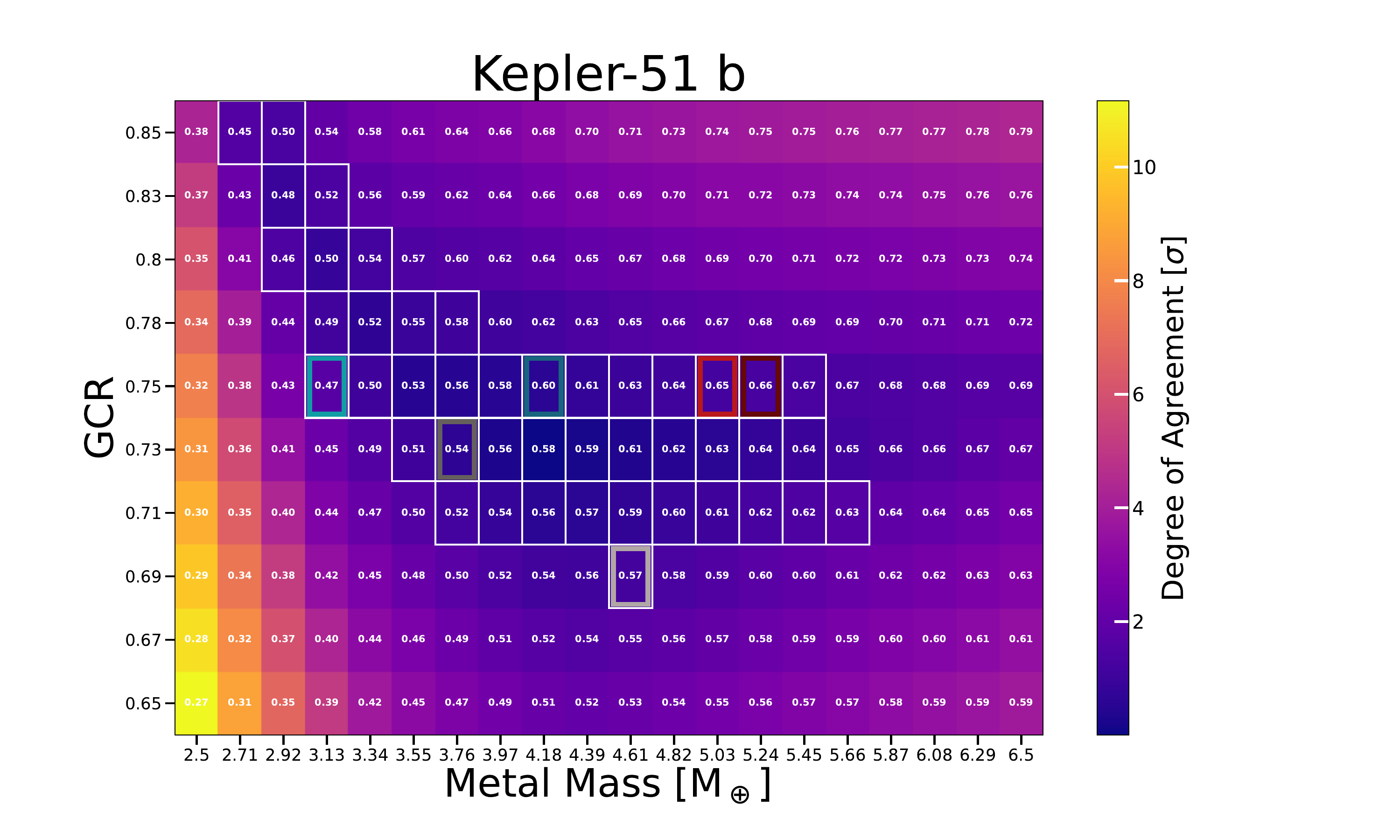}{0.5\textwidth}{}}
    \vspace{-1cm}    \gridline{\fig{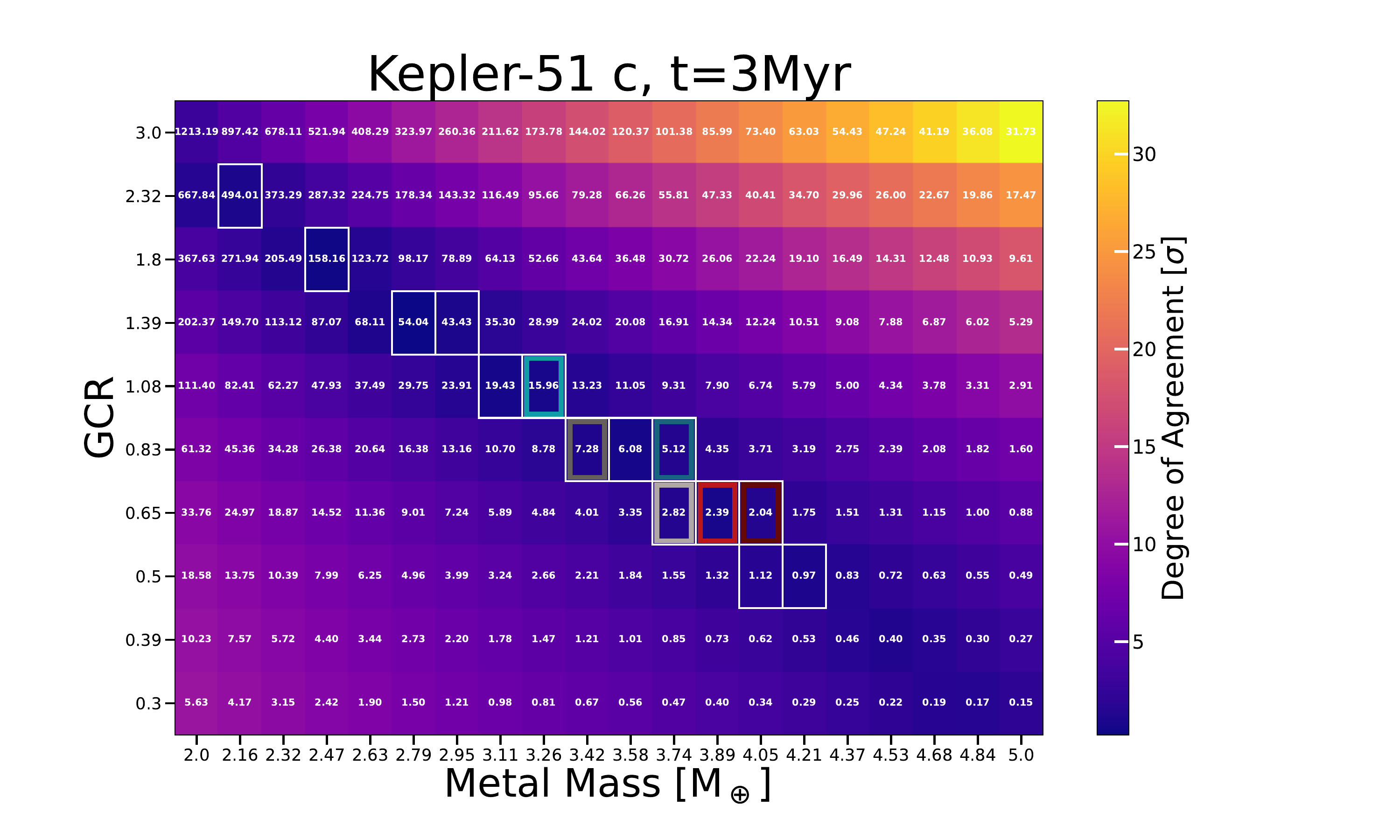}{0.5\textwidth}{}\fig{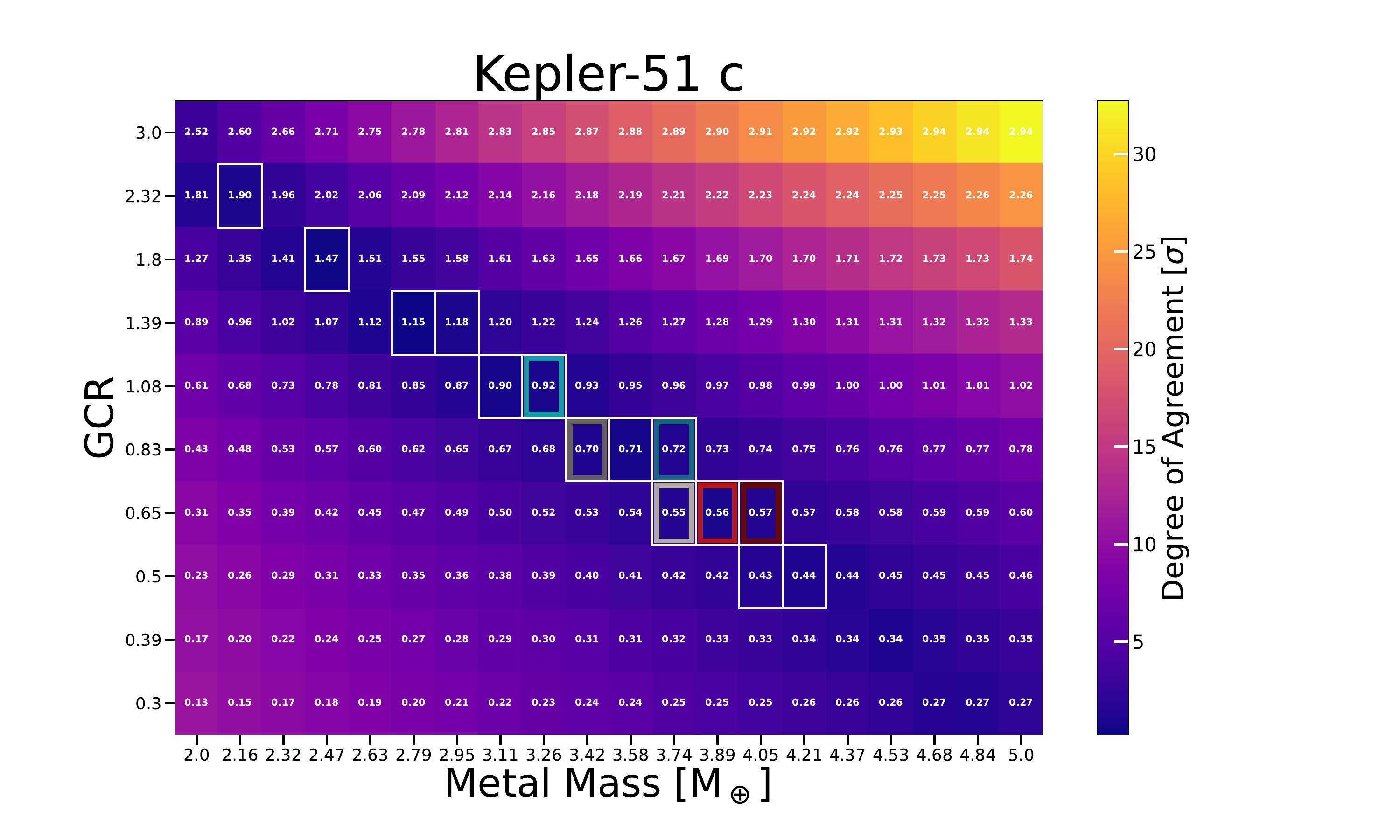}{0.5\textwidth}{}}
    \vspace{-1cm}
    \caption{Same as Figure \ref{fig:Kepler51} but at 750 Myr (upper 1-$\sigma$ limit of the estimated age).}
    \label{fig:Kepler51-ageuerr}
\end{figure*}

\begin{figure*}
    \gridline{\fig{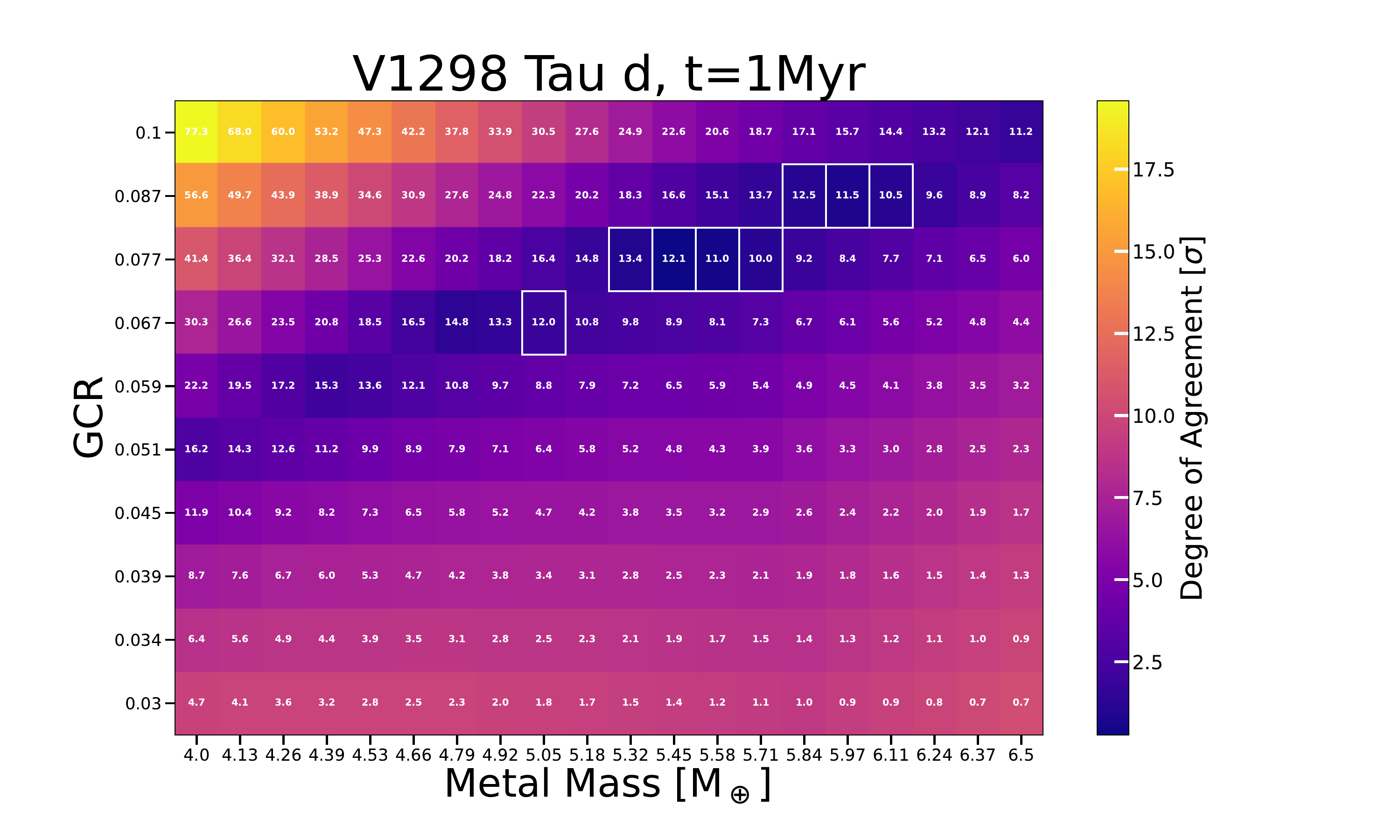}{0.5\textwidth}{}\fig{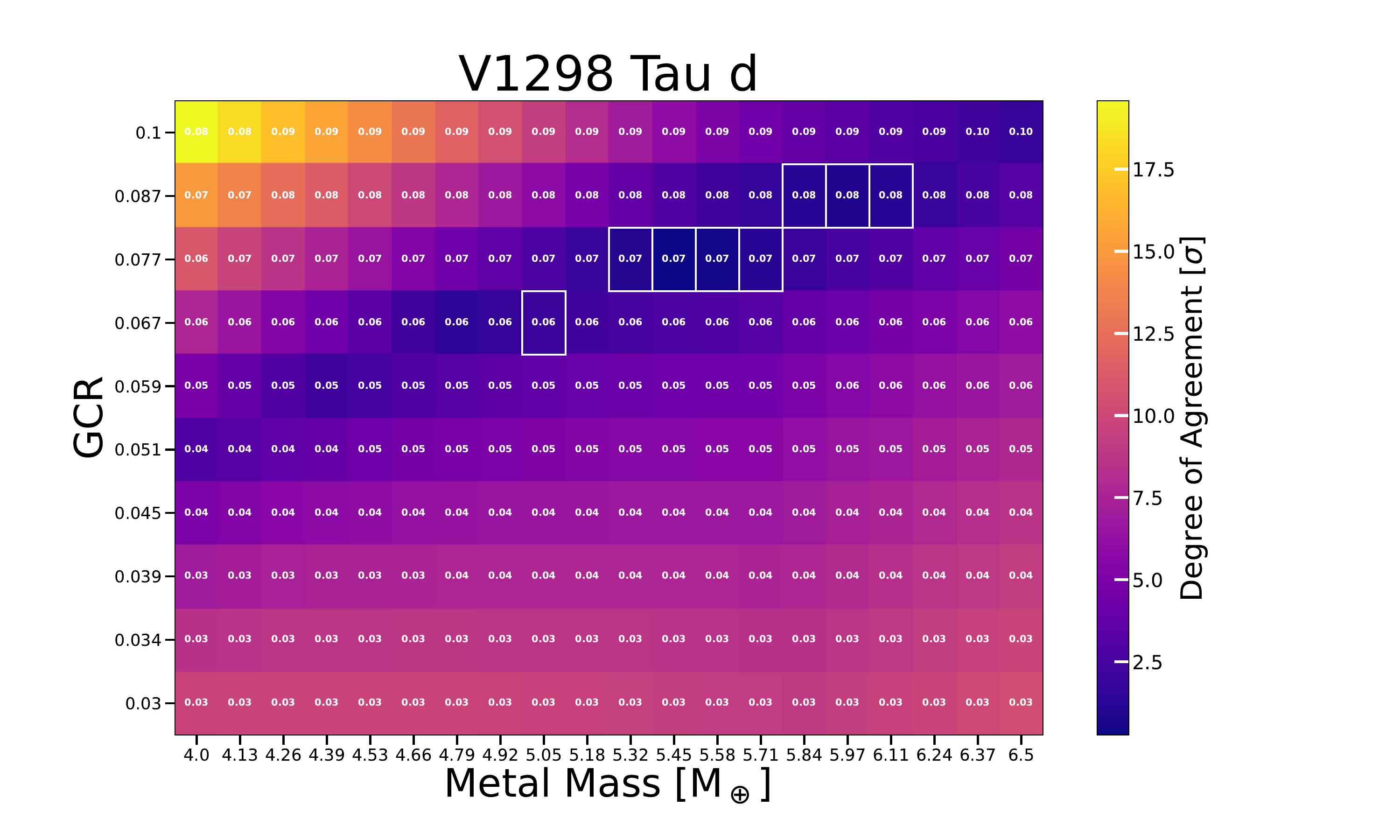}{0.5\textwidth}{}}
    \vspace{-0.9cm}
    \gridline{\fig{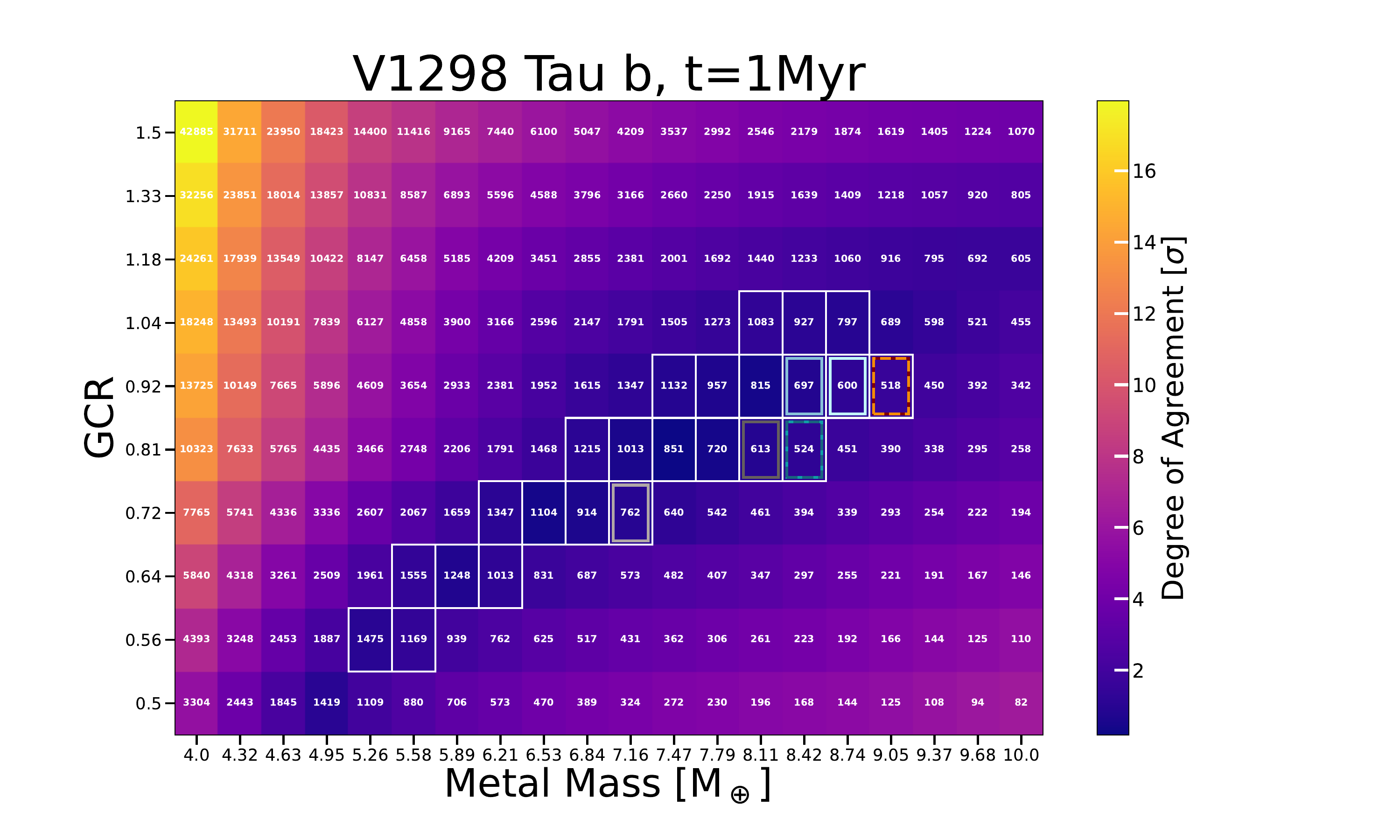}{0.5\textwidth}{}\fig{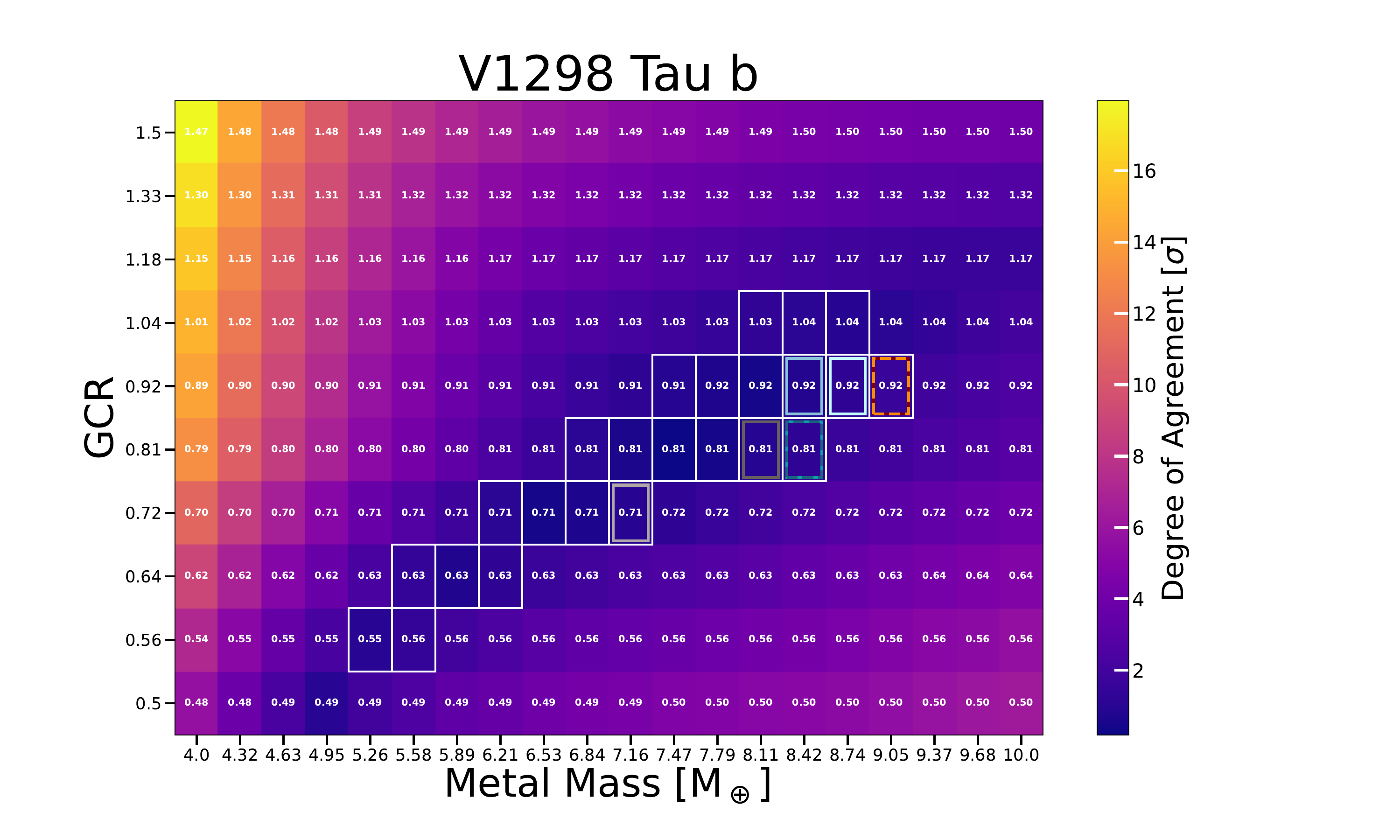}{0.5\textwidth}{}}
    \vspace{-0.9cm}
    \gridline{\fig{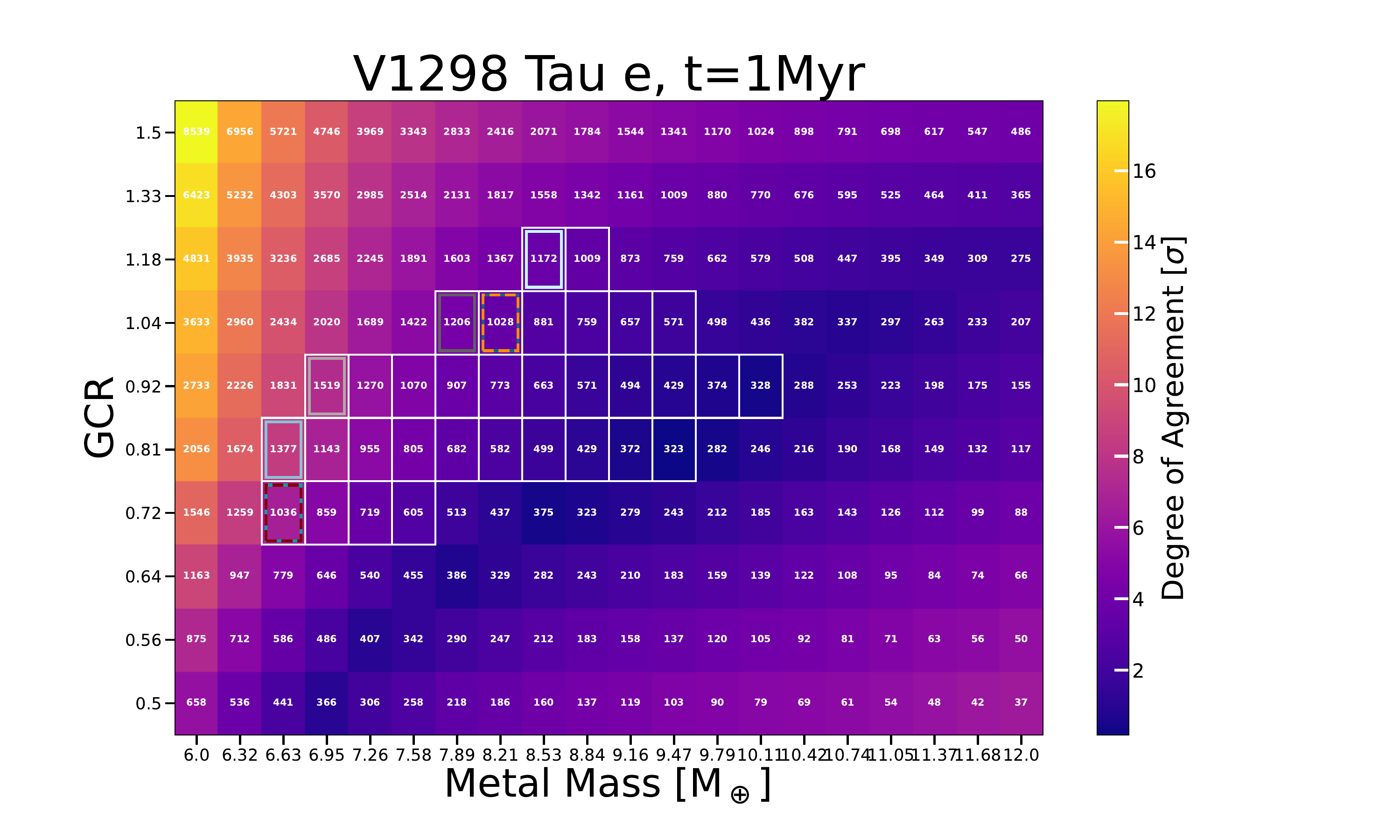}{0.5\textwidth}{}\fig{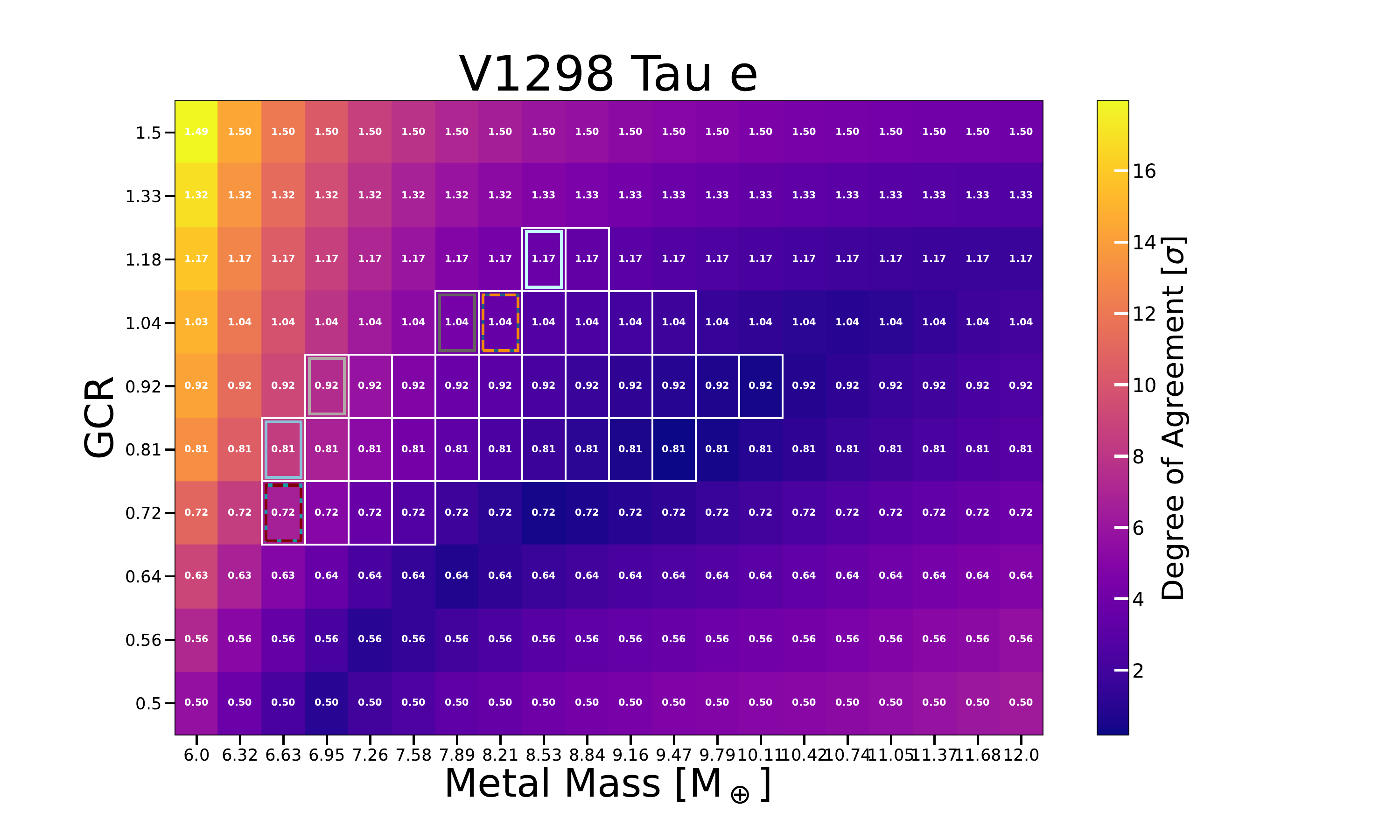}{0.5\textwidth}{}}
    \vspace{-0.9cm}
    \caption{Same as Figure \ref{fig:V1298Tau} but at 20 Myr (lower 1-$\sigma$ limit of the estimated age).}
    \label{fig:V1298Tau-agelerr}
\end{figure*}

\begin{figure*}
    \gridline{\fig{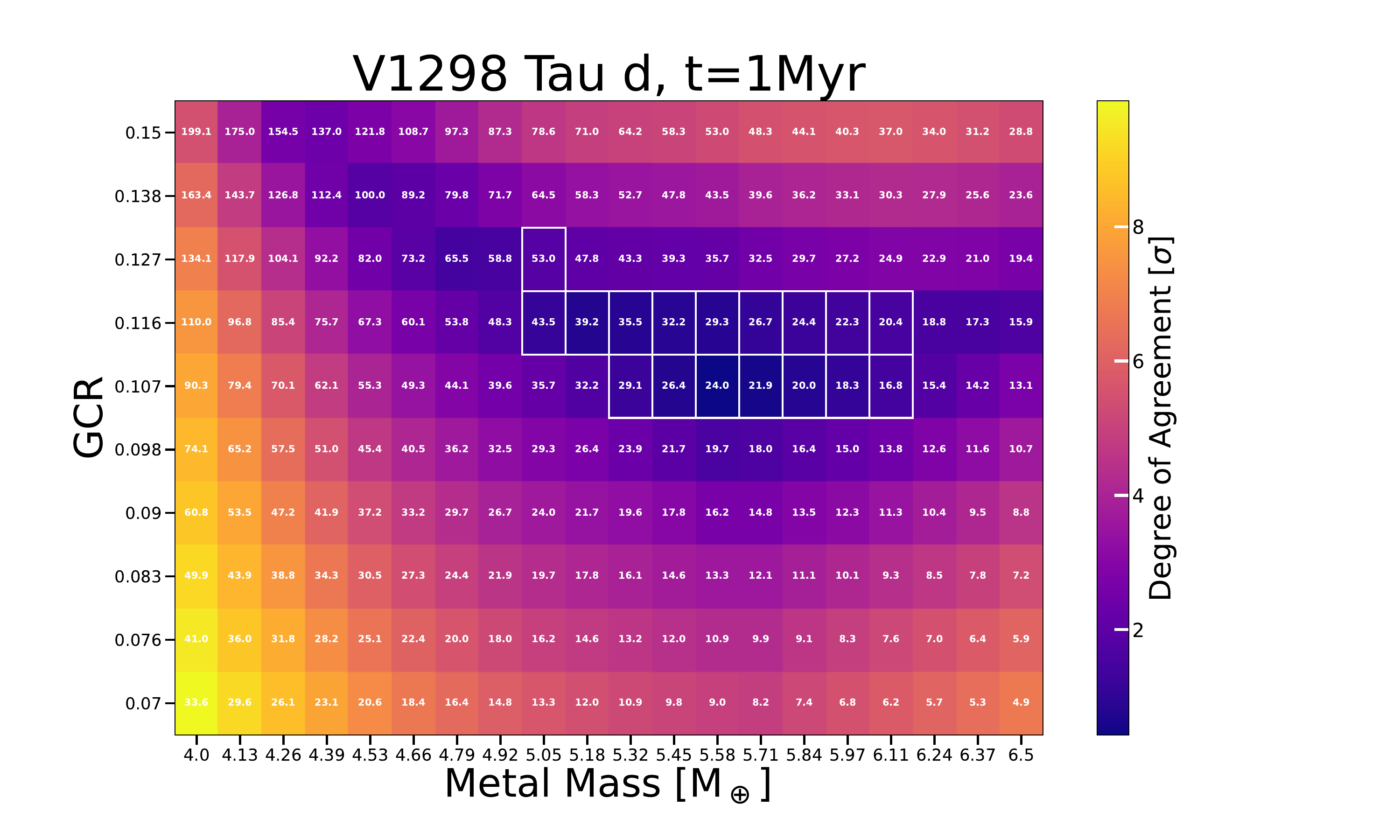}{0.5\textwidth}{}\fig{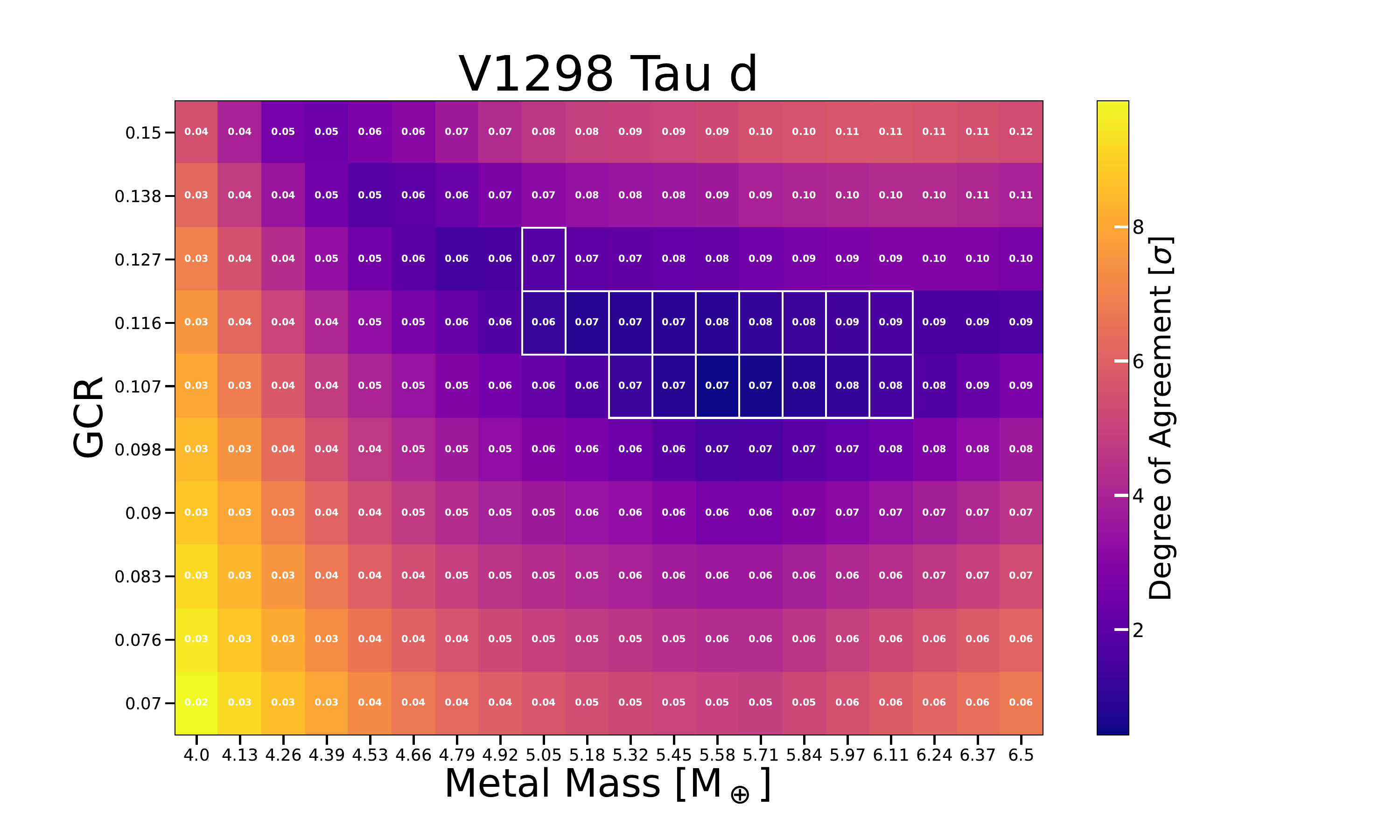}{0.5\textwidth}{}}
    \vspace{-0.9cm}
    \gridline{\fig{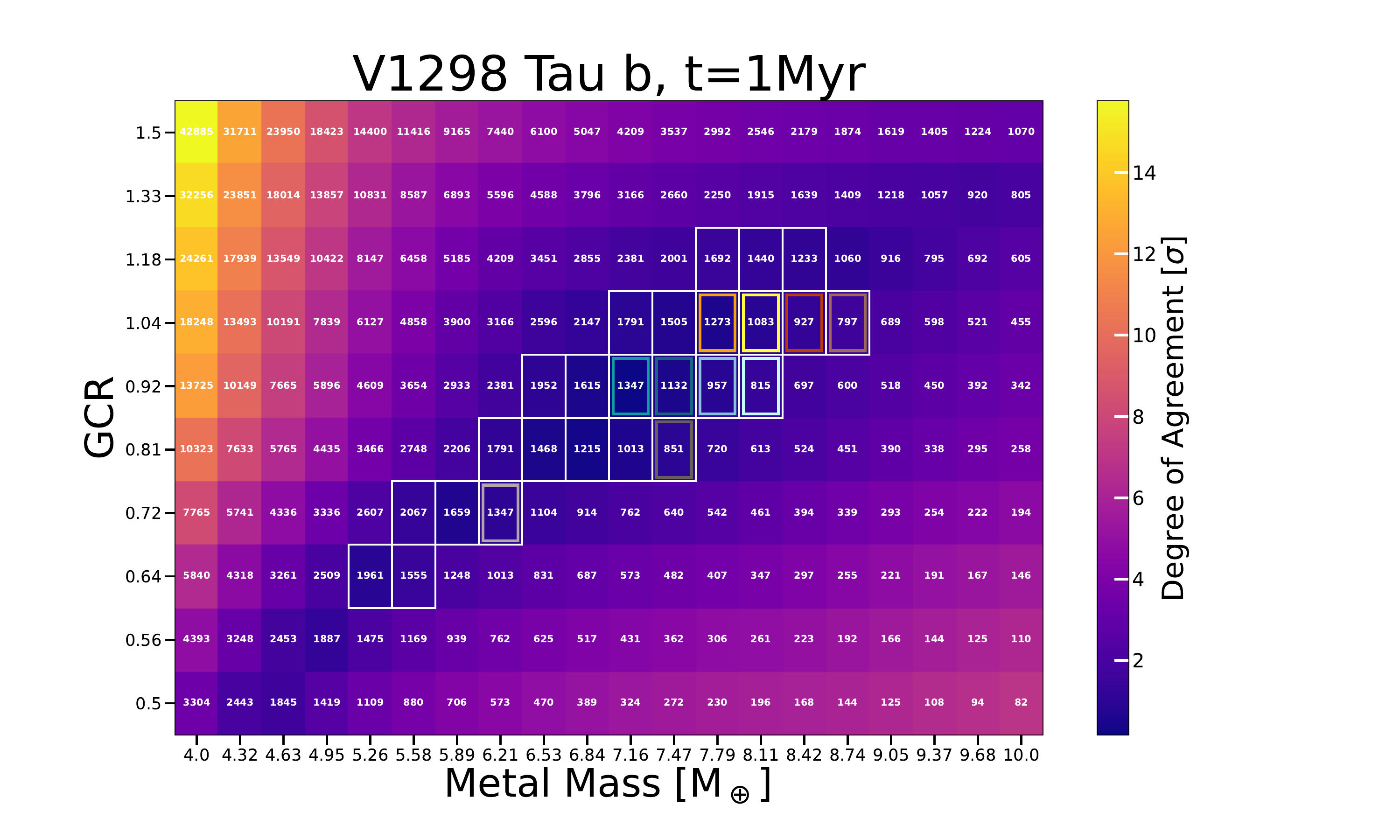}{0.5\textwidth}{}\fig{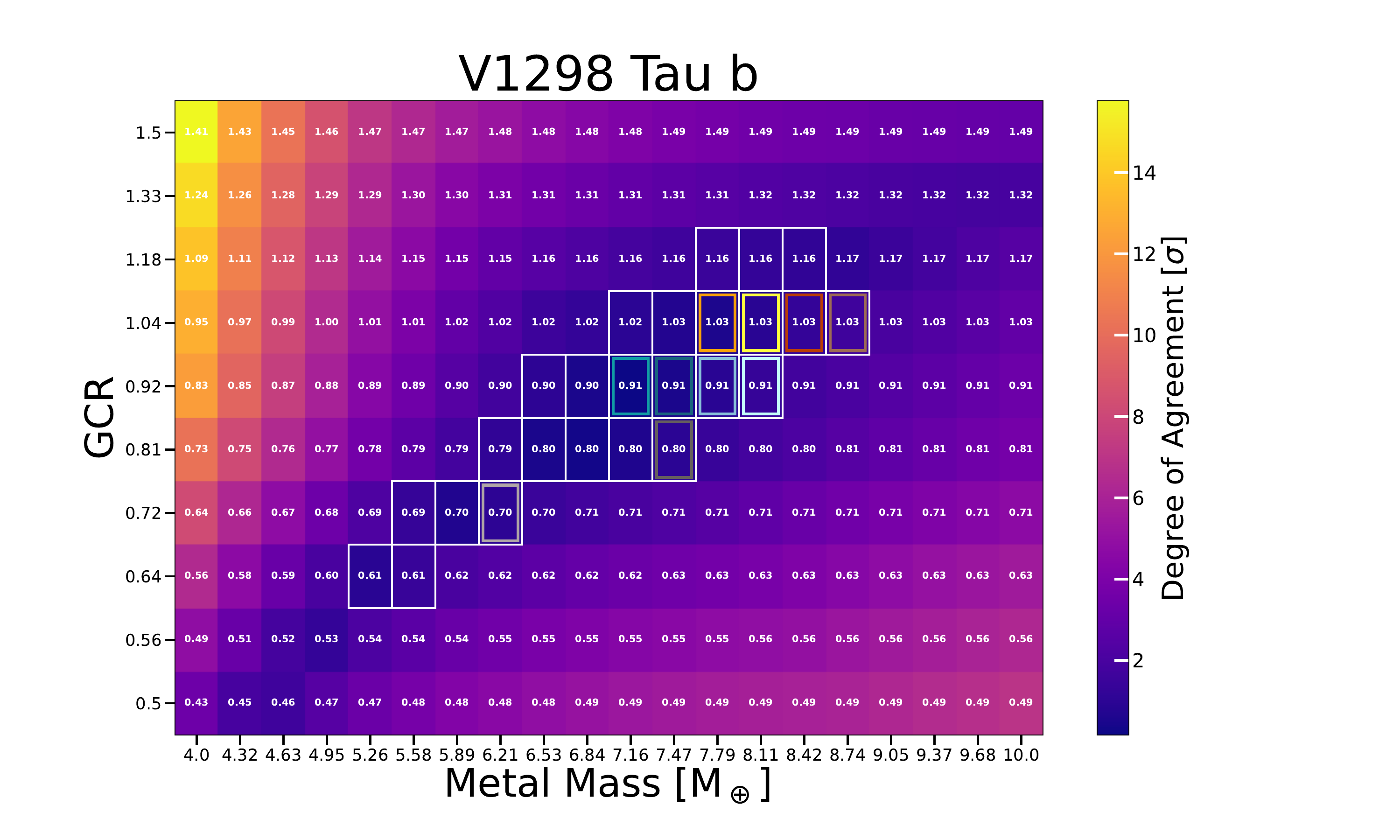}{0.5\textwidth}{}}
    \vspace{-0.9cm}
    \gridline{\fig{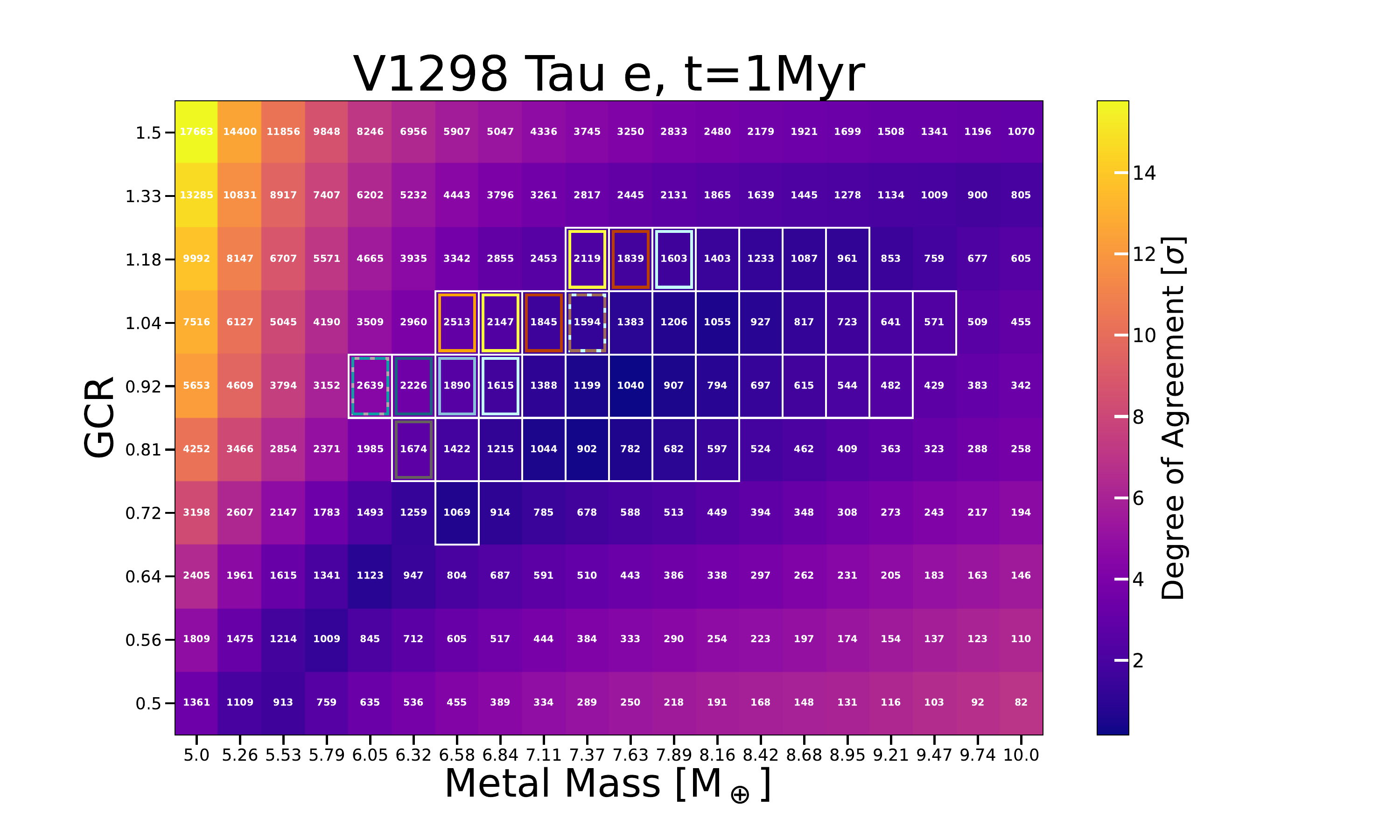}{0.5\textwidth}{}\fig{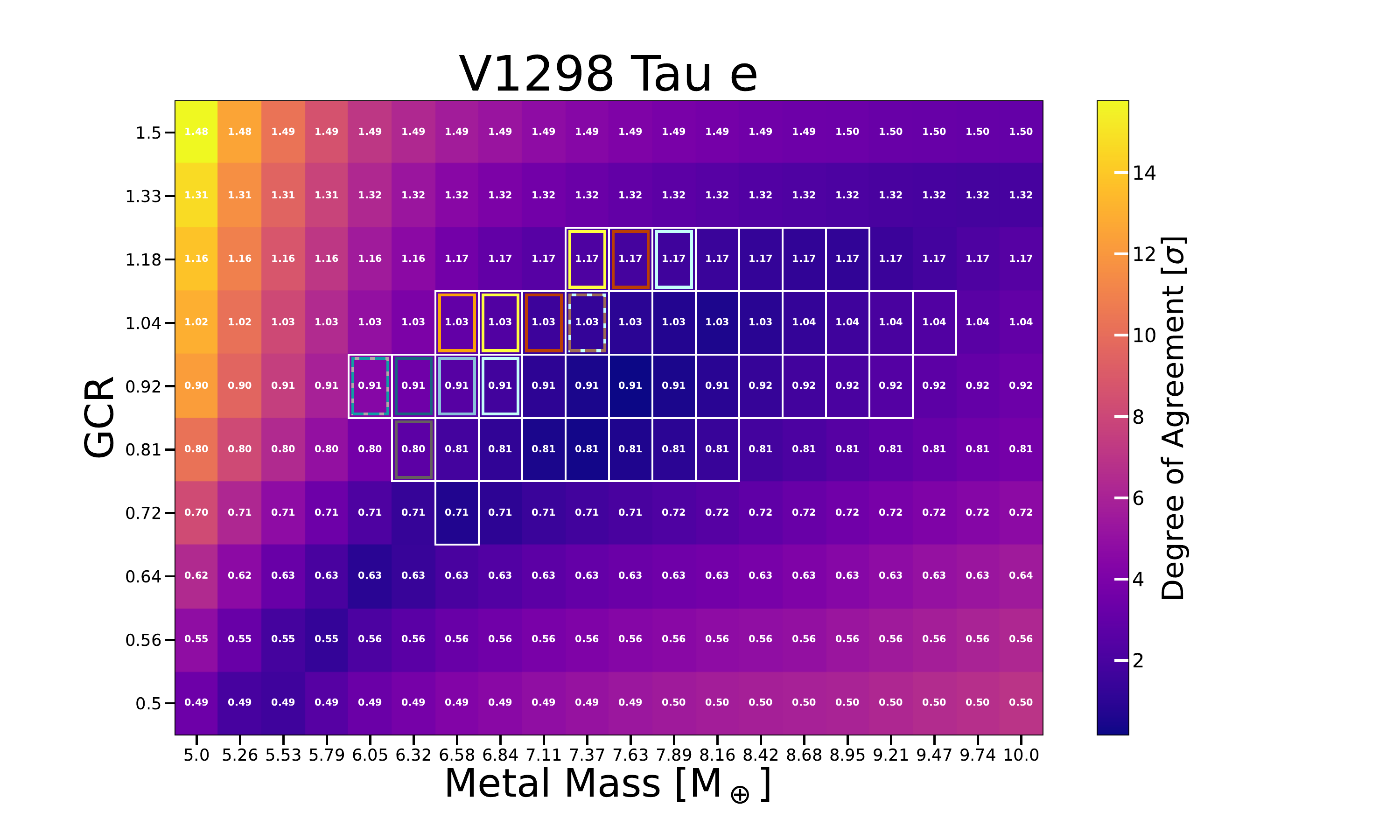}{0.5\textwidth}{}}
    \vspace{-0.9cm}
    \caption{Same as Figure \ref{fig:V1298Tau} but at 30 Myr (upper 1-$\sigma$ limit of the estimated age).}
    \label{fig:V1298Tau-ageuerr}
\end{figure*}

We show in this appendix the effect of age uncertainty in the formation-evolution solutions for Kepler-51 and V1298 Tau planets. We follow the methods described in Sections \ref{sec:Kepler51} and \ref{sec:V1298Tau} using the lower and upper 1-$\sigma$ limit on the estimated age of Kepler-51 (Figures \ref{fig:Kepler51-agelerr} and \ref{fig:Kepler51-ageuerr}) and V1298 Tau (Figure \ref{fig:V1298Tau-agelerr} and \ref{fig:V1298Tau-ageuerr}).

In general, younger age implies slightly smaller initial GCR (and therefore closer-in formation location) for a given metal mass, particularly for planets at shorter orbital periods which are more susceptible to mass loss. We also find more pronounced nonmonotonic behavior of GCR-metal mass pair for Kepler-51 b at younger age because the star is brighter accentuating the effect of extra heating puffing up the planet at lower metal mass. While the exact resonant solutions are different at different age estimates, the initial locations are all beyond 1 au for both Kepler-51 b,c and V1298 Tau b,e (V1298 Tau d still requires formation near its current location) and the initial and final masses differ by $\lesssim$10--20\% within the age uncertainty. Our conclusion of these planets having formed beyond the water ice line followed by a large-scale inward migration is robust against age uncertainty.

\bibliography{YoungPlanets}{}
\bibliographystyle{aasjournal}

\end{document}